\documentclass[conference,9pt]{IEEEtran}

\usepackage{import}
\usepackage{graphicx}
\usepackage{subcaption}
\usepackage{hyperref}
\usepackage{booktabs}
\usepackage{listings}
\usepackage{xcolor} 
\usepackage{nameref}

\definecolor{codegreen}{rgb}{0,0.6,0}
\definecolor{codegray}{rgb}{0.5,0.5,0.5}
\definecolor{codepurple}{rgb}{0.58,0,0.82}
\definecolor{backcolour}{rgb}{0.95,0.95,0.92}

\lstdefinestyle{mystyle}{
	backgroundcolor=\color{backcolour},   
	commentstyle=\color{codegreen},
	keywordstyle=\color{magenta},
	numberstyle=\tiny\color{codegray},
	stringstyle=\color{codepurple},
	basicstyle=\ttfamily\footnotesize,
	breakatwhitespace=false,         
	breaklines=true,                 
	captionpos=b,                    
	keepspaces=true,                 
	numbers=left,                    
	numbersep=5pt,                  
	showspaces=false,                
	showstringspaces=false,
	showtabs=false,                  
	tabsize=2
}
\usepackage{xspace}

\usepackage{xcolor}

\definecolor{bkgd}{RGB}{240,242,246}
\definecolor{eclipseBlue}{RGB}{42,0.0,255}
\definecolor{eclipseGreen}{RGB}{63,127,95}
\definecolor{eclipsePurple}{RGB}{127,0,85}
\definecolor{eclipseRed}{RGB}{223,26,65}
\definecolor{sublimegreen}{RGB}{112,200,0}
\definecolor{purple}{RGB}{255, 0, 255}

\definecolor{8ECFC9}{RGB}{142,207,201}
\definecolor{FFBE7A}{RGB}{255,190,122}
\definecolor{FA7F6F}{RGB}{250,127,111}
\definecolor{82BOD2}{RGB}{130,176,210}
\definecolor{0067D2}{RGB}{0,103,210}
\definecolor{codegreen}{rgb}{0,0.6,0}
\definecolor{yellow}{RGB}{195, 113, 0}

\definecolor{codegreen}{rgb}{0,0.6,0}
\definecolor{codegray}{rgb}{0.5,0.5,0.5}
\definecolor{codepurple}{rgb}{0.58,0,0.82}
\definecolor{backcolour}{rgb}{0.95,0.95,0.92}

\usepackage{amssymb}
\usepackage{amsmath}
\usepackage{multirow}
\usepackage{multicol}

\usepackage{realboxes}

\usepackage{enumitem}

\usepackage{algpseudocode}

\usepackage{braket}

\usepackage{url}

\usepackage{natbib}
\setcitestyle{numbers,square}
\usepackage{lipsum}

\usepackage{bm}

\usepackage{enumitem}

\usepackage{graphicx}
\usepackage{pythonhighlight}
\usepackage{listings}

\usepackage{tikz}
\usetikzlibrary{decorations.pathreplacing,decorations.pathmorphing}

\usepackage{siunitx}

\usepackage{parcolumns}

\usepackage{float}
\newfloat{lstfloat}{htbp}{lop}
\floatname{lstfloat}{Listing}

\input{qpanda_code_style.tex}

\begin{document}
\title{QPanda3: A High-Performance Software-Hardware Collaborative Framework for Large-Scale Quantum-Classical Computing Integration}

\author{
\IEEEauthorblockN{
    Tianrui Zou\textsuperscript{1}, Yuan Fang\textsuperscript{1}, Jing Wang\textsuperscript{1}, 
    Menghan Dou\textsuperscript{1,2}\textsuperscript{\dag}, Jun Fu\textsuperscript{1}, ZiQiang Zhao\textsuperscript{1},
    ShuBin Zhao\textsuperscript{1}, Lei Yu\textsuperscript{1}, Dongyi Zhao\textsuperscript{1} \\
    Zhaoyun Chen\textsuperscript{3}\textsuperscript{\dag}, Guoping Guo\textsuperscript{4}\textsuperscript{\dag}
}
\IEEEauthorblockA{
    \textsuperscript{1}Origin Quantum Computing Company Limited, Hefei, China \\
    \textsuperscript{2}Anhui Engineering Research Center of Quantum Computing, Hefei, China \\
    \textsuperscript{3}Institute of Artificial Intelligence (Hefei Comprehensive National Science Center), Hefei, China \\
    \textsuperscript{4}CAS Key Laboratory of Quantum Information, University of Science and Technology of China, Hefei, China\\
    \textsuperscript{\dag}Corresponding author: Menghan Dou (dmh@originqc.com), Zhaoyun Chen (chenzhaoyun@iai.ustc.edu.cn), Guoping Guo (gpguo@ustc.edu.cn)
}
}

\maketitle
\begin{abstract}
In emerging quantum–classical integration applications, the classical time cost—especially from compilation and protocol-level communication—often exceeds the execution time of quantum circuits themselves, posing a severe bottleneck to practical deployment.
To overcome these limitations, QPanda3 has been extensively optimized as a high-performance quantum programming framework tailored for the demands of the NISQ era and quantum–classical hybrid workflows.
It features optimized circuit compilation, a custom binary instruction stream (OriginBIS), and hardware-aware execution strategies to significantly reduce latency and communication overhead.
OriginBIS achieves up to 86.9$\times$ faster encoding and 35.6$\times$ faster decoding than OpenQASM 2.0, addressing critical bottlenecks in hybrid quantum systems.
Benchmarks show 10.7$\times$ compilation speedup and up to 597$\times$ acceleration in compiling large-scale circuits (e.g., a 118-qubit W-state) compared to Qiskit.
n high-performance simulation, QPanda3 excels in variational quantum algorithms, achieving up to 26$\times$ faster gradient computation than Qiskit, with minimal time-complexity growth across circuit depths.
These capabilities make QPanda3 well-suited for scalable quantum algorithm development in finance, materials science, and combinatorial optimization, while supporting industrial deployment and cloud-based execution in quantum–classical hybrid computing scenarios.
\end{abstract}
\maketitle

\begin{figure*}[htbp]
  \centering
  \includegraphics[width=0.9\textwidth]{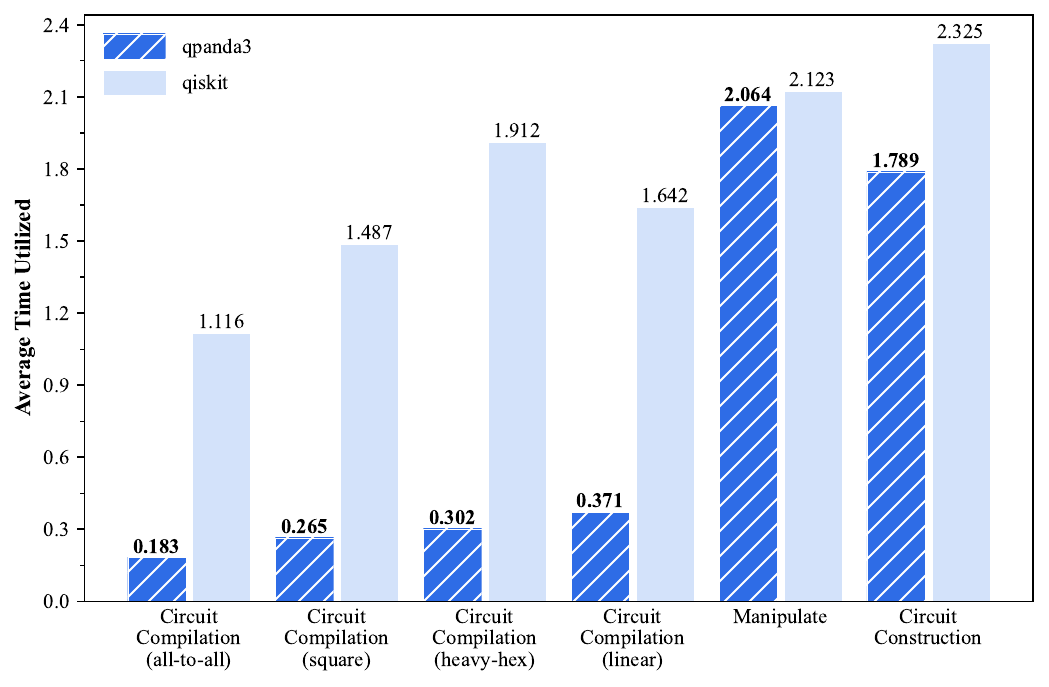}
  \caption{Comparison of average time utilized for various quantum circuit processing tasks between QPanda3 and Qiskit. 
  The tasks include circuit compilation on different topologies (all-to-all, square, heavy-hex, and linear), 
  circuit manipulation, and circuit construction. The results indicate that QPanda3 demonstrates significantly lower compilation time across all topologies compared to Qiskit,
  while the time difference in manipulation and construction tasks is relatively smaller.}
  \label{fig:circuits-construction-time}
\end{figure*}

\section{Introduction}

Quantum computing, as a transformative computational paradigm, has attracted significant attention from both academia and industry due to its exceptional parallelism and its potential to tackle problems intractable for classical computers. In recent years, substantial advances in quantum hardware have propelled the field from theoretical exploration into practical application. However, fully realizing the potential of quantum computing requires robust software frameworks that bridge the gap between end-users and quantum hardware. The effectiveness of such frameworks hinges on several critical factors, including user accessibility, computational efficiency, and seamless hardware integration. In this context, QPanda3~\cite{websiteqpanda}, a modern quantum programming framework, has undergone extensive engineering optimizations to improve usability, circuit compilation efficiency, and program transmission performance---accelerating the practical deployment of quantum computing across diverse application domains.

Quantum computing is currently undergoing a transition from theoretical exploration to engineering realization. Despite persistent hardware challenges—such as noise, decoherence, and limited qubit scalability—ongoing engineering advancements are steadily improving the usability of Noisy Intermediate-Scale Quantum (NISQ) devices. These improvements have enabled NISQ systems to demonstrate computational advantages over classical systems in specific problem domains. This evolution not only broadens the scope of near-term quantum applications but also lays the groundwork for scalable and fault-tolerant quantum computing in the future.

However, in contrast to the steady progress in hardware, most existing quantum programming frameworks exhibit a lack of engineering maturity, which poses significant barriers to the practical deployment of quantum–classical hybrid applications. In particular, the computational and communication overhead on the classical side often introduces substantial runtime latency, undermining the scheduling efficiency of quantum programs—especially in scenarios involving frequent quantum–classical interactions. These challenges are especially pronounced during quantum program compilation and in the implementation of quantum–classical communication protocols.

Many mainstream frameworks remain at a research-oriented stage, with limited attention paid to compilation efficiency, scalability for large-scale circuits, and deep integration with hardware architectures. For example, widely used plaintext-style quantum intermediate representations (QIRs), such as OpenQASM~\cite{cross2017open} and OriginIR~\cite{QPanda3-Tutorial-OriginIR}, while offering human readability, suffer from overly redundant instruction formats and lack structured expressiveness. These limitations complicate backend optimization and hinder efficient quantum–classical co-execution, exacerbating communication latency and creating scheduling bottlenecks. Moreover, plaintext QIRs introduce additional obstacles in real-world deployments, including increased parsing overhead, inefficient transmission, and security vulnerabilities. These issues collectively restrict the development of efficient, reliable, and scalable quantum compilation and execution infrastructures.

Addressing these limitations requires a fundamental rethinking of quantum programming frameworks from an engineering perspective. This includes optimizing the compilation pipeline, designing efficient intermediate representations, and enabling high-performance execution tailored to the demands of practical quantum–classical systems.

QPanda3, as a high-performance quantum programming framework, adopts an engineering-centric design philosophy aimed at enhancing the practical usability of quantum computing systems. Its core objectives include:

\begin{itemize}
    \item \textbf{Superior Compilation Performance:} \\
    QPanda3 employs a customized data structure tailored to the quantum compilation workflow, effectively reducing intermediate-stage information redundancy in instruction reordering, topology-constrained mapping, and gate fusion optimization. This design enables efficient and scalable compilation for quantum circuits at the scale of thousands of qubits. Benchmark results from Benchpress~\cite{nation2024benchmarking} demonstrate that, in fully connected topologies, QPanda3 achieves an average compilation speed 7.03$\times$ faster than Qiskit~\cite{websiteqiskit}, with peak speedups up to 123.33$\times$. For square, heavy-hexagon, and linear topologies, average and peak speedups reach 15.87$\times$/597.41$\times$, 12.45$\times$/313.86$\times$, and 7.43$\times$/374.48$\times$, respectively.

    \item \textbf{Quantum-Classical Protocol Optimization:} \\
    By introducing a concise and efficient binary stream format to represent quantum programs, QPanda3 significantly improves encoding and decoding speeds as well as compression ratios compared to traditional plaintext intermediate representations. Specifically, the encoding speed of OriginBIS surpasses that of OriginIR and OpenQASM 2.0 by factors of 18.7$\times$ and 86.9$\times$, respectively; decoding speed is enhanced by 18.9$\times$ and 35.6$\times$. This advancement effectively reduces communication latency in quantum hyper-converged applications.

    \item \textbf{Fast Simulation of Hybrid Quantum Programs:} \\
    QPanda3 provides modular support for efficient simulation of variational quantum circuits and quantum-classical hybrid algorithms, addressing the rapid iteration and feedback-driven development characteristic of the NISQ era. Comparative experiments with Qiskit, PennyLane~\cite{bergholm2018pennylane}, and DeepQuantum~\cite{webdeepquantum} confirm that QPanda3 achieves the shortest gradient computation times and overall superior performance.

    \item \textbf{Quantum Software and Hardware Profiling:} \\
    The framework incorporates a comprehensive performance analysis toolchain supporting benchmarking, calibration, and cross-platform performance comparison of both quantum software and underlying hardware, facilitating the validation and advancement of emerging quantum technologies.
\end{itemize}

The remainder of this paper is organized as follows: Section~2 reviews related work on quantum compilation frameworks. Section~3 outlines the architecture and key features of QPanda3. 
Section~4 presents experimental validation of QPanda3’s performance advantages. In addition, the Appendix provides a modular breakdown of QPanda3’s internal components along with illustrative code 
examples, enabling readers to gain a deeper understanding of its usage and extensibility in practical quantum programming scenarios.

\section{Related Work}

\subsection{Intermediate Representation of Quantum Programs}

The Intermediate Representation (IR) of quantum programs serves as a crucial 
bridge connecting high-level quantum algorithm descriptions with underlying 
hardware implementations in quantum computing. IR provides descriptions of 
quantum programs at various levels of abstraction, each suited for the design, 
processing, transmission, and execution of quantum programs. Lower-level 
abstractions tend to be closer to the machine instruction format of quantum 
computing devices, while higher-level abstractions are more easily understood 
by humans. Representing quantum programs based on the quantum circuit model is 
a common approach, and most IRs of quantum programs are built upon this 
foundation. Even for IRs of quantum programs that include classical instructions
 and complex control flows, they are typically extensions based on the quantum 
 circuit model.

Quantum Assembly Language (QASM)\cite{svore2006layered} is an important form of 
intermediate representation for quantum programs, primarily used to provide a 
low-level abstraction of quantum programs. It is important to note that QASM 
for quantum computing differs significantly from assembly languages in 
classical computer architectures. Although QASM is a low-level representation, 
it still requires the device to dynamically manage quantum registers and 
classical registers (logical-to-physical mapping) rather than directly 
operating on hardware registers.

Intermediate representations of quantum programs that resemble high-level 
programming languages offer good user readability. However, such representations 
tend to be relatively complex. For example, with the increasing complexity of 
quantum computing systems and application demands, IBM has continuously 
extended QASM to develop OpenQASM 2.0\cite{cross2017open} and OpenQASM 3.0
\cite{cross2022openqasm}. As widely adopted versions, OpenQASM 2.0 and OpenQASM 3.0 
not only extend quantum gate operations and qubit 
management but also introduce more complex syntactic structures and functional 
features, such as classical control flow and modular programming support. They 
are more akin to the style of the classical programming language C compared to 
the original QASM. F-QASM\cite{liu2018q} extends QASM with feedback instructions
, enhancing the efficiency of implementing measurement-based branch and loop 
statements. Similar intermediate representations include Scaffold
\cite{javadiabhari2015scaffcc}, QCL\cite{websiteqcl}, Quipper
\cite{green2013quipper}, ProjectQ\cite{steiger2018projectq}, and 
QIR\cite{websiteqsharp}. These types of intermediate representations for 
quantum programs sacrifice portability due to their complexity, hindering 
comparative experimental research by researchers.

Device-oriented intermediate representations of quantum programs exhibit strong 
dependencies on the specific device. OpenPulse\cite{mckay2018qiskit}, Pulser
\cite{silverio2022pulser}, and JaqalPaw\cite{lobser2023jaqalpaw} are examples of 
intermediate representations tailored for pulse-based devices, offering 
fine-grained control over quantum devices. QuMIs\cite{fu2017experimental} is an 
intermediate representation designed for distributed device control. eQASM
\cite{fu2019eqasm}, on the other hand, is a low-level intermediate 
representation that is directly linked to binary machine instructions. Clearly, 
these intermediate representations contain a significant amount of 
device-specific information within the code describing quantum programs, which 
provides precise control over the execution details of quantum programs but 
also reduces their device independence.

The intermediate representation of quantum programs can also serve as a 
transmission protocol for transferring quantum programs between devices. 
Currently, there is relatively little research discussing this topic. It is 
important to note that this type of protocol differs from NetQASM
\cite{dahlberg2022netqasm} and InQuIR\cite{nishio2023inquir}. NetQASM and 
InQuIR are quantum network-oriented intermediate representations based on 
quantum communication protocols, used to control the devices and processes 
involved in quantum communication.

\subsection{Compilation, Optimization, Qubit Mapping, and Routing of Quantum 
Circuits}

Quantum logic circuits can represent the quantum computing component of 
quantum programs. The quantum circuit model can also depict the actual 
physical operation steps in a quantum processor. The quantum software stack 
provides high-level programming languages for designing the former. The 
latter typically corresponds to a sequence of machine instructions that 
directly operate the quantum device. Quantum circuit compilation is the 
process of converting quantum logic circuits into sequences of machine 
instructions. The transformation from an initial quantum logic circuit to a 
sequence of machine instructions involves multiple structural conversions of 
the quantum circuit. These conversions include steps such as unitary matrix 
decomposition, qubit mapping, qubit routing, optimization, and compilation into 
a sequence of machine instructions.

Quantum computing software such as Qiskit and QASMTrans\cite{hua2023qasmtrans} 
provides modern support for quantum circuit compilation. Paulihedral
\cite{li2022paulihedral} is a compiler designed specifically for the VQE 
algorithm. Similarly, application-oriented compilers also include Twoqan
\cite{lao20222qan}, which is dedicated to QAOA circuits. CaQR\cite{chen2023pulse}
 focuses on the 
generation of dynamic circuits, is a compiler specifically 
for pulse-based devices. AutoComm\cite{wu2022autocomm} and QuComm
\cite{wu2022collcomm}, on the other hand, provide support for distributed 
devices.

Qubit mapping and routing are essential steps for quantum logic circuits to be 
executed by quantum processors. Qubit mapping assigns corresponding physical 
qubit resources to each logical qubit, while qubit routing ensures the 
connectivity constraints between physical qubits. Optimization aims to obtain 
compiled circuits with good computational performance. Sabre
\cite{li2019tackling} proposes a method to minimize the number of ancillary 
qubits. TOQM\cite{zhang2021time}, on the other hand, aims to reduce the depth 
of the translated circuits\cite{shi2019optimized}, among others, investigates 
optimization schemes related to quantum gate aggregation.

\subsection{Quantum Program Profiling}

Quantum program profiling is of great significance for fully harnessing the 
potential of quantum computing. With the development of quantum software, 
quantum programs often contain not only pure quantum gate operations but also 
other classical instruction execution processes. Therefore, quantum program 
profiling involves the analysis of quantum circuits and operational 
subprocesses.

Quantum computing performance analysis benchmarks can be utilized for quantum 
circuit analysis. Currently, numerous proposed benchmarks are employed for 
quantum computing performance analysis, primarily focusing on evaluating the 
performance of quantum processors when executing specific circuits. These 
metrics include quantum volume\cite{cross2019validating}, Q-score
\cite{martiel2021benchmarking}, quantum LINPACK\cite{dong2021random}, and 
quantum process tomography\cite{chuang1997prescription}, as well as using 
quantum applications like VQE\cite{peruzzo2014variational} and QAOA
\cite{farhi2014quantum} to analyze the performance of quantum processors. 
SupermarQ further proposes multiple analysis benchmarks. Clearly, these 
benchmarks also reflect the performance of the quantum circuits used for 
testing on the execution device. The benchmarks proposed by SupermarQ
\cite{tomesh2022supermarq}, such as inter-qubit communication volume, critical 
depth, coherence rate, parallelism, and qubit activity, can be calculated based 
on available device information and quantum circuit structure, and thus can 
also be applied to device-performance-oriented quantum circuit analysis.

Ideas from classical program analysis can be applied to analyze the operational 
processes in quantum programs. Qprof\cite{suau2022qprof}, based on gprof
\cite{websitegprof}, can be used for quantum program analysis to obtain metrics 
such as subprocess invocation rates, qubit occupancy rates, and time 
consumption. In this paper, QPanda3 is introduced, which applies process 
analysis methods to the interconnectivity analysis of quantum logic gates and 
qubit utilization analysis.

\subsection{Variational Quantum Circuit}

Variational quantum algorithms \cite{cerezo2021variational} and quantum machine 
learning \cite{biamonte2017quantum}, as two important research fields, represent 
two key directions and milestones in the practical application of quantum computing.  
The Variational Quantum Eigensolver (VQE) has been widely applied in quantum chemistry, 
condensed matter physics, and optimization problem-solving. The Quantum Approximate 
Optimization Algorithm (QAOA) \cite{farhi2014quantum} can be employed for combinatorial 
optimization problems such as the Max-Cut problem \cite{goemans1995improved}, the Traveling 
Salesman Problem, portfolio optimization, and scheduling. The Variational Quantum 
Classifier \cite{schuld2020circuit} is a significant application of variational quantum 
algorithms in the field of machine learning. Meanwhile, other quantum machine learning 
techniques, such as quantum neural networks, may also utilize non-variational quantum 
algorithms. In quantum computing, most of these critical algorithms rely on parameterized 
quantum circuits. Therefore, parameterized quantum circuits serve as essential components 
for the practical implementation of quantum computing.

Variational quantum algorithms adopt the term Ansatz \cite{borin2020approximating}. In 
algorithms such as VQE \cite{peruzzo2014variational}, QAOA, and VQC, the term Ansatz can 
unambiguously refer to a parameterized quantum circuit while also implying the optimization 
of parameters to solve a target problem. However, Ansatz models that are not based on quantum 
circuits also exist. QPanda2 \cite{dou2022qpanda} explicitly uses variational quantum circuit 
to refer to parameterized quantum circuits employed in variational algorithms \cite{QPanda2-Var}. 
It is worth noting that parameterized quantum circuits designed for variational algorithms can 
also be repurposed as general parameterized circuits and applied in quantum machine learning 
algorithms that do not rely on variational methods. QPanda3 retains the term *variational 
quantum circuit* from QPanda2, with parameterized quantum circuits at its core. It provides 
functionalities such as quantum circuit generation, Hamiltonian expectation value calculation, 
and gradient computation, thereby supporting the use of parameterized quantum circuits in both 
variational quantum algorithms and quantum machine learning algorithms. This differs from 
frameworks like Qiskit \cite{websiteqiskit}, PennyLane \cite{bergholm2018pennylane}, and 
MindQuantum \cite{xu2024mindspore}, which do not explicitly distinguish between standard 
quantum circuits (without tunable parameters) and parameterized quantum circuits. In contexts 
related to QPanda3, where no ambiguity arises, the terms Ansatz, parameterized quantum circuit, 
and variational quantum circuit are used interchangeably without explicit distinction.

QPanda3 is a foundational quantum programming framework designed for cross-disciplinary research. 
While quantum machine learning-focused frameworks such as PennyLane, MindQuantum, and 
pyVQNet \cite{bian2023vqnet} have been optimized for various aspects of machine learning, they often 
struggle to support quantum computing tasks in other domains. A similar limitation exists with 
ChemiQ \cite{wang2021chemiq}, which provides comprehensive support for quantum chemistry but is 
specialized for that field. QPanda3 maintains good compatibility with pyVQNet and ChemiQ. However, 
unlike these frameworks-which incorporate extensive application-specific features tailored to their 
respective domains-QPanda3's variational quantum circuits do not extend many application-dependent 
functionalities. Instead, they focus on providing efficient implementations of essential core operations. 
By offering more fundamental building blocks, QPanda3 enables researchers to use a single software 
platform to customize algorithmic implementations across different application areas, thereby facilitating 
cross-disciplinary research. These essential core functionalities include: Constructing parameterized 
quantum circuits, Obtaining parameterized quantum states for given parameter values, Reusing variational 
quantum circuit structures for modular programming, Computing Hamiltonian expectation values, Calculating 
gradients, and Supporting expressions as parameters in quantum gates. This design philosophy allows 
QPanda3 to serve as a versatile and adaptable tool for researchers exploring diverse quantum computing 
applications.

QPanda3 is a quantum programming framework designed for high-performance computing. While parameterized 
quantum circuits can be implemented by combining function definitions in programming languages with the 
general quantum circuit construction mechanisms provided by programming frameworks, this approach imposes 
limitations on certain high-performance extensions. For instance, when computing Hamiltonian expectation 
values and gradients, more efficient implementations exist on classical computers-some of which rely on 
direct manipulation of quantum states stored in classical memory \cite{jones2020efficient}. Currently, 
quantum programming SDKs such as Qiskit, PennyLane, pyQuil, MindQuantum, and QPanda all offer Python-based 
interfaces. However, constructing parameterized circuits using Python functions and generic circuit-building 
mechanisms prevents fine-grained control over memory operations, introduces redundant instructions, and 
ultimately restricts performance improvements in variational quantum circuit functionalities. To address this, 
QPanda3 provides dedicated data structures and processing logic specifically optimized for variational quantum 
circuits, aiming to enhance the performance of variational quantum computing tasks.
\section{Overview of QPanda3}

\begin{figure*}[htbp]
    \centering
    \includegraphics[width=0.6\textwidth]{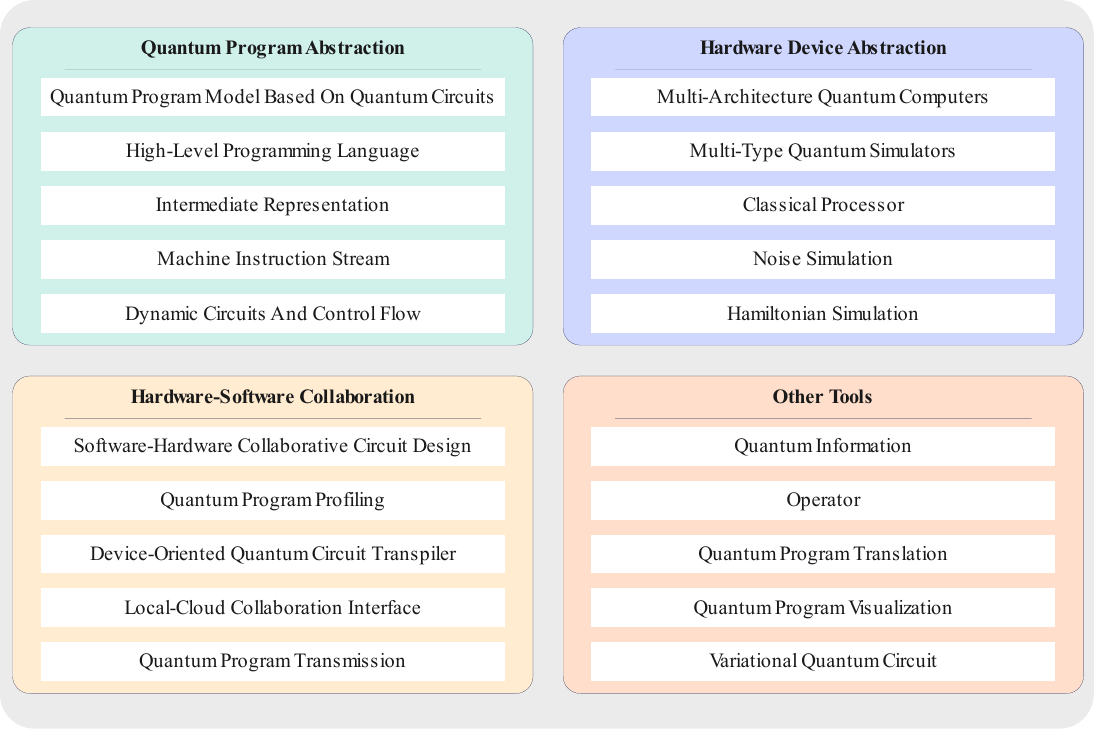}
    \caption{Architecture of the QPanda3 Framework}
    \label{fig:framework}
\end{figure*}

QPanda3 is an advanced quantum programming framework that provides comprehensive support for quantum computing across both software and hardware domains. As illustrated in Figure~\ref{fig:framework}, QPanda3 offers abstractions for quantum programs and computing devices, along with a wide array of supporting components and tools.

\subsection{Quantum Computing Orientation}

At the core of quantum software lies the quantum program, for which QPanda3 provides multi-level abstractions. Specifically, QPanda3 adopts a quantum gate-based circuit model, distinguishing between circuits with and without measurements. This distinction emphasizes the differing roles of measurement operations and quantum gate operations and is directly embodied in the \texttt{QProg} and \texttt{QCircuit} structures provided by QPanda3's high-level programming interface.

Quantum programs are expressed as instruction sequences in Python, allowing users to design quantum circuits using QPanda3’s APIs. While Python enhances usability, it introduces complexities in hardware instruction mapping and transmission inefficiency. QPanda3 addresses these issues by introducing a streamlined intermediate representation (IR) that supports efficient encoding and transmission, and by compiling programs into hardware-level instruction formats.

For hardware abstraction, QPanda3 encompasses real quantum processors, quantum simulators, and classical computing units, enabling users to leverage diverse resources. Additionally, it supports noise modeling and Hamiltonian simulation to facilitate physical-system-level experimentation.

Surrounding the quantum program, QPanda3 incorporates a rich set of components designed to enable hardware–software co-design. These include modules for circuit optimization and compilation, device-specific quantum program analysis, and cloud-based execution interfaces. Other auxiliary tools include quantum program translation for portability, visualization tools for circuit design, and a variational circuit module for efficient circuit generation and post-processing.

Variational quantum circuits—key elements of quantum-classical hybrid computing—are also natively supported in QPanda3.

To further support quantum development, QPanda3 provides tools for quantum state and channel representation, operator simulation, and testing. These include utilities for state transformation, information-theoretic analysis, and random circuit generation, catering to diverse testing and benchmarking needs.

\subsection{Advancements from QPanda2}

QPanda3 builds on the successful design principles of QPanda2. It continues to support Python, offering a concise syntax and rich ecosystem, thereby lowering the entry barrier and accelerating development—particularly for interdisciplinary researchers.

A key innovation in QPanda3 is the introduction of OriginIR, an efficient intermediate representation, along with conversion tools for interoperability with OpenQASM. This enables seamless migration across platforms and supports program optimization and analysis. The framework also supports multiple simulator types, noise and Hamiltonian simulations, and classical register management, thereby offering a robust hardware abstraction layer.

To ease the transition from QPanda2, QPanda3 maintains compatibility with its circuit construction paradigm, allowing users to reuse familiar interfaces while benefiting from enhanced performance and functionality.

\subsection{End-to-End Software-Hardware Co-Design in QPanda3}

\begin{figure*}[htbp]
\centering
\includegraphics[width=0.6\textwidth]{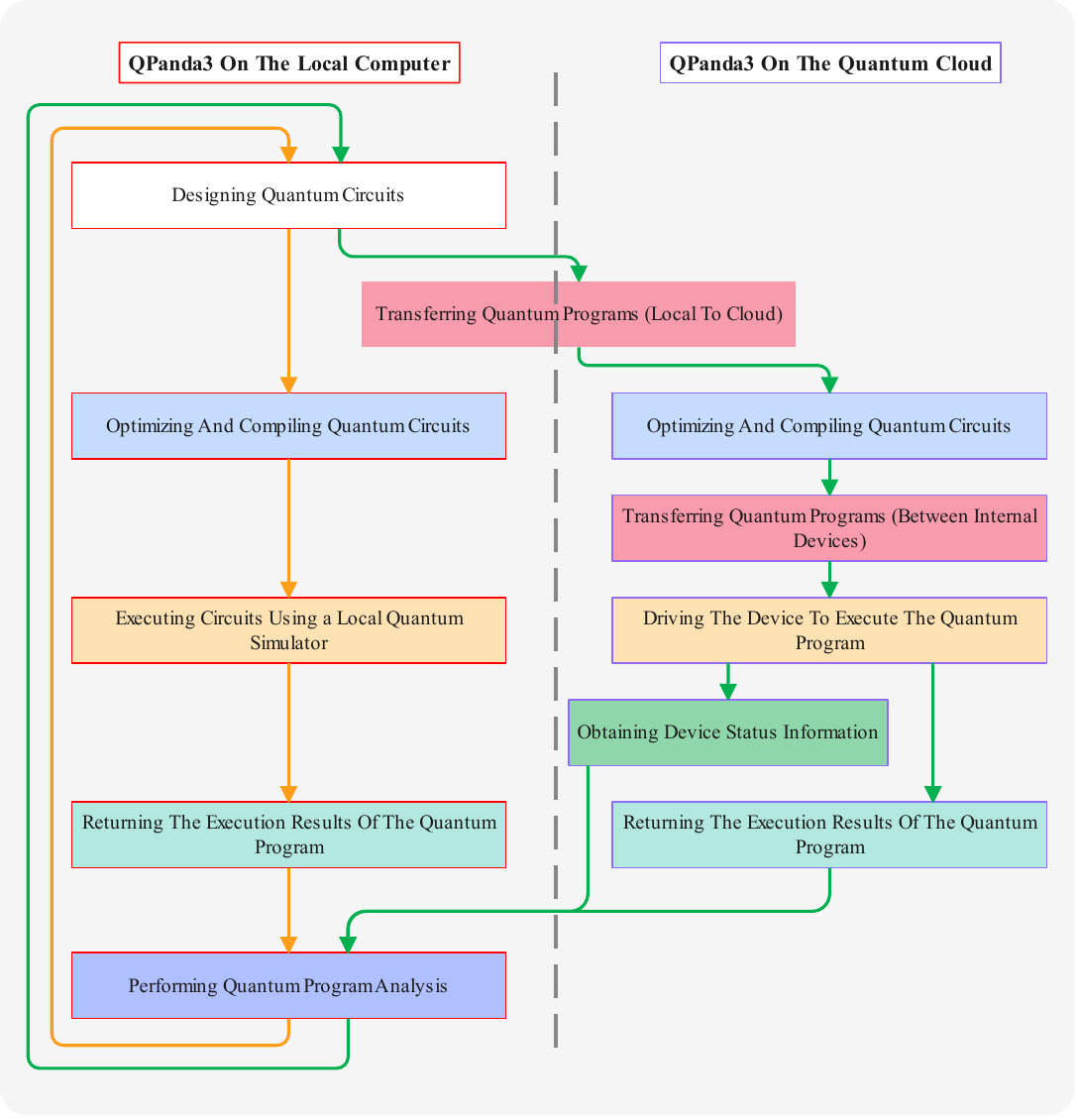}
\caption{Workflow of QPanda3 emphasizing software-hardware co-design. The framework realizes an integrated closed-loop encompassing program construction, cloud-based transmission, device-oriented compilation, and adaptive performance analysis.}
\label{fig:software-hardware}
\end{figure*}

To address the various challenges encountered during the execution of quantum algorithms on real hardware, QPanda3 adopts a software-hardware co-design paradigm that spans the entire workflow including program design, transmission, compilation, analysis, and execution. Figure~\ref{fig:software-hardware} illustrates the framework's full-stack integration process in both local and cloud environments.

\subsubsection{Cloud Execution and Cross-Device Program Transmission}

With the increasing adoption of the ``local design, cloud execution'' paradigm in quantum computing, QPanda3 offers the following capabilities to support this scenario:

\begin{itemize}
    \item Employs a compact intermediate representation (IR) that supports efficient, compressed, and encrypted transmission;
    \item Provides APIs and CLI tools to enable seamless interoperability with cloud backend systems;
    \item Supports device introspection interfaces capable of retrieving target quantum device topology, calibration data, and error rates.
\end{itemize}

These capabilities empower users to efficiently deploy quantum programs across heterogeneous platforms, thereby enabling truly cloud-native quantum computing operations.

\subsubsection{Hardware-Oriented Collaborative Optimization Compilation Mechanism}

QPanda3 utilizes a dynamic feedback-driven compilation architecture that constructs quantitative evaluation models based on real-time hardware performance metrics (e.g., gate fidelity, decoherence times, crosstalk noise) to drive multi-level compilation optimizations. The core innovation lies in establishing a closed-loop optimization system of compilation–execution–feedback:

\paragraph{Hardware-Aware Real-Time Tuning Mechanism}
\begin{itemize}
    \item Dynamically monitors chip topology, gate error rates, and other parameters at runtime to generate adaptive mapping strategies;
    \item Supports dynamic instruction set redirection for heterogeneous quantum chips (e.g., superconducting, trapped ion).
\end{itemize}

\paragraph{Performance-Driven Layered Compilation}
\begin{itemize}
    \item \textbf{Physical Layer:} Gate decomposition strategies dynamically adjusted based on chip T1/T2 parameters, optimizing parallel gate scheduling;
    \item \textbf{Logical Layer:} Incorporates backend feedback on chip information to optimize layout and routing;
    \item \textbf{System Layer:} Orchestrates pipeline scheduling of hybrid classical-quantum instructions to enhance quantum resource utilization.
\end{itemize}

\paragraph{Heterogeneous Computing Adaptation Framework}
\begin{itemize}
    \item Constructs a unified intermediate representation layer (QIR) supporting cross-platform instruction translation;
    \item Continuously updates a chip characteristics database (QCDB) including gate sets, topology, and noise models;
    \item Employs adaptive backend code generators optimized respectively for control timing on superconducting and trapped ion chips.
\end{itemize}

\subsubsection{Quantum Program Analysis and Optimization}

QPanda3 provides a comprehensive quantum program analysis toolchain supporting multi-dimensional program characteristic evaluation and optimization from the logical layer down to the physical layer. The main analysis functions include:

\paragraph{Program Structure Analysis}
\begin{itemize}
    \item Quantum gate-level dependency analysis;
    \item Identification of parallelism and critical paths;
    \item Control flow analysis of quantum-classical hybrid programs.
\end{itemize}

\paragraph{Performance Evaluation}
\begin{itemize}
    \item Supports fidelity evaluation methods such as cross-entropy benchmarking (XEB) and randomized benchmarking;
    \item Complexity analysis including quantum circuit depth and gate count.
\end{itemize}

\paragraph{Resource Bottleneck Analysis}
\begin{itemize}
    \item Statistics on gate-level parallelism and qubit utilization;
    \item Detection of routing conflicts under qubit topology constraints;
    \item Pulse-level scheduling and timing analysis.
\end{itemize}

\subsection{Enhanced Performance}

QPanda3 delivers notable performance enhancements, especially in quantum program transmission and high-efficiency circuit compilation.

\textbf{(1) Efficient Quantum Program Transmission}

To address the data representation and communication bottlenecks inherent in large-scale and distributed quantum–classical hybrid computing workflows, QPanda3 introduces a unified, high-efficiency quantum program transmission protocol. This protocol enables robust interaction between user clients, cloud-based orchestration layers, and quantum backends.

At its core lies the \textit{Unified Quantum Program Representation Module}, which abstracts quantum instructions and metadata into a compact, binary-encoded intermediate representation (IR). This structure-preserving design facilitates compatibility across heterogeneous hardware and execution environments, while also enabling fast serialization and deserialization.

\begin{figure}[htbp]
    \centering
    \includegraphics[width=0.5\textwidth]{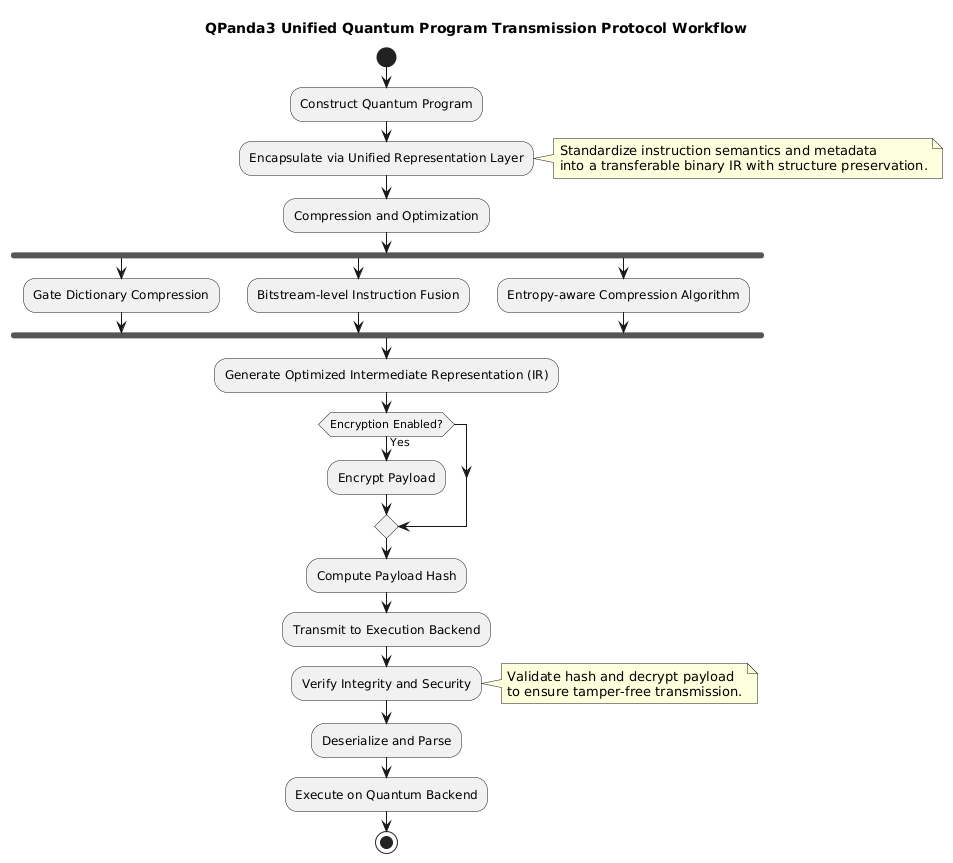}
    \caption{Architecture of the compilation module with lightweight data structures and module-level transformations.}
    \label{fig:transmission_workflow}
\end{figure}

As illustrated in Figure~\ref{fig:transmission_workflow}, the IR undergoes multi-level compression, including gate dictionary encoding, instruction fusion at the bitstream level, and entropy-aware payload optimization. These strategies collectively reduce the size of the transmitted data by up to 80\% compared to text-based formats such as OpenQASM, thereby enhancing throughput and minimizing latency in cloud-device quantum computing pipelines.

Optional encryption and integrity verification mechanisms—via hashing and secure payload encapsulation—are incorporated to safeguard against tampering in multi-tenant or federated quantum systems.

\textbf{(2) High-Performance Quantum Circuit Compilation}

A customized data structure design method tailored to the characteristics of the quantum compilation workflow is proposed to address the issue of redundant intermediate information management in instruction reordering, topology-constrained mapping, and gate fusion optimization. The core of this approach lies in the construction of the quantum circuit's directed acyclic graph (DAG), where lightweight operation units and contextual annotation mechanisms are introduced. Logical gate operations are organized into pipelined fragments with localized context, enabling structured phase buffers and operation dependency chains to support efficient access and dynamic updates of critical dependencies throughout gate-level optimization, scheduling, and topology mapping. 
This method significantly enhances compilation speed while preserving compilation quality, thereby meeting the rapid deployment requirements for large-scale quantum systems with 1000+ qubits.

\begin{figure*}[htbp]
    \centering
    \includegraphics[width=0.8\textwidth]{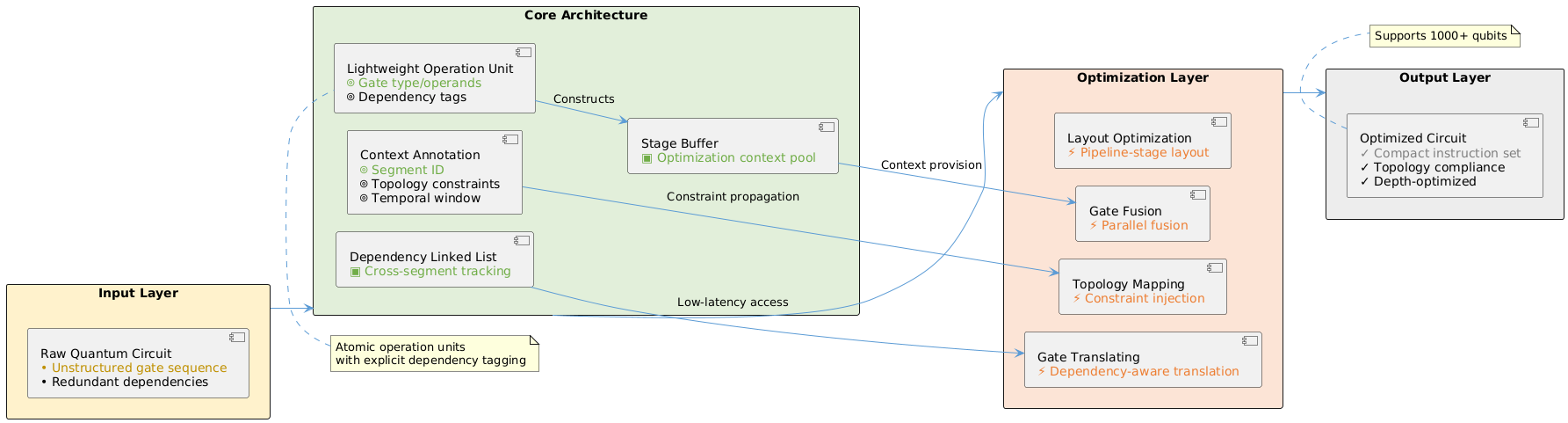}
    \caption{Architecture of the compilation module with lightweight data structures and module-level transformations.}
    \label{fig:compilation module}
\end{figure*}

The Figure~\ref{fig:compilation module} introduces a contextual annotation mechanism that segments the circuit into pipelined fragments with localized context. 
Each segment is annotated with metadata such as segment identifiers, topological constraints, and temporal execution windows, enabling fine-grained control over optimization boundaries. 
These annotated fragments are managed within a stage buffer, which acts as an optimization context pool for subsequent compilation stages.

A dependency-linked list structure is maintained to support cross-segment dependency tracking. This linked representation allows for low-latency traversal and update of gate dependencies during optimization. 
The integrated design of the DAG, stage buffer, and dependency tracking supports structured construction of operation dependency chains, which facilitates efficient updates and localized scheduling 
decisions throughout various optimization passes.

The optimization layer leverages this structured data representation to enable a sequence of specialized transformations, including:
\begin{itemize}
\item Pipeline-stage layout optimization, which improves spatial coherence of gate execution,

\item Parallel gate fusion, which merges compatible gates across segments to reduce circuit depth,

\item Constraint-driven topology mapping, which ensures hardware compliance, and

\item Dependency-aware gate translation, which refines gate translating under strict causal constraints.
\end{itemize}

This architecture demonstrates significant improvements in compilation throughput, supporting compilation for systems with 1000+ qubits. 
By integrating operation structure, temporal context, and dependency linkage into a unified data-driven framework, this approach not only accelerates the compilation pipeline but also ensures 
high-quality output circuits that are compliant with the constraints of realistic NISQ hardware.

\subsubsection{Gate Fusion}

Gate fusion is performed in a two-stage iterative pipeline. First, adjacent two-qubit gates are fused in parallel across pipeline segments to form localized composite gate blocks. 
Each fused block is then analyzed for potential decomposition into a tensor product of single-qubit gates. If such decomposition is feasible, the resulting single-qubit gates are merged into the surrounding context, enabling further compression of circuit depth. 

This fusion--decomposition--merging loop, driven by explicit dependency tracking and localized segment annotations, yields a structurally optimized two-qubit gate layer that significantly reduces overall circuit depth.

\subsubsection{Layout Optimization}

Layout optimization begins with a subgraph reconstruction algorithm that attempts to map logical circuit substructures onto hardware topologies within a bounded runtime. This mapping process considers spatial gate alignment, dependency constraints, and SWAP minimization.

If an exact subgraph match is not found within the allowed time frame, the compiler switches to a noise-aware mapping strategy. In this fallback mode, qubit blocks are selected based on backend calibration data, prioritizing hardware regions with higher gate fidelity to improve the robustness and execution fidelity of the final circuit.

\subsubsection{Topology Mapping}

Topology mapping is built upon the LightSabre algorithm, which offers superior scalability for deep circuits and large hardware graphs. To improve mapping efficiency, a lightweight data structure is designed to represent qubit connectivity and cost metrics. 

This structure maintains sparse neighborhood graphs augmented with local path caches and dynamic edge costs, allowing fast constraint evaluation and route selection without the overhead of global matrix traversal. The result is a high-throughput mapping process that scales effectively to devices with 1000+ qubits.

\subsubsection{Gate Translation}

Gate translation enables backend-agnostic circuit generation by converting logical gate sets into native hardware instructions. A graph-based dependency model is used to capture gate relationships, allowing flexible decomposition of complex operations into target-compatible primitives.

To mitigate the depth overhead introduced by redundant single-qubit gates---often a side effect of two-qubit gate decomposition---the compiler adopts a \textit{fused translation--optimization strategy}, where gate synthesis and simplification are co-executed in a single pass.

For large-scale quantum programs, this process is further accelerated via parallelism, allowing multiple gate segments to be translated and optimized concurrently.

\begin{figure*}[t]
    \centering
    \begin{subfigure}[t]{0.25\textwidth}
        \centering
        \includegraphics[width=\linewidth]{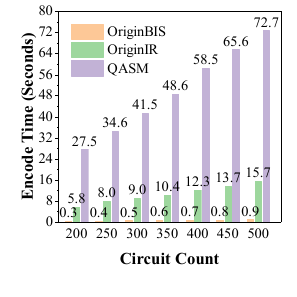}
        \caption{OriginIRs' Encode Time}
    \end{subfigure}
    \hfill
    \begin{subfigure}[t]{0.25\textwidth}
        \centering
        \includegraphics[width=\linewidth]{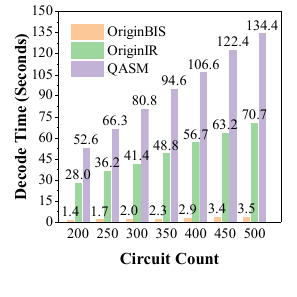}
        \caption{OriginIRs' Decode Time}
    \end{subfigure}
    \hfill
    \begin{subfigure}[t]{0.25\textwidth}
        \centering
        \includegraphics[width=\linewidth]{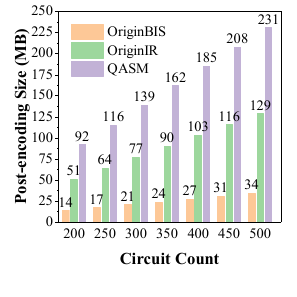}
        \caption{OriginIRs' Post-encoding Size}
    \end{subfigure}

    \begin{subfigure}[t]{0.25\textwidth}
        \centering
        \includegraphics[width=\linewidth]{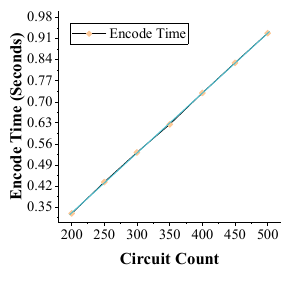}
        \caption{OriginBIS' Encode Time}
    \end{subfigure}
    \hfill
    \begin{subfigure}[t]{0.25\textwidth}
        \centering
        \includegraphics[width=\linewidth]{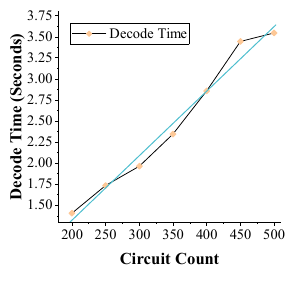}
        \caption{OriginBIS' Decode Time}
    \end{subfigure}
    \hfill
    \begin{subfigure}[t]{0.25\textwidth}
        \centering
        \includegraphics[width=\linewidth]{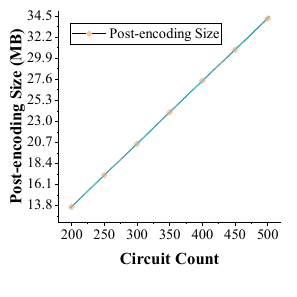}
        \caption{OriginBIS' Post-encoding Size}
    \end{subfigure}
    
    \caption{For Circuit Count}
    \label{fig:for-circuit-count}
\end{figure*}
\section{Experiments}
\subsection{Performance Experiments of OriginBIS}

QPanda3 has specifically designed OriginBIS for quantum program transmission. A 
typical application scenario involves users designing quantum programs on local 
devices and executing them using abundant computing resources on quantum clouds. 
In this scenario, quantum programs exist as memory objects in high-level 
programming languages on the source device. After transmission, they exist as 
directly executable machine instructions on the target device. This scenario 
comprises three stages: firstly, the source device converts the quantum program 
in the form of a high-level programming language memory object into an 
intermediate representation of the quantum program; secondly, the intermediate 
representation is transmitted to the target device using a communication link; 
and finally, the target device converts the intermediate representation into 
directly executable machine instructions. It should be noted that the scope of 
the concept of quantum program transmission may vary depending on specific 
business requirements. If the business only requires the quantum program on the 
source device to appear on the target device, then transmitting the data storing 
the intermediate representation of the quantum program from the source device to 
the target device completes the quantum program transmission, which only includes 
the second stage. If the business requires quantum program transmission from a 
memory object on the source device to executable machine instructions on the 
target device, then the quantum program transmission should encompass all three 
stages. For convenience of description, in the experiments of this paper, the 
quantum program transmission process refers to the second stage, the first stage 
is termed the Encode stage, and the third stage is termed the Decode stage. The 
experiments in this paper primarily focus on testing and comparing the performance 
of OriginBIS across these three stages.

This paper tests the performance of OriginBIS in quantum program transmission and 
quantum program format conversion through multiple comparative experiments. The 
experimental results demonstrate that OriginBIS significantly improves the 
processing efficiency of the corresponding processes.
\begin{figure*}[htbp]
    \centering
    \begin{subfigure}[t]{0.25\textwidth}
        \centering
        \includegraphics[width=\linewidth]{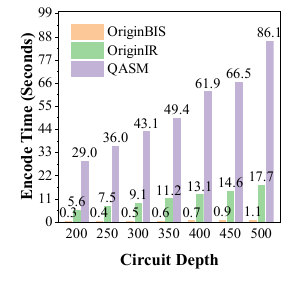}
        \caption{OriginIRs' Encode Time}
    \end{subfigure}
    \hfill
    \begin{subfigure}[t]{0.25\textwidth}
        \centering
        \includegraphics[width=\linewidth]{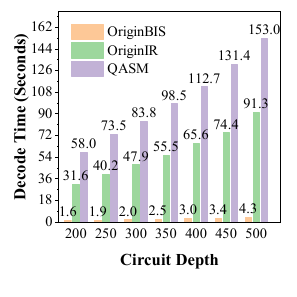}
        \caption{OriginIRs' Decode Time}
    \end{subfigure}
    \hfill
    \begin{subfigure}[t]{0.25\textwidth}
        \centering
        \includegraphics[width=\linewidth]{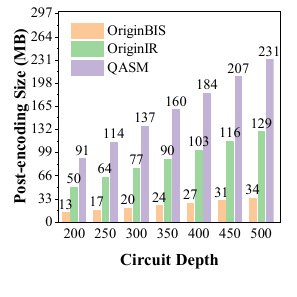}
        \caption{OriginIRs' Post-encoding Size}
    \end{subfigure}

    \begin{subfigure}[t]{0.25\textwidth}
        \centering
        \includegraphics[width=\linewidth]{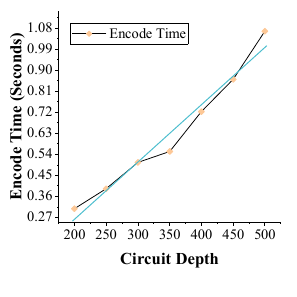}
        \caption{OriginBIS' Encode Time}
    \end{subfigure}
    \hfill
    \begin{subfigure}[t]{0.25\textwidth}
        \centering
        \includegraphics[width=\linewidth]{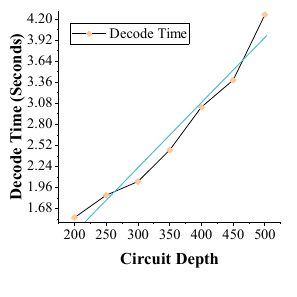}
        \caption{OriginBIS' Decode Time}
    \end{subfigure}
    \hfill
    \begin{subfigure}[t]{0.25\textwidth}
        \centering
        \includegraphics[width=\linewidth]{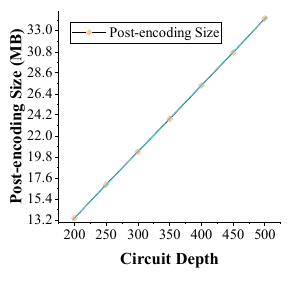}
        \caption{OriginBIS' Post-encoding Size}
    \end{subfigure}
    
    \caption{For Circuit Depth}
    \label{fig:for-circuit-depth}
\end{figure*}

\begin{figure*}[htbp]
    \centering
    \begin{subfigure}[t]{0.25\textwidth}
        \centering
        \includegraphics[width=\linewidth]{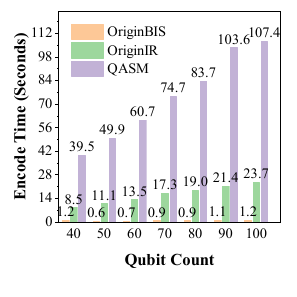}
        \caption{OriginIRs' Encode Time}
    \end{subfigure}
    \hfill
    \begin{subfigure}[t]{0.25\textwidth}
        \centering
        \includegraphics[width=\linewidth]{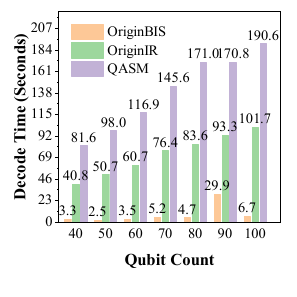}
        \caption{OriginIRs' Decode Time}
    \end{subfigure}
    \hfill
    \begin{subfigure}[t]{0.25\textwidth}
        \centering
        \includegraphics[width=\linewidth]{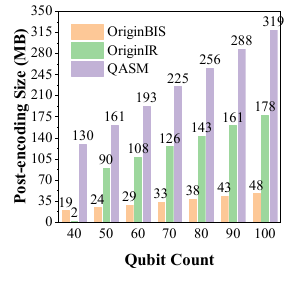}
        \caption{OriginIRs' Post-encoding Size}
    \end{subfigure}

    \begin{subfigure}[t]{0.25\textwidth}
        \centering
        \includegraphics[width=\linewidth]{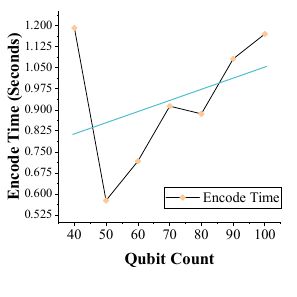}
        \caption{OriginBIS' Encode Time}
    \end{subfigure}
    \hfill
    \begin{subfigure}[t]{0.25\textwidth}
        \centering
        \includegraphics[width=\linewidth]{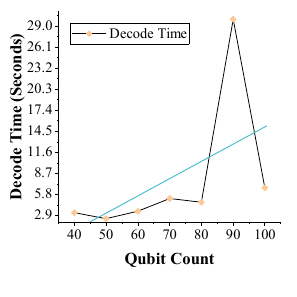}
        \caption{OriginBIS' Decode Time}
    \end{subfigure}
    \hfill
    \begin{subfigure}[t]{0.25\textwidth}
        \centering
        \includegraphics[width=\linewidth]{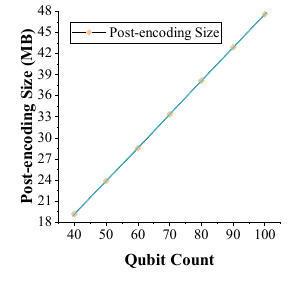}
        \caption{OriginBIS' Post-encoding Size}
    \end{subfigure}
    
    \caption{For Qubit Count}
    \label{fig:for-qbit-count}
\end{figure*}
\subsubsection{Evaluation Metrics}

\textbf{\\ (1) Encode Time and Decode Time}

For the Encode and Decode stages, the speed at which the device completes the 
entire processing operations of the respective stages reflects the processing 
efficiency of the corresponding processes. Considering that it is difficult to 
quantify the amount of processing operations performed by the device into specific 
numerical values, this paper uses the time taken to complete the same conversion 
task to reflect the impact of the intermediate representation on processing 
efficiency during the Encode and Decode stages. Specifically, for the same 
conversion task, the longer the time taken, the lower the processing efficiency; 
the shorter the time taken, the higher the processing efficiency.

\textbf{\\ (2) Post-encoding Size}

Unlike the Encode and Decode stages, where time is used as a metric, this paper 
employs the data size of the quantum program after encoding (Post-encoding Size) 
to measure the impact of the intermediate representation on transmission 
efficiency during the quantum program transmission stage. The quantum program 
transmission stage constitutes a classical communication process, where latency 
is a crucial performance evaluation metric. However, in practical communication 
environments, latency is influenced by various factors, including the bandwidth 
and distance of physical communication links and devices, as well as the classical 
network protocols used for data transmission. Quantum program transmission 
primarily involves transmitting the intermediate representation of the quantum 
program as pure data. The data size of the intermediate representation is the only 
component of the quantum program intermediate representation that is directly 
related to communication latency, excluding the influences of physical 
communication links, devices, and classical network protocols. Specifically, for 
the same quantum program transmission task, a larger Post-encoding Size results 
in lower transmission efficiency, while a smaller Post-encoding Size leads to 
higher transmission efficiency.

\subsubsection{Experimental Setup}

This paper conducts comparative experiments from three aspects: the number of 
circuits, circuit depth, and the number of qubits. Batch transmission of a large 
number of circuits is one of the real and essential practical application 
requirements. Circuit depth is a crucial factor that affects the execution results 
of quantum circuits. The number of qubits measures the amount of quantum computing 
resources occupied by quantum circuits. These three metrics serve as important 
references for users when submitting quantum programs and for devices when 
processing quantum programs for execution. The experiments in this paper are 
carried out using the scheme shown in Table 
\ref{tab:post-encoding-size-experiment-setup}. For the experiment on the number 
of quantum circuits, the number of quantum circuits is taken as the independent 
variable, while the circuit depth is controlled at a fixed value of 500 and the 
number of qubits is controlled at a fixed value of 72. For the experiment on 
circuit depth, the circuit depth is taken as the independent variable, with 
the number of circuits controlled at a fixed value of 500 and the number of qubits 
controlled at a fixed value of 72. For the experiment on the number of qubits, the 
number of qubits is taken as the independent variable, with the circuit depth 
controlled at a fixed value of 500 and the number of circuits controlled at a 
fixed value of 500. It should be noted that the parameters related to circuits 
are not limited to the number of circuits, circuit depth, and the number of 
qubits. Moreover, there may be numerical constraints among these parameters. For 
example, when the number of quantum gates is fixed, there is a certain negative 
correlation between circuit depth and the number of qubits. To minimize the 
influence of other parameters and the relationships among them, this paper adopts 
a randomized technical approach, generating random circuits based on the 
conventions shown in Table \ref{tab:post-encoding-size-experiment-setup} to test 
the performance of quantum program intermediate representations in quantum program 
transmission. Relevant experiments are conducted on OriginBIS, OriginIR, and QASM 
to compare the quantum program transmission performance of these intermediate 
representations.

\begin{table*}[htbp]
\centering
\caption{Experimental Setup For Quantum Program Transmission}
\label{tab:post-encoding-size-experiment-setup}
\begin{tabular}{cccc}
\toprule
\textbf{Experiment} & \textbf{Circuit Count} & \textbf{Circuit Depth} & \textbf{Qubit Count} \\
\midrule
For Circuit Count & Independent Variable & 500 & 72 \\
For Circuit Depth & 500 & Independent Variable & 72 \\
For Qubit Count & 500 & 500 & Independent Variable \\
\bottomrule
\end{tabular}
\end{table*}

\subsubsection{Experimental Results and Analysis}

\textbf{\\ (1) For Circuit Count}

The experimental results for the number of circuits are shown in Figure 
\ref{fig:for-circuit-count}. The grouped columns chart in subplots (a), (b), 
and (c) demonstrates that OriginBIS is significantly smaller than OriginIR and 
QASM in terms of Encode Time, Decode Time, and Post-encoding Size. This indicates 
that in these experiments, OriginBIS outperforms OriginIR and QASM in the 
efficiency of quantum program transmission. Through simple numerical estimation, 
it can be found that for Encode Time, OriginIR is approximately 17 times that of 
OriginBIS, and QASM is approximately 80 times that of OriginBIS; for Decode Time, 
OriginIR is approximately 20 times that of OriginBIS, and QASM is approximately 
40 times that of OriginBIS; for Post-encoding Size, OriginIR is approximately 3.5 
times that of OriginBIS, and QASM is approximately 6.5 times that of OriginBIS. 
The point-line charts in subplots (d), (e), and (f) show that for OriginBIS, 
Encode Time, Decode Time, and Post-encoding Size increase with the growth in the 
number of circuits.

\textbf{\\ (2) For Circuit Depth}

The experimental results regarding circuit depth are presented in Figure 
\ref{fig:for-circuit-depth}. The grouped columns chart in subplots (a), (b), 
and (c) demonstrate that OriginBIS exhibits significantly lower values in terms 
of Encode Time, Decode Time, and Post-encoding Size compared to both OriginIR and 
QASM. This indicates that in these experiments, OriginBIS outperforms OriginIR and 
QASM in terms of efficiency for quantum program transmission. Upon simple numerical 
estimation, it can be observed that for Encode Time, OriginIR is approximately 18 
times that of OriginBIS, while QASM is about 88 times; for Decode Time, OriginIR 
is roughly 21 times that of OriginBIS, and QASM is approximately 39 times; for 
Post-encoding Size, OriginIR is around 3.5 times that of OriginBIS, and QASM is 
about 6.5 times. The point-line charts in subplots (d), (e), and (f) reveal that 
for OriginBIS, Encode Time, Decode Time, and Post-encoding Size increase with the 
growth of circuit depth.

\begin{figure*}[htbp]
    \centering
    \begin{subfigure}[t]{0.25\textwidth}
        \centering
        \includegraphics[width=\linewidth]{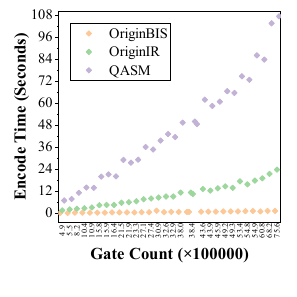}
        \caption{OriginIRs' Encode Time}
    \end{subfigure}
    \hfill
    \begin{subfigure}[t]{0.25\textwidth}
        \centering
        \includegraphics[width=\linewidth]{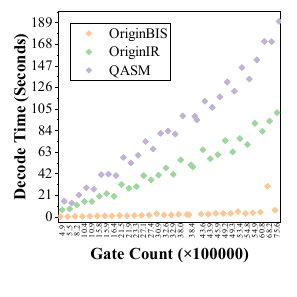}
        \caption{OriginIRs' Decode Time}
    \end{subfigure}
    \hfill
    \begin{subfigure}[t]{0.25\textwidth}
        \centering
        \includegraphics[width=\linewidth]{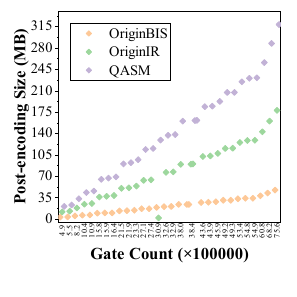}
        \caption{OriginIRs' Post-encoding Size}
    \end{subfigure}

    \begin{subfigure}[t]{0.25\textwidth}
        \centering
        \includegraphics[width=\linewidth]{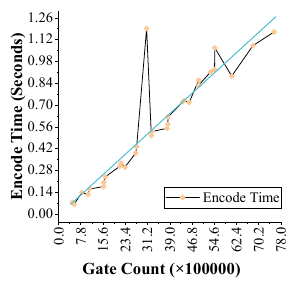}
        \caption{OriginBIS' Encode Time}
    \end{subfigure}
    \hfill
    \begin{subfigure}[t]{0.25\textwidth}
        \centering
        \includegraphics[width=\linewidth]{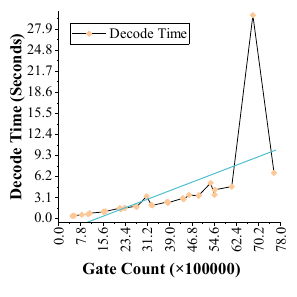}
        \caption{OriginBIS' Decode Time}
    \end{subfigure}
    \hfill
    \begin{subfigure}[t]{0.25\textwidth}
        \centering
        \includegraphics[width=\linewidth]{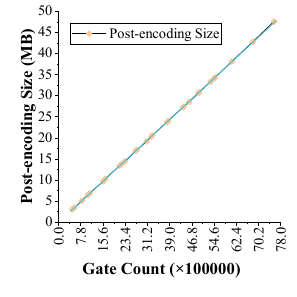}
        \caption{OriginBIS' Post-encoding Size}
    \end{subfigure}
    
    \caption{For Gate Count}
    \label{fig:for-gate-count}
\end{figure*}
\textbf{\\ (3) For Qubit Count}

The experimental results concerning the number of qubits are illustrated in 
Figure \ref{fig:for-qbit-count}. The grouped columns chart in subplots (a), (b), 
and (c) indicates that OriginBIS exhibits significantly lower Encode Time and 
Decode Time compared to both OriginIR and QASM. Additionally, for qubit counts of 
50 or more, OriginBIS also demonstrates lower Post-encoding Time compared to 
OriginIR and QASM. This suggests that in the majority of the samples tested, 
OriginBIS outperforms OriginIR and QASM in terms of efficiency for quantum 
program transmission. Excluding the 90-qubit data in subplot (b) and the 40-bit 
data in subplot (c), upon simple numerical estimation, it can be observed that 
for Encode Time, OriginIR is approximately 21 times that of OriginBIS, while 
QASM is about 93 times; for Decode Time, OriginIR is roughly 20 times that of 
OriginBIS, and QASM is approximately 39 times; for Post-encoding Size, OriginIR 
is around 3.5 times that of OriginBIS, and QASM is about 6.5 times. The 
point-line charts in subplots (d), (e), and (f) reveal that for OriginBIS, 
there is a trend of increasing Encode Time and Decode Time with the growth 
in the number of qubits, albeit with considerable data fluctuation. However, 
Post-encoding Time consistently increases with the increase in the number of 
qubits.

\textbf{\\ (4) Gate Count}

\begin{figure*}[htbp]
    \centering
    \begin{subfigure}[t]{0.25\textwidth}
        \centering
        \includegraphics[width=\linewidth]{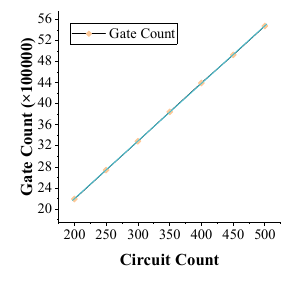}
        \caption{}
    \end{subfigure}
    \hfill
    \begin{subfigure}[t]{0.25\textwidth}
        \centering
        \includegraphics[width=\linewidth]{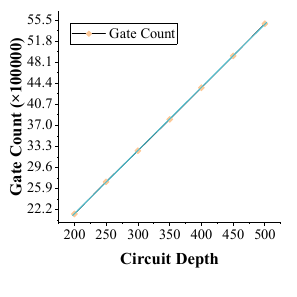}
        \caption{}
    \end{subfigure}
    \hfill
    \begin{subfigure}[t]{0.25\textwidth}
        \centering
        \includegraphics[width=\linewidth]{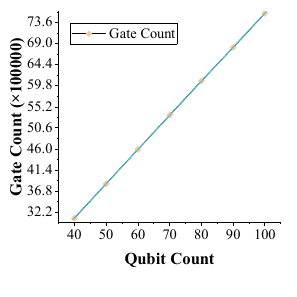}
        \caption{}
    \end{subfigure}
    \caption{Gate Count for Circuit Count, Circuit Depth and QBit Count}
    \label{fig:gate-and-circuit-count}
\end{figure*}

During the experiments examining circuit count, circuit depth, and qubit count, 
it was observed that the number of quantum gates varies with changes in these 
parameters. Figure \ref{fig:gate-and-circuit-count} illustrates the trend in the 
number of quantum gates, demonstrating a strong positive linear correlation with 
the value of the third parameter when the values of any two of these three 
quantities are held constant.

The number of gates is also an important metric for evaluating quantum programs. 
By combining the gate count data from Figure \ref{fig:gate-and-circuit-count} 
with the corresponding Encode Time, Decode Time, and Post-encoding Size data, the 
results presented in Figure \ref{fig:for-gate-count} were obtained. Subplots (a), 
(b), and (c) indicate that the Encode Time, Decode Time, and Post-encoding Size 
for OriginBIS are significantly lower than those for OriginIR and QASM. Subplots 
(d), (e), and (f) reveal that there is a trend of increasing Encode Time and 
Decode Time with the growth in the number of quantum gates, albeit with noticeable 
data fluctuations. However, Post-encoding Time increases with the increase in the 
number of quantum gates, exhibiting a strong linear characteristic.

The experimental results concerning the number of qubits are illustrated in 
Figure \ref{fig:for-qbit-count}. The grouped columns chart in subplots (a), (b), 
and (c) indicates that OriginBIS exhibits significantly lower Encode Time and 
Decode Time compared to both OriginIR and QASM. Additionally, for qubit counts 
of 50 or more, OriginBIS also demonstrates lower Post-encoding Time compared to 
OriginIR and QASM. This suggests that in the majority of the samples tested, 
OriginBIS outperforms OriginIR and QASM in terms of efficiency for quantum 
program transmission. Excluding the 90-qubit data in subplot (b) and the 40-qubit 
data in subplot (c), upon simple numerical estimation, it can be observed that for 
Encode Time, OriginIR is approximately 21 times that of OriginBIS, while QASM is 
about 93 times; for Decode Time, OriginIR is roughly 20 times that of OriginBIS, 
and QASM is approximately 39 times; for Post-encoding Size, OriginIR is around 3.5 
times that of OriginBIS, and QASM is about 6.5 times. The point-line charts in 
subplots (d), (e), and (f) reveal that for OriginBIS, there is a trend of 
increasing Encode Time and Decode Time with the growth in the number of qubits, 
albeit with considerable data fluctuation. However, Post-encoding Time 
consistently increases with the increase in the number of qubits.

\begin{figure*}[htbp]
    \centering
    \includegraphics[width=0.7\textwidth]{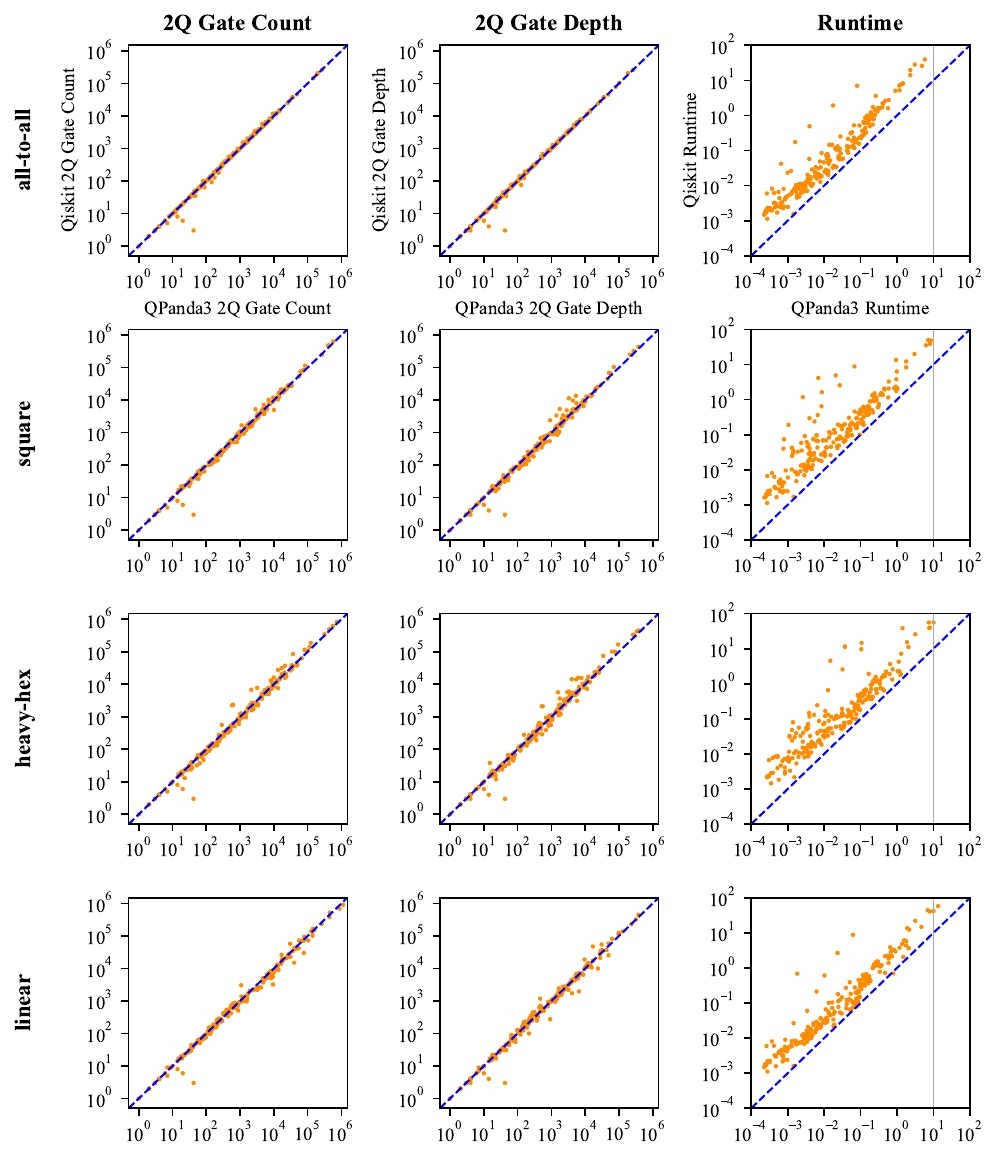}
    \caption{Experimental Results About Circuit Compilation}
    \label{fig:manipulate-time-2}
\end{figure*}
\subsection{Compilation Efficiency Comparison Experiment}

\subsubsection{Experiment Setup}

In this paper, a comparative experiment related to compilation efficiency was 
conducted on multiple mainstream quantum programming frameworks based on 
Benchpress. Benchpress is an evaluation suite for multi-quantum computing 
software development kits, containing thousands of test cases. It allows for 
the uniform testing of multiple quantum software packages across various 
performance and functional indicators, with the results reflecting the cost 
of processing quantum circuits on quantum computing devices using different 
software packages. All relevant experiments for each software package mentioned 
in the Benchpress literature were replicated in this study, with the addition 
of tests for QPanda3. All software packages tested are listed in Table 
\ref{tab:software-packages}. It should be noted that a timeout limit of 180 
seconds was imposed for executing each test case.

The testing of these software packages was conducted in the same software and 
hardware environments. The environmental parameters used in our experiments are 
presented in Table \ref{tab:environmental-parameters}.

\begin{table}[htbp]
\centering
\caption{SDKs}
\label{tab:software-packages}
\begin{tabular}{cc}
\toprule
\textbf{SDK} & \textbf{Version} \\
\midrule
amazon-braket-sdk(braket) & 1.91.0 \\
bqskit & 1.2.0 \\
cirq & 1.4.1 \\
pytket(tket) & 2.0.1 \\
qiskit & 2.0.1 \\
qiskit\_ibm\_transpiler & 0.39.0 \\
pystaq & 3.5 \\
pyqpanda3 & 0.3.0 \\
\bottomrule
\end{tabular}
\end{table}

\begin{table}[htbp]
\centering
\caption{Software and Hardware Environments}
\label{tab:environmental-parameters}
\begin{tabular}{cc}
\toprule
\textbf{Type} & \textbf{Info} \\
\midrule
CPU & Intel(R) Xeon(R) Platinum 8336C CPU @ 2.30GHz \\
Memory & 2.0TB \\
Operating System & Ubuntu 20.04.5 LTS \\
Python & 3.11.11 \\
\bottomrule
\end{tabular}
\end{table}

\subsubsection{Evaluation Metrics}

This paper adopts the evaluation metrics introduced by Benchpress. They are 
Standard Pytest Output Type, circuit construction time, manipulate time, and 
circuit compilation time.

\subsubsection{Experimental Results And Analysis}

\begin{table}[H]
\centering
\caption{Experimental Results Evaluated Based on The Criterion of Standard Pytest Output Type}
\label{tab:table4}
\begin{tabular}{cccccc}
\toprule
\textbf{SDK} & \textbf{Passed} & \textbf{Failed} & \textbf{XFailed} & \textbf{Skipped} & \textbf{Total} \\
\midrule
braket & 7 & 2 & 0 & 1057 & 1066 \\
cirq & 10 & 2 & 0 & 1054 & 1066 \\
qiskit & 1044 & 0 & 0 & 22 & 1066 \\
QPanda3 & 1016 & 28 & 0 & 22 & 1066 \\
\bottomrule
\end{tabular}
\end{table}

\begin{sloppypar}
    In our experiments, the results of the \lstinline{test_status_counts} are 
    presented in Table \ref{tab:table4}. It was observed that many SDKs failed 
    to satisfy the requirements of numerous test cases in Benchpress. 
    Specifically, bqskit was only capable of performing tests related to 
    circuit construction. pytket and pystaq exhibited compatibility issues 
    with the environments of other SDKs, making it impossible to conduct fair 
    testing and comparison using a unified test suite and environment. 
    Furthermore, \lstinline{qiskit_ibm_transpiler} demonstrated strong dependency 
    on the IBM computing platform, rendering it untestable under our experimental 
    environment. As evident from Table \ref{tab:table4}, both braket and cirq 
    skipped a significant number of test cases. According to the Benchpress 
    literature, qiskit skipped 22 test cases and passed 1,044 test cases. 
    Tabel \ref{tab:table4} indicates that our experiments actually tested 
    multiple SDKs using 1,066 test cases. The performance of qiskit in our 
    experiments was consistent with the reports in the Benchpress literature. 
    QPanda successfully passed $95.3\%$ of all test cases, with the number of 
    skipped cases equal to that of qiskit. This table showcases the excellent 
    versatility of QPanda3 within the unified test suite, Benchpress.    
\end{sloppypar}

\begin{sloppypar}
The performance of various SDKs in circuit construction is illustrated in 
Figure \ref{fig:circuits-construction-time}. It can be observed that QPanda3, 
qiskit, and cirq have successfully passed all seven test cases. For the four 
test cases, namely \lstinline[breaklines=true]{test_multi_control_circuit}, 
\lstinline[breaklines=true]{test_bigint_qasm2_import}, 
\lstinline[breaklines=true]{test_param_circSU2_100_build}, and 
\lstinline[breaklines=true]{test_clifford_build}, QPanda3 demonstrated the 
shortest execution time. In the case of the three test cases 
\lstinline[breaklines=true]{test_QV100_qasm2_import}, 
\lstinline[breaklines=true]{test_param_circSU2_100_build}, and 
\lstinline{test_QV100_build}QPanda3 took longer than qiskit but significantly 
less time compared to cirq, braket, and bqskit. For the test case 
\lstinline[breaklines=true]{test_DTC100_set_build}, QPanda3's execution time was 
shorter than that of qiskit but longer than cirq's, ranking as the second 
shortest. This figure highlights the advantages of QPanda3 in circuit 
construction.
\end{sloppypar}

\begin{figure}[htbp]
    \centering
    \includegraphics[width=0.5\textwidth]{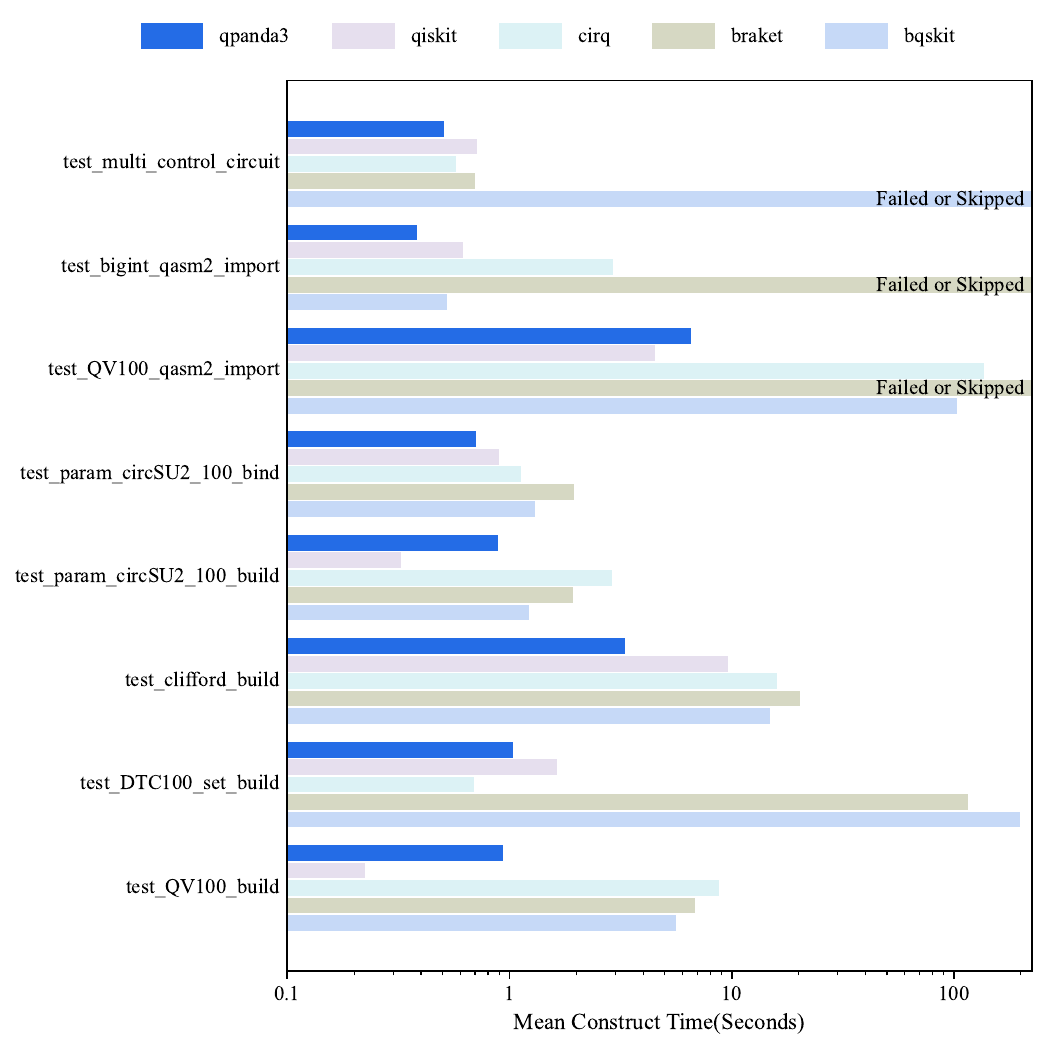}
    \caption{Experimental Results About Circuit Construction(shorter is better)}
    \label{fig:circuits-construction-time}
\end{figure}
\begin{figure}[htbp]
    \centering
    \includegraphics[width=0.5\textwidth]{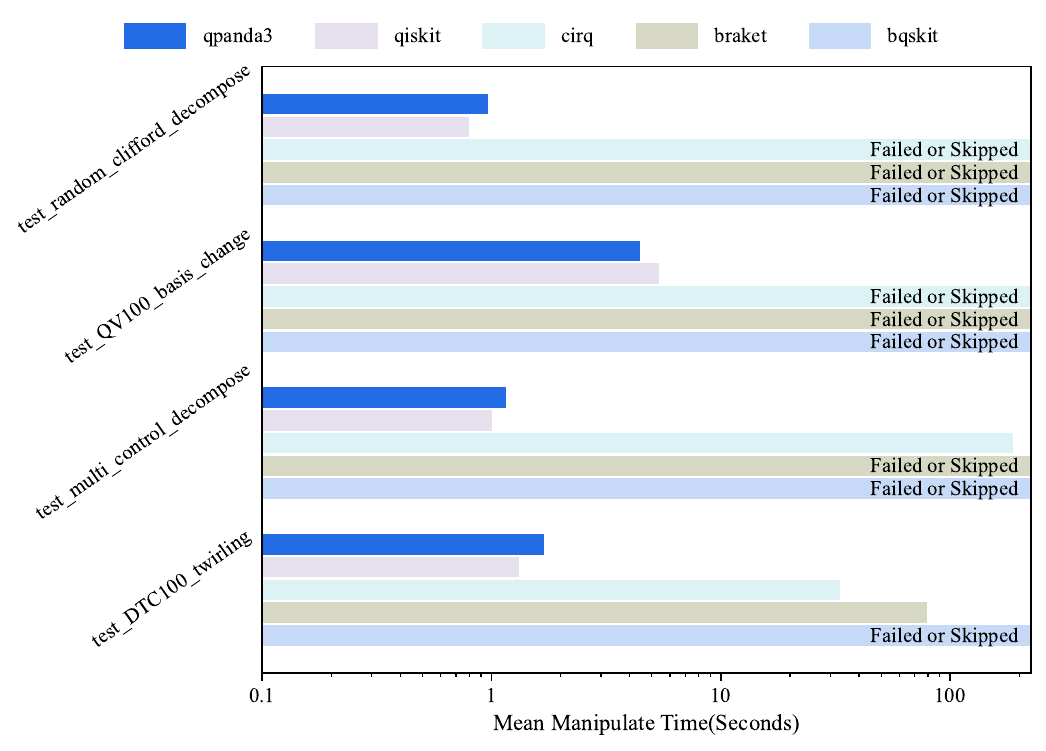}
    \caption{Experimental Results About Circuit Manipulation(shorter is better)}
    \label{fig:manipulate-time}
\end{figure}

The Benchpress literature indicates that Manipulate is a crucial metric for 
evaluating SDKs. As illustrated in Figure \ref{fig:manipulate-time}, only 
QPanda3 and qiskit have successfully passed all four test cases. In the 
\lstinline{test_QV100_basis_change} test case, QPanda3 demonstrated the 
shortest execution time. For the three test cases: 
\lstinline{test_random_clifford_decompose}, 
\lstinline{test_multi_control_decompose}, and \lstinline{test_DTC100_twirling}, 
QPanda3's execution time slightly exceeded that of qiskit, with a minimal 
difference. This figure suggests that QPanda3 holds an advantage in terms of 
the Manipulate metric.

In our experiments, only QPanda3 and qiskit passed the majority of the test 
cases related to compilation. We have plotted the corresponding experimental 
results into multiple scatter plots, as shown in Figure 
\ref{fig:manipulate-time-2}, which illustrate the differences between QPanda3 
and qiskit in terms of 2Q Gate Count, 2Q Gate Depth, and compilation time for 
various topological structures. In each plot, the blue dashed line represents 
the point where the values for QPanda3 and qiskit are equal for the corresponding 
metric. Points above the blue dashed line indicate that the corresponding value 
for QPanda3 is less than that for qiskit. As observed from Figure 
\ref{fig:manipulate-time-2}, for both 2Q Gate Count and 2Q Gate Depth, the 
majority of the scatter points are densely distributed along or very close to the 
blue dashed line, suggesting that the performance of QPanda3 and qiskit is 
essentially the same. However, regarding compilation time, all scatter points 
in the four subplots of Figure \ref{fig:manipulate-time-2} are located above the 
blue dashed line, indicating that QPanda3 exhibits higher compilation efficiency 
than qiskit. This characteristic is not limited by topological structure and is 
effective for a large number of different circuits. Additionally, it can be 
observed that only a very few test cases result in compilation times exceeding 
10 seconds when using QPanda3. These observations collectively demonstrate the 
high compilation efficiency of QPanda3.

\subsection{Performance Comparison Experiment for Gradient Computation in Variational Quantum Circuits}

\paragraph{Analysis of Gradient Computation Performance and Circuit Depth in 12-Qubit Systems}

\begin{figure}[htbp]
    \centering
    \includegraphics[width=0.5\textwidth]{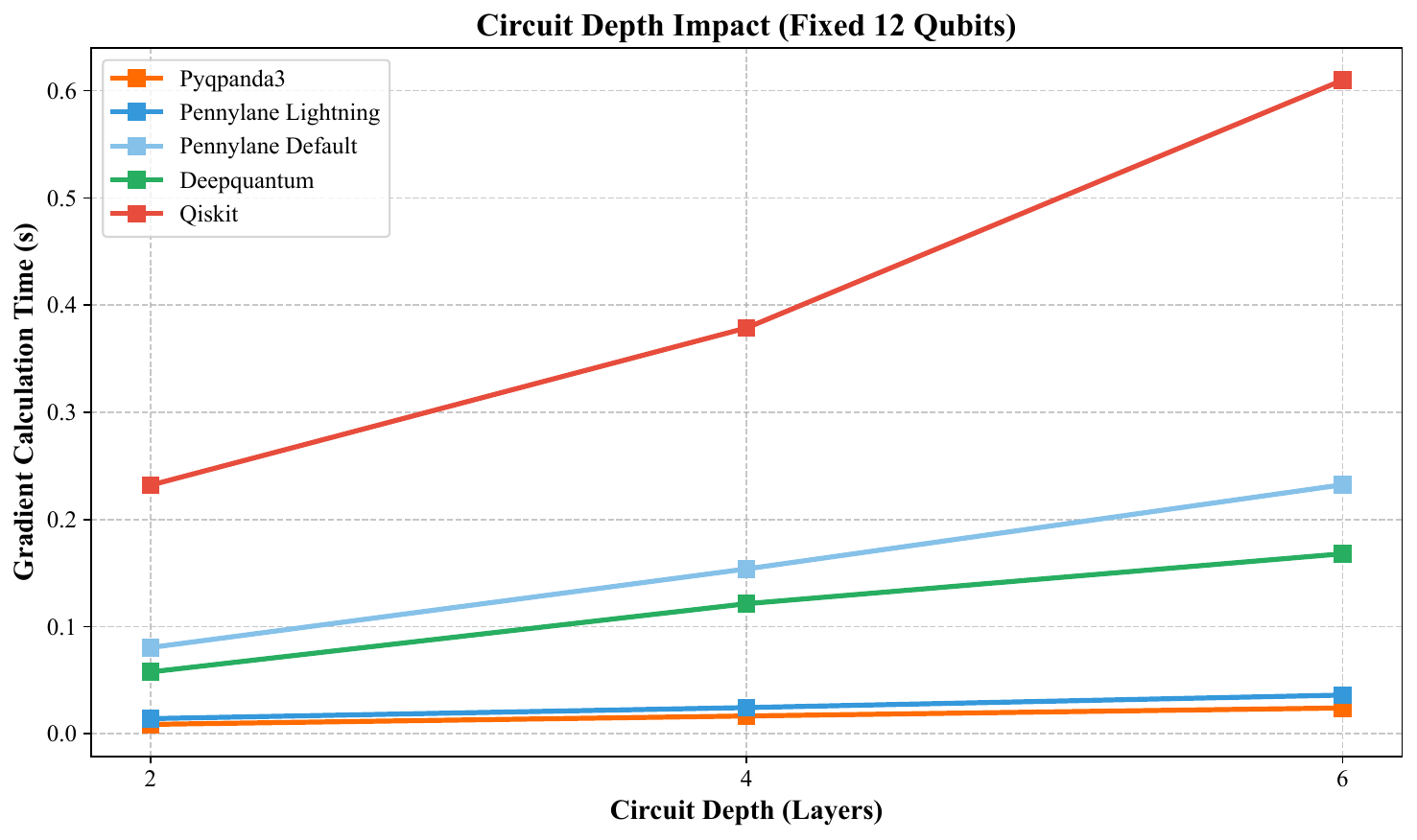}
    \caption{Circuit Depth Impact (Fixed 12 Qubits)}
    \label{fig:exp_cal_gradient_depth_impact_12qubits}
\end{figure}

The performance analysis of gradient computation in 12-qubit systems corresponding 
to Figure \ref{fig:exp_cal_gradient_depth_impact_12qubits} demonstrates that quantum 
framework selection critically determines computational efficiency. PyQPanda3 exhibits 
significant time advantages, requiring only 0.0088 seconds average time at 2-layer 
circuit depth compared to Qiskit's 0.232 seconds - a 26-fold difference. When depth 
increases to 6 layers, PyQPanda3's time merely grows to 0.0241 seconds (2.7x increase), 
substantially lower than Qiskit's 0.6097 seconds (2.6x) and DeepQuantum's 0.1680 seconds 
(2.9x). Notably, frameworks maintain a consistent performance hierarchy: PyQPanda3 
consistently delivers optimal performance, with computation times 40\%-50\% lower 
than second-ranked PennyLane-Lightning (0.0241s vs 0.0362s at 6 layers), while Qiskit 
persistently trails. During depth scaling, PyQPanda3 shows the smallest absolute time 
increment (+0.0153s), equivalent to just 4\% of Qiskit's increment (+0.3777s), 
enabling 97.6\% time savings in deep circuits. Experimental data confirms PyQPanda3's 
superior depth-scaling stability, with significantly lower time-complexity growth rates 
than comparable frameworks, providing an effective computational efficiency solution 
for deep quantum circuits.

\paragraph{Comparative Analysis for Fixed Qubit Count and Circuit Depth}

\begin{figure}[htbp]
    \centering
    \includegraphics[width=0.5\textwidth]{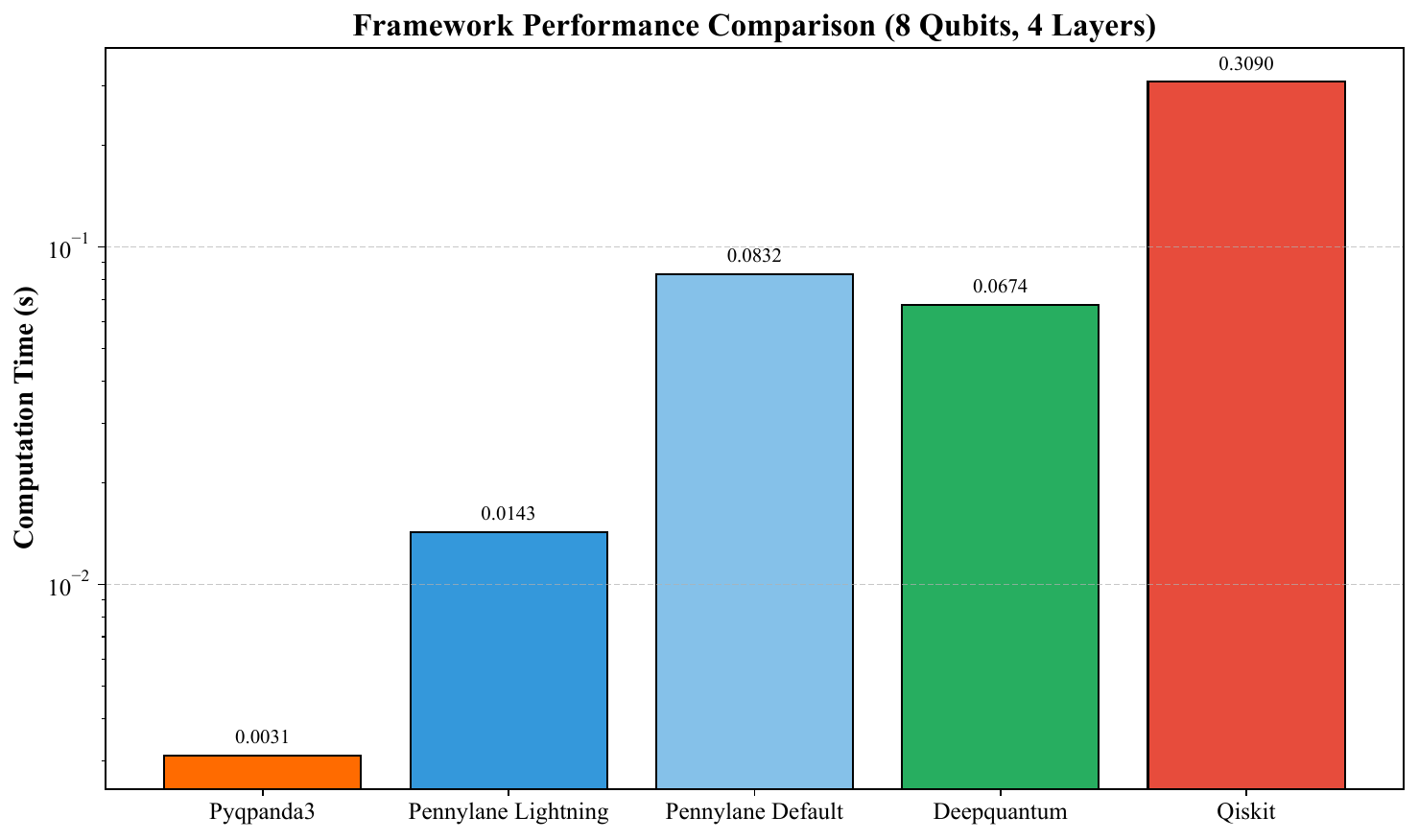}
    \caption{Framework Comparison 8qubits 4layers}
    \label{fig:exp_cal_gradient_framework_comparison_8qubits_4layers}
\end{figure}

\begin{figure}[htbp]
    \centering
    \includegraphics[width=0.5\textwidth]{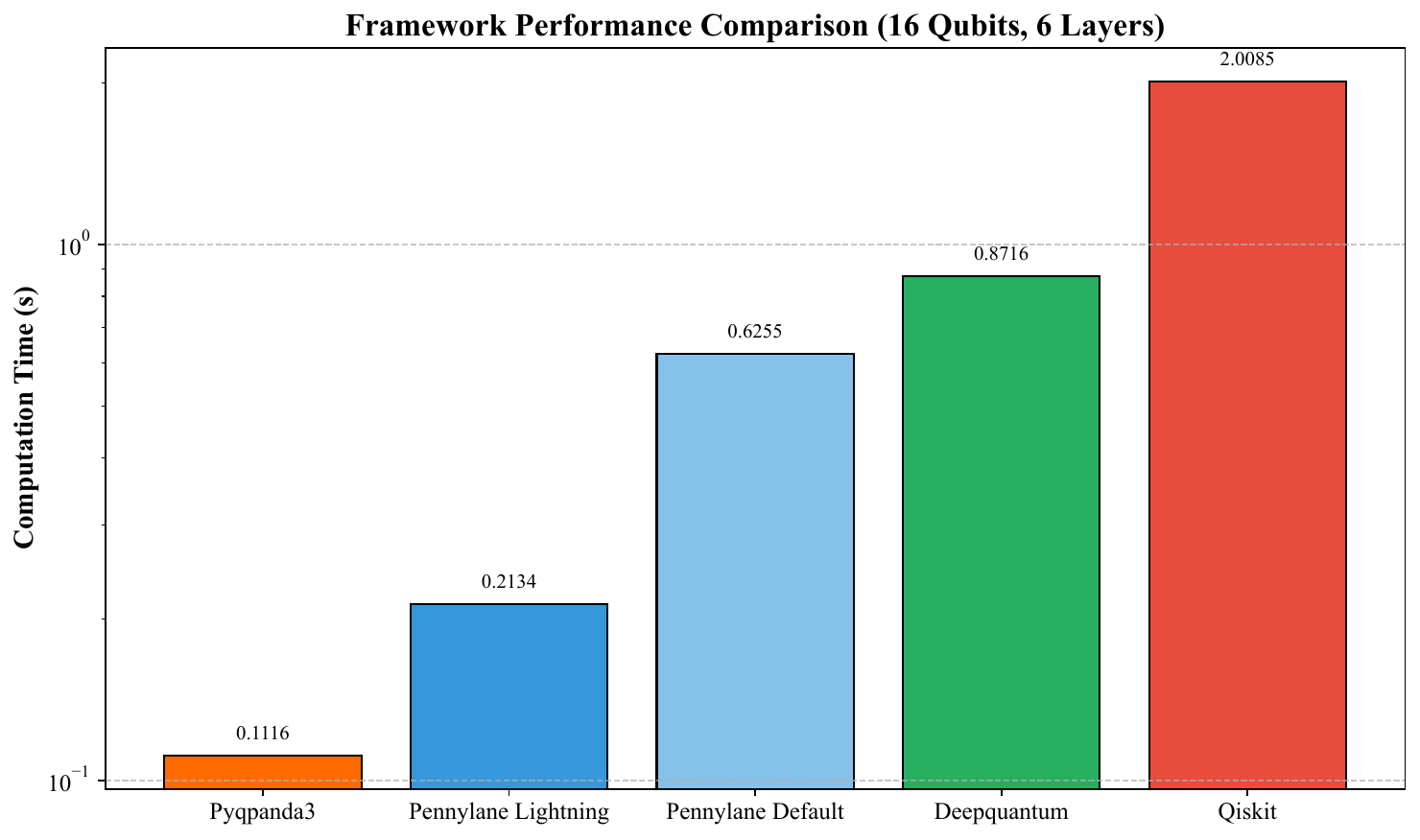}
    \caption{Framework Comparison 16qubits 6layers}
    \label{fig:exp_cal_gradient_framework_comparison_16qubits_6layers}
\end{figure}

Based on the comparative analysis of Figure 
\ref{fig:exp_cal_gradient_framework_comparison_8qubits_4layers} and Figure 
\ref{fig:exp_cal_gradient_framework_comparison_16qubits_6layers}, a significant 
and stable statistical pattern emerges in gradient computation times across quantum 
frameworks. Under 8-qubit/4-layer conditions, PyQPanda3 demonstrates exceptional 
performance (0.0031s), requiring only 21.7\% of the time needed by second-ranked 
PennyLane-Lightning (0.0143s), while Qiskit's duration (0.3090s) reaches 99.7x that 
of PyQPanda3. When scaling to 16-qubit/6-layer configurations, the performance 
ranking remains strictly consistent: PyQPanda3 (0.1116s) maintains the lowest time, 
outperforming PennyLane-Lightning (0.2134s) by 47.7\%, with Qiskit (2.0085s) requiring 
18x more time than PyQPanda3.

Notably, the performance differences exhibit systematic hierarchical characteristics: 
PyQPanda3 and PennyLane-Lightning consistently form an efficient cluster, collectively 
accounting for 16.1\% and 7.5\% of total computation time under the two respective scales; 
DeepQuantum and PennyLane-Default comprise a mid-tier cluster; while Qiskit independently 
constitutes an inefficient unit, dominating 82.9\% (8-qubit) and 68.2\% (16-qubit) of total
 time. As computational scale increases, PyQPanda3's absolute advantage over Qiskit expands 
 from 0.3059s to 1.8969s, though its relative advantage multiplier decreases from 99x to 18x, 
 indicating nonlinear evolution of performance differentials with scaling. These statistical 
 patterns confirm that quantum framework performance disparities maintain intrinsic stability 
 regardless of computational scale.

\paragraph{Time Analysis of Gradient Computation for Fixed 4-Layer Quantum Circuits}

\begin{figure}[htbp]
    \centering
    \includegraphics[width=0.5\textwidth]{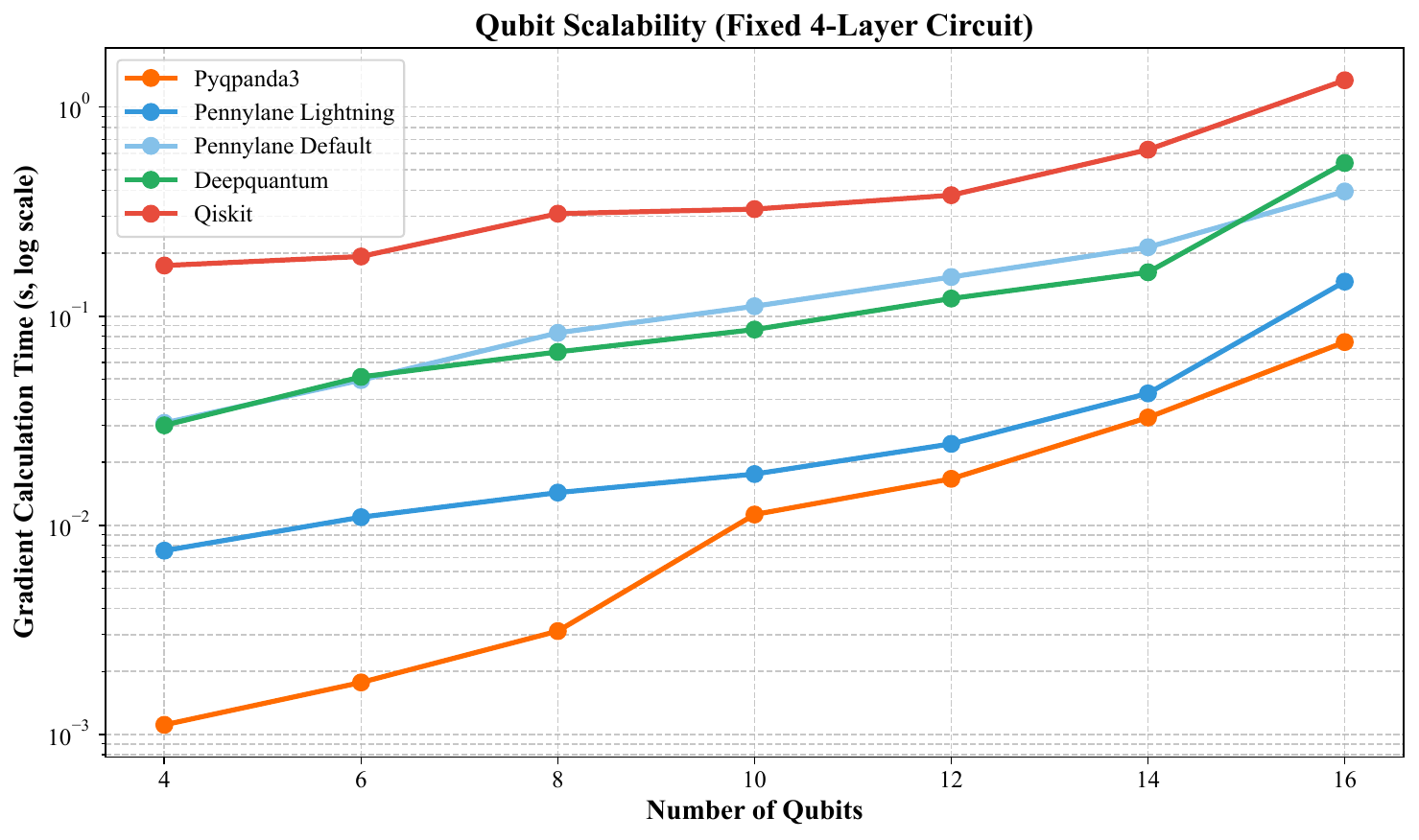}
    \caption{Qubit Scalability 4layers}
    \label{fig:qubit_scalability_4layers}
\end{figure}

Analysis of gradient computation time statistics for fixed 4-layer quantum circuits 
reveals the following patterns: As qubit count increases from 4 to 16, all frameworks 
exhibit exponential time growth (manifesting as linear trends in logarithmic 
coordinates), yet with significantly different growth slopes. PyQPanda3 demonstrates 
the mildest growth (approximately 100x increase from 4 to 16 qubits), while Qiskit 
shows the steepest escalation (approximately 10x increase). PennyLane-Default and 
DeepQuantum exhibit intermediate growth slopes with stable differentials, maintaining 
a consistent time ratio of $\approx$1:1.2 throughout the scaling process.

The performance differentials between frameworks exhibit dual-scale dependency: 
Regarding absolute differences, the maximum time gap continuously widens with 
increasing qubits (from 0.099s between PyQPanda3 and Qiskit at 4 qubits to 0.9s 
at 16 qubits); whereas in relative terms, the advantage of efficient frameworks 
diminishes with scale (the PyQPanda3/Qiskit time ratio decreases from 1:100 at 
4 qubits to 1:10 at 16 qubits).

\paragraph{Comparison with PennyLane-Lightning}

\begin{figure}[htbp]
    \centering
    \includegraphics[width=0.5\textwidth]{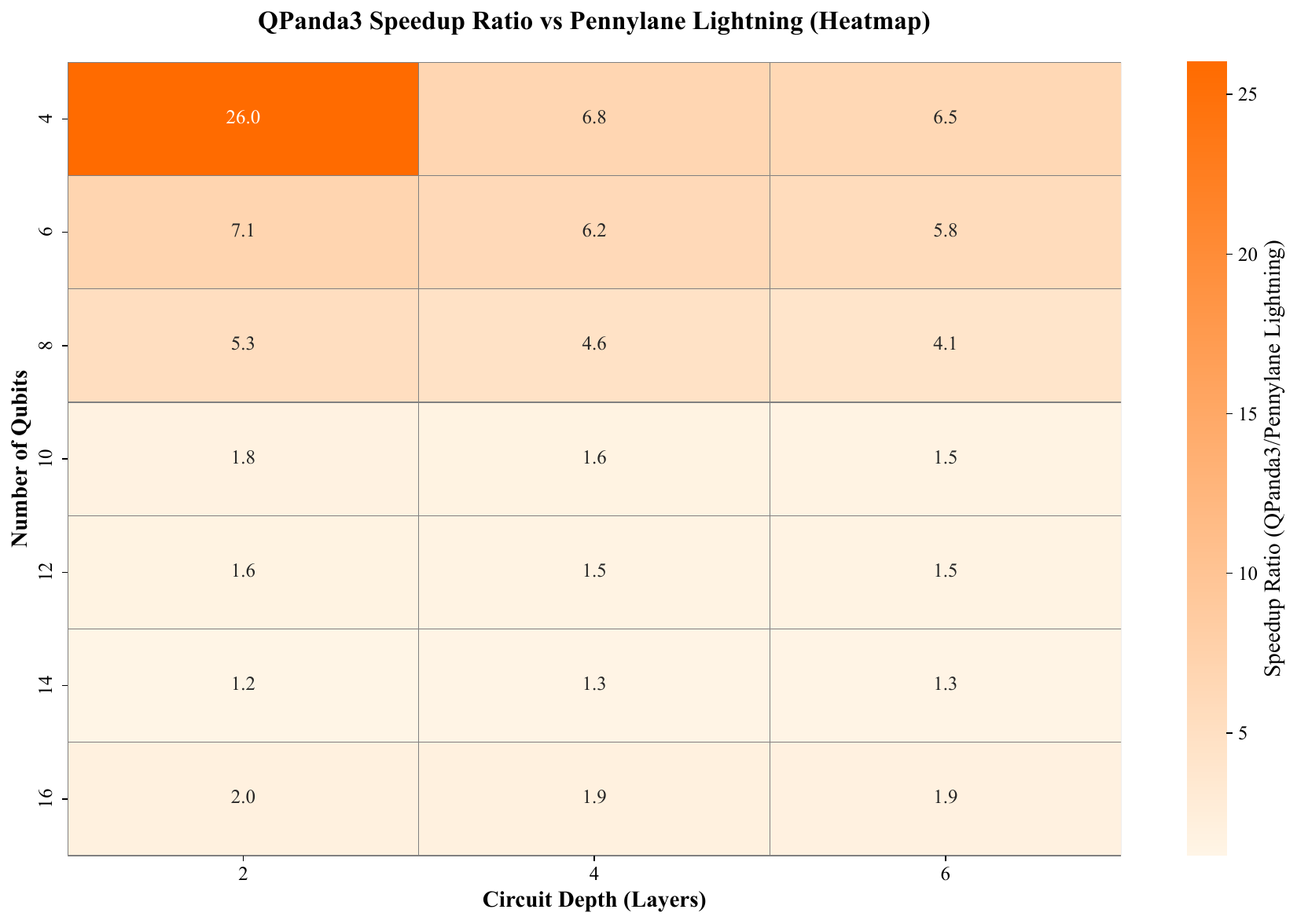}
    \caption{Speedup Heatmap vs Pennylane-lightning}
    \label{fig:speedup_heatmap_vs_pennylane-lightning}
\end{figure}

Based on the analysis of the speedup heatmap 
\ref{fig:speedup_heatmap_vs_pennylane-lightning}, QPanda3 exhibits significant 
systematic performance advantages over Pennylane Lightning. In shallow small-scale 
quantum circuits (depth $\le$6 layers, qubits $\le$8), QPanda3 demonstrates prominent 
acceleration effects, reaching up to 26.0x (2-layer/4-qubit configuration). 
However, this advantage exhibits dual-decay characteristics with parameter variations: 
when fixing qubit count, the speedup ratio decreases sharply with increasing depth 
(e.g., for 4-qubit circuits: from 26.0 at 2 layers to 1.2 at 12 layers); when 
fixing depth, the speedup ratio decays progressively with qubit scaling (e.g., at 
2-layer depth: from 26.0 at 4 qubits to 6.5 at 16 qubits).

The impact of circuit depth on acceleration effects is particularly significant: 
a depth of 8 layers constitutes a critical turning point, where speedup ratios 
are generally $>$4.0 below this threshold (e.g., 5.3 for 6-layer/4-qubit circuits) 
but typically $<$2.0 above it (e.g., only 1.6 for 10-layer/4-qubit circuits). A 
qubit count of 12 emerges as another demarcation point-below this scale, speedup 
ratios exhibit drastic fluctuations (maximum 73\% decrease from 4 to 8 qubits), 
while above this scale they tend to stabilize ($<$5\% variation between 14-16 qubits).

The spatial distribution of performance advantages exhibits a clear hierarchical 
structure: strong acceleration zones ($>$5.0) are concentrated in the upper-left 
triangular region (depth $\le$6 layers \& qubits $\le$8), weak acceleration zones (1.2-2.0) 
occupy the lower-right triangular region (depth $\ge$10 layers or qubits $\ge$14), and transition 
zones (1.6-4.9) are distributed along the diagonal. This distribution pattern indicates 
that the inhibitory effect of circuit depth on acceleration surpasses that of qubit 
scale, converging to a 1.2-2.0x performance gap between the two frameworks when 
depth $\ge$10 layers or qubits $\ge$14.

\paragraph{Comparison with Qiskit}

\begin{figure}[htbp]
    \centering
    \includegraphics[width=0.5\textwidth]{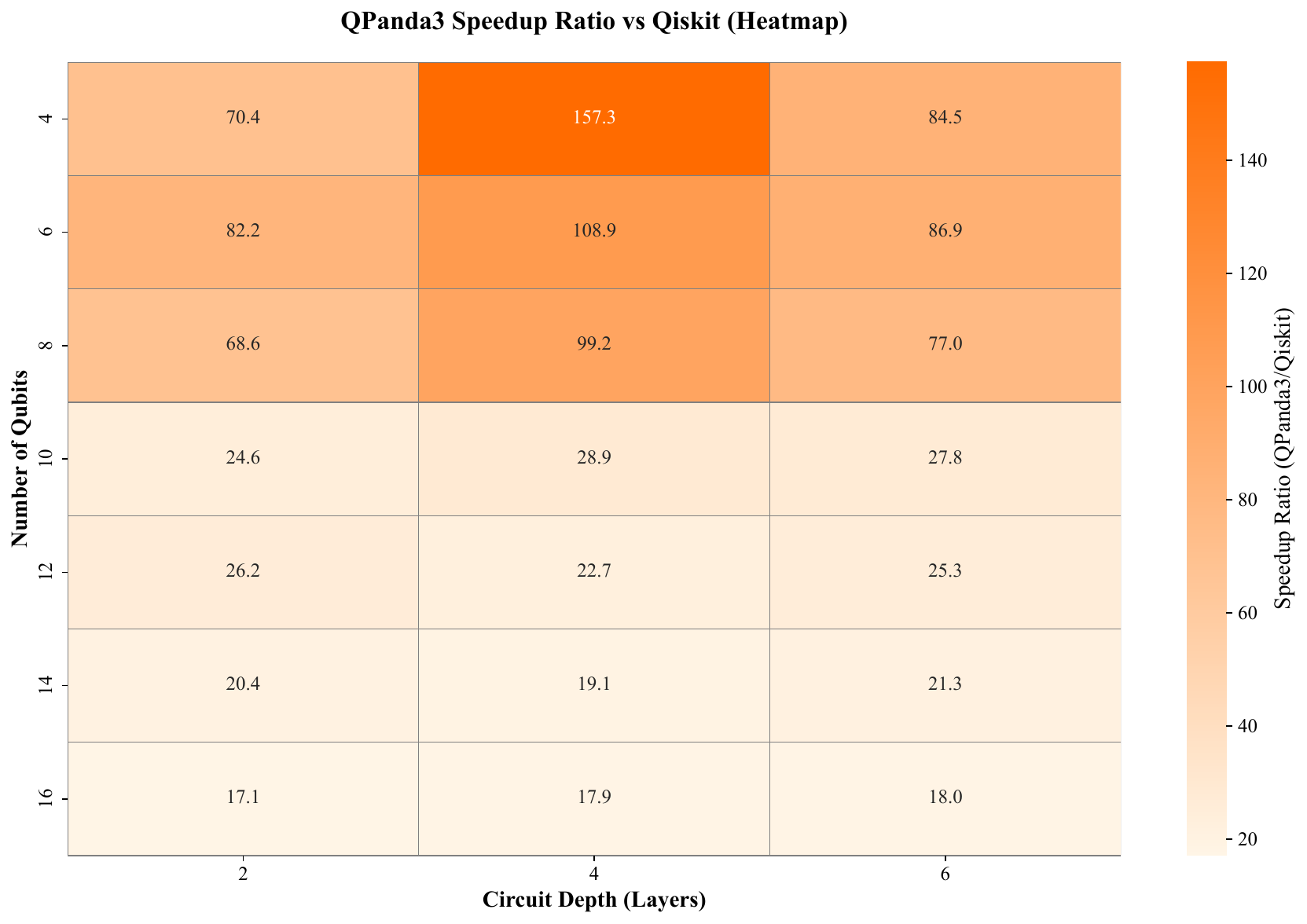}
    \caption{Speedup Heatmap vs Qiskit}
    \label{fig:speedup_heatmap_vs_qiskit}
\end{figure}

Analysis based on heatmap \ref{fig:speedup_heatmap_vs_qiskit} data indicates 
that QPanda3 demonstrates significant performance advantages compared to Qiskit, 
with its acceleration effects exhibiting clear parameter-dependent patterns. 
In medium-depth small-scale quantum circuits (4\-8 layers depth, 6\-10 qubits), 
QPanda3 achieves super-strong acceleration, peaking at 157.3 times (6 layers 
depth/6 qubits). This advantage shows asymmetric attenuation with circuit 
parameter changes: the attenuation intensity caused by increasing depth (up to 
83\%) is significantly higher than the impact of increasing qubit count (up to 
45\%). Notably, a depth of 8 layers constitutes a critical threshold-below this 
value, the speedup ratio is generally $>$68.6 times (e.g., 99.2 times for 6 layers 
depth/8 qubits), while above this value it sharply decreases to $<$27.8 times 
(e.g., 24.6 times for 10 layers depth/4 qubits).

The impact of qubit scale on acceleration exhibits phased characteristics: when 
the number of qubits $\le$10, the speedup ratio fluctuates drastically with increasing 
depth (e.g., from 70.4 to 24.6 for 4-qubit circuits when depth increases from 2 to 10 
layers); whereas when qubits $\ge$12, the speedup ratio shows blunted responses to 
parameter variations (e.g., 16-qubit circuits maintain only 17.1\-18.0x acceleration 
across different depths). This differential response forms three characteristic 
spatial distribution zones: super-strong acceleration zone ($>$80x) concentrates 
in the rectangular region with depth $\le$6 layers and qubits $\le$10; strong acceleration 
zone (40-99x) extends to depth 8 layers / qubits 14; stable acceleration zone 
(17-28x) covers the lower-right triangular region where depth $\ge$10 layers or qubits 
$\ge$12.

Even under deep-layer large-qubit-scale conditions (16 layers depth/16 qubits), 
QPanda3 maintains stable acceleration advantages exceeding 17x, with fluctuation 
amplitude $<$20\%. This phenomenon indicates that while circuit depth exerts stronger 
inhibitory effects on acceleration than qubit scale, QPanda3 consistently demonstrates 
significant performance advantages across the entire parameter space, particularly 
exhibiting groundbreaking acceleration capabilities in shallow-to-medium quantum 
circuits.

\section{Conclusion}
This paper focuses on efficient compilation and execution in quantum computing, 
introducing OriginIR and OriginBIS as intermediate representations to enhance 
programmability, transmission efficiency, and execution performance. Experimental 
results demonstrate that OriginBIS significantly outperforms OpenQASM 2.0 in 
encoding and decoding speed as well as information capacity. Additionally, QPanda3 
surpasses Qiskit in quantum circuit construction, operation execution, and 
compilation speed, exhibiting exceptional performance improvements, particularly 
in large-scale quantum circuit compilation tasks. QPanda3 
provides fundamental and efficient functionalities related to variational quantum 
circuits, offering robust support for quantum-classical hybrid computation. 
These advancements lay a solid 
foundation for the engineering applications of future quantum computing. 
\section{Future Work}
The quantum programming framework QPanda3 is poised to enhance its capabilities 
for future "quantum-HPC-AI" integrated computing by focusing on several key 
advancements. Building on its existing strength of compiling quantum circuits 
14.97 times faster than Qiskit in Benchpress tests, QPanda3 will develop hybrid 
task schedulers to enable seamless coordination between quantum computations and 
classical supercomputing resources. It aims to integrate AI-driven optimization 
tools, such as AutoML modules, to automatically reduce quantum gate counts and 
circuit depths while mitigating noise impacts. The framework will expand its 
hybrid programming interfaces to support classical HPC technologies like MPI and 
CUDA, facilitating efficient quantum-classical algorithm interoperability. 
Additionally, QPanda3 plans to implement a unified resource management platform 
for dynamic allocation of quantum hardware, supercomputing clusters, and AI 
accelerators. To strengthen ecosystem integration, it will promote standardization 
through open-source collaboration, enabling deeper compatibility with classical AI 
libraries (e.g., TensorFlow) while maintaining its low-latency compilation 
advantages. These upgrades will position QPanda3 as a core enabler for 
cross-paradigm applications in finance, drug discovery, and materials science.

\bibliographystyle{ieeetr}
\bibliography{main.bib}

\newpage
\begin{appendices}
\section{Representation and Transmission of Quantum Programs}

\subsection{User-Friendly Intermediate Representation-OriginIR}

The significant role of high-level intermediate representations of quantum programs 
in the quantum software stack cannot be overstated. QPanda3 employs OriginIR as a 
lower-level representation of quantum programs compared to high-level programming 
languages.

\subsubsection{Compatibility with QPanda2}

QPanda3 largely retains all the features of OriginIR from QPanda2, while also 
introducing appropriate extensions. Specifically, the syntax for declaring 
classical bits, declaring quantum bits, separating adjacent instruction statements, 
representing block statement scopes, and declaring quantum logic gates and 
quantum-related operations remains consistent with QPanda2. QPanda3 also supports 
all the quantum logic gates available in QPanda2. OriginIR instruction strings 
generated by QPanda2 can be correctly parsed and processed by QPanda3. Conversely, 
OriginIR instruction strings generated by QPanda3, which do not include auxiliary 
information such as comments, can also be correctly parsed and processed by 
QPanda2. OriginIR enables seamless portability of quantum programs between QPanda2 
and QPanda3.

\subsubsection{Geared Towards Researchers}

\textbf{\\(1) Complexity and Conversion}

The QASM series represents intermediate representations of quantum programs. To 
investigate the performance of quantum programs, researchers often test and compare 
the same quantum program on different platforms. QASMBench\cite{li2023qasmbench}, 
based on OpenQASM-2, serves as a relevant benchmark for quantum computing-related 
tests and has been widely adopted in numerous research studies.

Unlike the universality of the quantum circuit model, intermediate representations 
(IRs) of quantum programs exhibit significant variations due to differences in 
quantum computing devices, abstraction levels, and syntactic structures. Even IRs 
of quantum programs targeting the same quantum computing device and operating at 
similar abstraction levels can still have considerable discrepancies. Taking the 
QASM series as an example, OpenQASM-2 and OpenQASM-3 differ in many syntactic 
aspects, to the extent that even Qiskit itself requires separate conversion tools 
for OpenQASM-2 and OpenQASM-3. When other software platforms provide auxiliary 
conversion tools, they often only support a subset of the main content within the 
QASM series. The more complex the syntactic rules of an intermediate representation 
of a quantum program are, the greater the difficulty in converting it to other 
quantum program IRs. This limits researchers' ability to use quantum program IRs 
for porting and experimental research of quantum programs across multiple 
platforms. This issue has given rise to research on more universal quantum program 
IRs, such as cQASM\cite{khammassi2018cqasm}.

OriginIR offers a perspective for reducing the difficulties associated with 
intermediate conversions of quantum programs, from the standpoint of low 
complexity. OriginIR provides a concise representation of quantum programs based 
on a quantum logic gate model, with minimal details unrelated to the circuit. 
Furthermore, OriginIR adopts a unified format for describing classical registers, 
as opposed to the multiple formats supported by the QASM series. Additionally, 
the syntactic rules of OriginIR are straightforward. These characteristics enable 
researchers to achieve the transplantation of quantum programs from OriginIR 
through simple processing.

To meet the needs of researchers, QPanda3 provides a tool for converting QASM to 
OriginIR. It should be noted that QPanda3 only supports a subset of the syntactic 
rules of OpenQASM-2. However, this tool is already capable of handling the majority 
of use cases in QASMBench.

\textbf{\\(2) Readability}

To enhance the readability of OriginIR, compared to QPanda2, QPanda3 allows users 
to freely add line comments and block comments to OriginIR instruction strings, and 
also permits appropriate indentation of OriginIR line instructions using spaces. 
Although these comments and indentations are filtered out during the process of 
parsing OriginIR instructions in QPanda3, the information in the comments provides 
important reference for users, and the indentation significantly improves the 
readability of the OriginIR instruction sequence. The improvement in readability 
will greatly facilitate researchers in conducting quantum computing-related 
research based on OriginIR.

\subsection{Intermediate Representation for Quantum Program Transmission - OriginBIS}

Quantum program transmission is a crucial component of the quantum computing 
software stack. Although user-friendly intermediate representations (IRs) such as 
OriginIR can be used to transmit quantum programs in their entirety, these IRs 
carry a considerable amount of redundant information. Aspects like readability are 
unnecessary for mere inter-machine quantum program transmission, and excessive 
redundant information significantly reduces the efficiency of quantum program 
transmission and increases unnecessary transmission costs. Compressing 
user-friendly IRs like OriginIR before transmission is an optional solution to 
reduce quantum program transmission costs. However, this approach adds compression 
and decompression steps, increasing the time cost of the conversion process. 
QPanda3 has designed a binary instruction stream (BIS), named OriginBIS, as an 
intermediate representation for quantum program transmission. This IR operates at 
the binary machine instruction level, enabling efficient quantum program 
transmission without adding unnecessary conversion time costs.

OriginBIS employs a stream-based approach for organization and transmission. On 
one hand, consecutive quantum program instructions are concatenated in data 
packets and transmitted over communication links in a streaming manner. On the 
other hand, data within individual quantum program instructions is concatenated 
in data packets using a streaming approach. This stream-based method allows 
OriginBIS to provide efficient binary representations tailored to different 
quantum program instructions and also supports the application of specialized 
optimization techniques for data transmission in quantum programs, thereby 
enhancing the efficiency of quantum program transmission.

OriginBIS employs a classification-based binary instruction format alignment 
scheme to enhance the speed at which devices process OriginBIS instructions. 
Specifically, OriginBIS first classifies the various instructions in a quantum 
program based on their function, the type and number of associated data. Then, 
it designs a binary instruction format with a unified length and bit fields for 
each category of instructions. This provides efficient support for the 
instruction dispatch and execution of quantum program instructions within the 
device. The classification strategy allows OriginBIS to provide adaptive and 
efficient binary instruction formats for complex instruction types, and the fact 
that instructions within the same category share the same format enables 
efficient processing of each category of instructions by the device. In this way, 
OriginBIS effectively balances the richness of quantum program instructions and 
the efficiency of device processing of these instructions.

In some quantum program transmission scenarios, the time cost of transmission may 
be a primary concern. To address this, OriginBIS incorporates an adaptive 
compression scheme. This scheme is a lightweight data compression approach based 
on variable-length integer encoding. It applies limited compression to the 
quantum program.

\section{Optimized Compilation of Quantum Circuits}

The compilation process of quantum circuits involves the transformation from 
high-level abstract descriptions to hardware-executable instructions. It 
encompasses the entire process from initialization to final translation, with 
each step closely interconnected, collectively forming a complete, efficient, 
and reliable compilation chain. Achieving high-quality execution of quantum 
circuits and obtaining desirable results relies on effective adaptation and 
utilization of quantum computing resources. High-quality compilation of quantum 
circuits is a crucial approach to breaking through the performance bottlenecks 
in their execution. Efficient compilation of quantum circuits aims to reduce the 
time and economic costs associated with processing large-scale quantum circuits. 
The practical application of quantum software is inseparable from efficient 
quantum circuit compilation schemes.

\subsection{Transpiler of QPanda3}

QPanda3 features a device-oriented modern quantum circuit transpiler. Significant 
differences exist among various quantum processors. Regardless of the disparities 
in implementation technologies, there are substantial variations in the topological 
structures of physical qubits among different quantum processors. Not only do 
different product series exhibit marked differences, but different versions 
within the same series may also vary due to factors such as the number of qubits. 
Additionally, dynamic allocation of physical qubit resources can lead to changes 
in the topological structure of available physical qubits on the same quantum 
processor. These factors pose challenges for quantum circuit transpilation. 
Therefore, modern quantum circuit transpilers need to generate high-quality 
transpiled circuits that are tailored to the current resource status of the 
quantum computing device. The quantum circuit transpiler built into QPanda3 takes 
the topological structure of available physical qubits of the quantum computing 
device as an input parameter and transpiles a specific quantum logic circuit into 
a sequence of machine instructions that can be efficiently executed on the 
corresponding device.

QPanda3 decouples the key steps in the internal implementation of its quantum 
circuit transpiler, thereby enhancing the flexibility for functional extensions 
of the compiler module. These key steps include preprocessing, qubit mapping, 
qubit routing, optimization, and machine instruction generation. The preprocessing 
step, as the starting point of the compilation process, is responsible for 
initializing the quantum circuit, including identifying the qubits, quantum gates, 
and their dependencies within the circuit, laying the foundation for subsequent 
operations. The objective of qubit mapping is to map logical qubits to the 
available physical qubits in the quantum processor. This process involves matching 
the quantum circuit with the corresponding topological structure of the physical 
qubits. After qubit mapping, the allocated physical qubits may not satisfy the 
connectivity constraints originally imposed by the quantum circuit. Qubit routing, 
as the immediate subsequent step to qubit mapping, is specifically designed to 
address this issue. Optimization is a crucial part of the quantum circuit 
compilation process, and reasonable optimization strategies can improve the speed 
and quality of compilation to varying degrees. Unlike other steps that have a 
strict sequential order among them, optimization may occur throughout various 
other steps, enhancing compilation efficiency from different aspects. Some 
targeted and independent optimization strategies can also be abstracted into a 
separate step. QPanda3 employs such a strategy, using OptimizationPass for 
abstraction and management, thereby facilitating multiple optimizations before 
qubit mapping and after qubit routing.

To promote the development of quantum computing-related research and industries, 
QPanda3 provides external access to its built-in transpiler through dedicated 
interfaces. These interfaces enable users to perform device-based quantum circuit 
compilation tasks. Specifically, users are required to generate the edge set of 
an undirected graph that corresponds to the topological structure of the available 
physical qubits. Apart from a few specific constraints that need to be satisfied, 
users can complete this step based on the information of any quantum computing 
device. Subsequently, users can utilize the interfaces provided by QPanda3 to 
generate the corresponding machine instruction sequences based on the designed 
quantum programs and the edge set data. Quantum programs designed using QPanda3 
can seamlessly leverage these interfaces. For quantum programs developed on other 
platforms, users can employ the intermediate representation conversion tool 
provided by QPanda3 to facilitate quantum program migration. Furthermore, to 
further enhance user convenience, QPanda3 offers an interface for randomly 
generating the edge set of an undirected graph. This interface can also generate 
classic square, fully connected, and linear topological structures based on the 
number of qubits.

To achieve efficient quantum circuit compilation, QPanda3 has introduced several 
improvements over its predecessor, QPanda2. In addition to making the built-in 
transpiler accessible externally for the first time, QPanda3 has also optimized 
other components used in quantum program design and the quantum circuit 
compilation process. QPanda3 has undergone a comprehensive upgrade in its 
internal conversion mechanisms. During the construction of quantum circuits, 
extensive internal data structure and algorithm conversions are often required 
to ensure the correctness and executability of the circuits. QPanda3 employs more 
efficient data structures and algorithms, optimizing these internal conversion 
processes and significantly enhancing the speed of circuit construction. Whether 
it is the merging of quantum gates, simplification of circuits, or allocation of 
resources, QPanda3 can accomplish these tasks at a faster pace, thereby improving 
overall compilation efficiency.

\subsection{Starting with an Example}

This section demonstrates the use of the built-in transpiler in QPanda3 through 
a simple example. In the following explanation, we will also introduce the basic 
steps for designing quantum programs using QPanda3, as well as useful tools for 
quantum circuit visualization and topology data generation.

\textbf{(1) A Quantum Circuit Compilation Task}

\begin{figure}[h]
    \centering
    \begin{subfigure}[t]{0.3\textwidth}
        \centering
        \includegraphics[width=\linewidth]{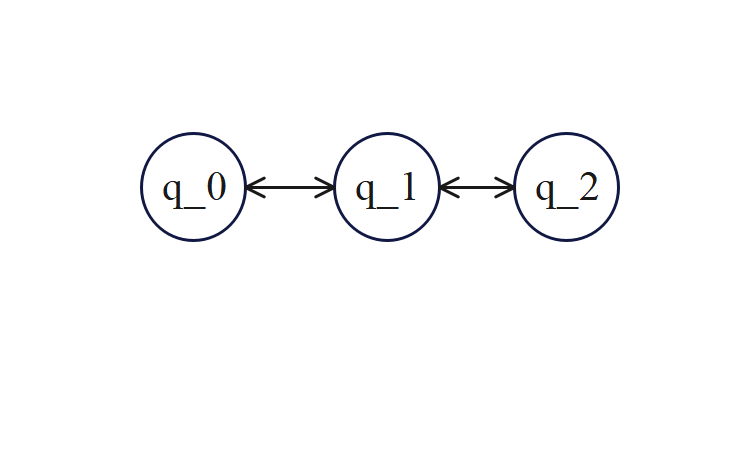}
        \caption{Qubits' Topology}
    \end{subfigure}
    \hfill
    \begin{subfigure}[t]{0.3\textwidth}
        \centering
        \includegraphics[width=\linewidth]{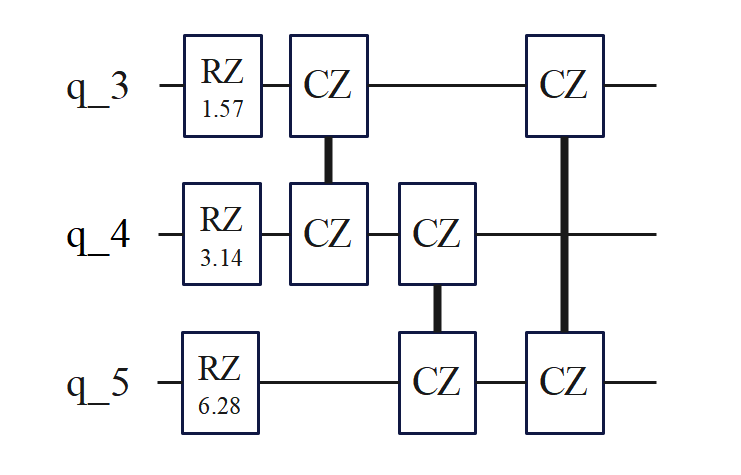}
        \caption{Quantum Logic Circuit}
    \end{subfigure}
    \hfill
    \begin{subfigure}[t]{0.3\textwidth}
        \centering
        \includegraphics[width=\linewidth]{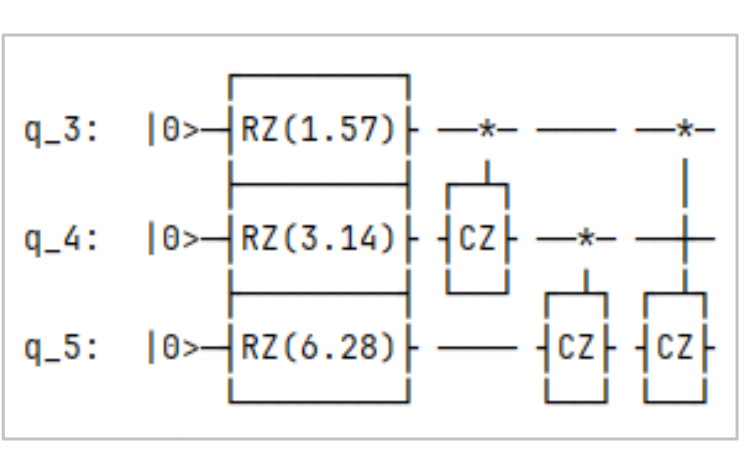}
        \caption{Visualization Result of The Circuit}
    \end{subfigure}
    \caption{Quantum circuit compilation task}
    \label{fig:quantum-circuit-compilation-task}
\end{figure}

Figure \ref{fig:quantum-circuit-compilation-task}, subgraph (a) illustrates the 
topology of available qubits on a certain quantum computing device using an 
undirected graph, with the corresponding physical qubits labeled as $q_0$, $q_1$, 
and $q_2$. Subgraph (b) depicts a quantum logic circuit. The compilation task in 
this section is to obtain the machine instruction sequence for this circuit that 
adapts to the given topology. Subgraph (c) presents the visualization result of 
the circuit in subgraph (b) using QPanda3. It is worth noting that the CZ gate is 
a symmetric gate.

QPanda3 uses "pyqpanda3" as its Python package name, which differs from the package 
name "pyqpanda" used in QPanda2. This allows users to import both the "pyqpanda3" 
and "pyqpanda" packages simultaneously, thereby ensuring compatibility with 
different versions. It should be noted that QPanda3 shares most of its naming 
conventions with QPanda2, so care should be taken to avoid issues related to name 
conflicts during the import process.

The core components of QPanda3 are exported by the package pyqpanda3.core. These 
core components include, but are not limited to, natively supported quantum logic 
gates such as RZ, X1, and SWAP. They also encompass the abstract object QProg for 
quantum programs. \lstinline{draw_qprog} and \lstinline{set_print_options} are two 
interfaces related to quantum circuit visualization tools. Specifically, 
\lstinline{set_print_options} is used to control the number of decimal places 
displayed for gate parameters when visualizing parametric gates, while 
\lstinline{draw_qprog} is used for visualizing quantum circuits.

\begin{sloppypar}
The components related to the built-in circuit transpiler in QPanda3 are exported 
by pyqpanda3.transpilation. The Transpiler class is used for managing circuit 
compilation. The \lstinline{generate_topology} function is used for randomly 
generating topology data.
\end{sloppypar}


\begin{lstlisting}[language=Python, caption=An example Python program that evaluates the properties of quantum states., label=alg_char]
from pyqpanda3.core import QProg, draw_qprog, RZ, X1, CZ, SWAP, set_print_options
from pyqpanda3.transpilation import Transpiler, generate_topology
\end{lstlisting}

\textbf{(2) Constructing Quantum Logic Circuits}


\begin{lstlisting}[language=Python, caption=An example Python program that evaluates the properties of quantum states., label=alg_char]
prog_1 = QProg()
prog_1 << RZ(3,1.57) << RZ(4,3.14) << RZ(5,6.28) << CZ(3,4) << CZ(4,5) << CZ(3,5)
set_print_options(2)
\end{lstlisting}

\begin{figure}[h]
    \centering
    \includegraphics[width=0.5\textwidth]{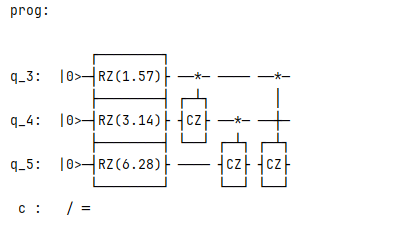}
    \caption{Result of Constructing Quantum Logic Circuits}
    \label{fig:visualization-of-the-quantum-circuit}
\end{figure}

The output is show as Fig \ref{fig:visualization-of-the-quantum-circuit}

\textbf{(3) Setting the Topology of Physical Qubits}

This line of code defines a two-dimensional Python list. This list stores the edge 
set of an undirected graph, which can represent the topology of the currently 
available physical qubits on a quantum computing device. Users can also generate 
similar data using the \lstinline{generate_topology} interface.

\begin{lstlisting}[language=Python, caption=An example Python program that evaluates the properties of quantum states., label=alg_char]
topo = [[0,1],[1,2]]
\end{lstlisting}

\textbf{(4) Compiling with Different Optimization Levels}

The transpile method of the Transpiler object is used to perform the quantum 
circuit compilation step. Thanks to the generality of the quantum circuit model, 
a quantum circuit composed of quantum gates and other operations that correspond 
one-to-one with machine instructions can be mapped to an executable sequence of 
machine instructions. This brings many conveniences, including the ability to 
observe the results of quantum circuit compilation using quantum circuit 
visualization methods.

Due to the dynamic nature of the quantum circuit compilation process, different 
but equivalent compiled circuits may be generated each time. The output content 
is extensive and will not be shown here. In the subsequent sections, some key 
output results of this program will be extracted and analyzed.


\begin{lstlisting}[language=Python, caption=An example Python program that evaluates the properties of quantum states., label=alg_char]
transpiler = Transpiler()
prog_level_0 = transpiler.transpile(prog_1, topo, {}, 0)
print('Transpiler lavel 0: \n', draw_qprog(prog_level_0, param_show=True)) 
\end{lstlisting}

\subsection{Qubit Mapping and Routing}

\textbf{(1) Qubit Mapping}

Mapping logical qubits in a quantum circuit to physical qubits, taking into 
account the topology and connectivity of the quantum hardware, to ensure the 
executability of the circuit.

\begin{figure}[h]
    \centering
    \includegraphics[width=0.5\textwidth]{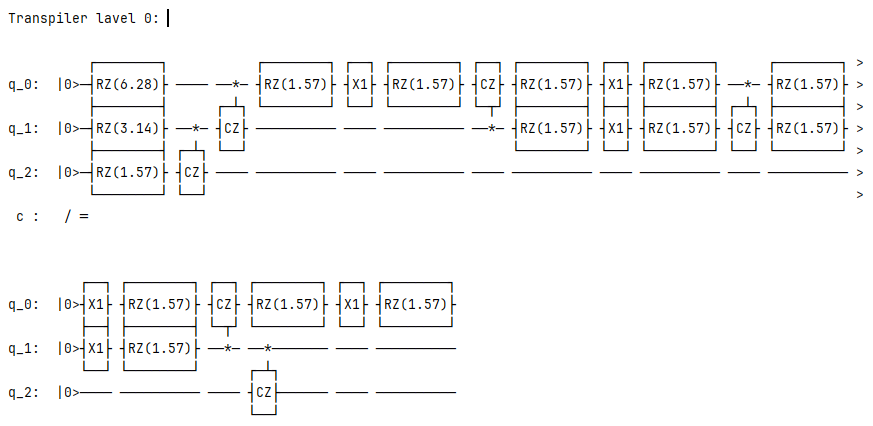}
    \caption{Qubit Mapping Result}
    \label{fig:transpilation-result-of-the-quantum-circuit}
\end{figure}

An example of a compilation result from the code in Section 5.2 is shown in 
Figure \ref{fig:transpilation-result-of-the-quantum-circuit}. Based on the 
parameters of the first three RZ gates added to the quantum circuit, it can be 
determined that logical qubit $q_5$ is mapped to physical qubit $q_0$, logical 
qubit $q_4$ is mapped to physical qubit $q_1$, and logical qubit $q_3$ is mapped 
to physical qubit $q_2$.

\textbf{(2) Qubit Routing}

Inserting necessary SWAP gates between physical qubits to address connectivity 
limitations in quantum hardware and ensure that quantum gates can be executed 
smoothly on physical qubits in the predetermined order.

In the quantum logic circuit example in Section 5.2, there are direct CZ gate 
connections between each pair of the three logical qubits, including a direct 
connection between logical qubit $q_5$ and logical qubit $q_3$. However, in the 
aforementioned qubit mapping result, physical qubit $q_0$ and physical qubit 
$q_1$ are not directly connected, which violates the corresponding connectivity 
constraint.

In the optimization compilation at level 0 in QPanda3, the SWAP gate is 
implemented equivalently using the following sub-circuit.

\begin{figure}[h]
    \centering
    \includegraphics[width=0.5\textwidth]{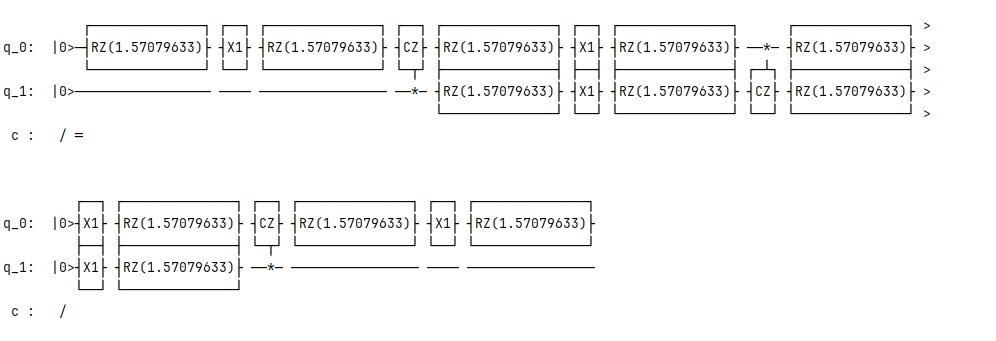}
    \caption{Qubit Routing Result}
    \label{fig:fig6}
\end{figure}

Comparing the two figures below, it can be observed that the compilation effect 
is equivalent to using SWAP gates to satisfy the connectivity constraints. The 
equivalent circuit corresponding to the aforementioned compilation result is 
shown in the figure below. Clearly, this circuit is equivalent to the original 
quantum logic circuit. It should be noted that the example uses gates such as 
RZ and CZ, which are not decomposed or transformed during the preprocessing 
stage.

\begin{figure}[h]
    \centering
    \includegraphics[width=0.5\textwidth]{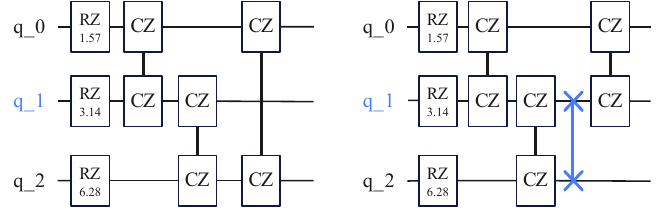}
    \caption{Using SWAP Gates to Satisfy The Connectivity Constraints}
    \label{fig:fig7}
\end{figure}

It should be noted that there are multiple ways to perform layout and routing for 
this quantum logic circuit. When compiling this quantum logic circuit, QPanda3 
may not always use this specific layout and routing scheme, as other layout and 
routing solutions can still satisfy the target requirements.

\subsection{Optimization}

QPanda3 adopts a strategy that combines local and global optimization. Local 
optimization primarily focuses on optimizing small regions within the quantum 
circuit, such as merging adjacent quantum gates and eliminating redundant 
operations, to reduce the complexity and depth of the circuit. Global 
optimization, on the other hand, considers the optimization of the entire 
quantum circuit by adjusting the order of quantum gates, reallocating resources, 
and other means to further improve the execution efficiency of the circuit.

\subsubsection{Sabre}

Sabre\cite{li2019tackling} is an advanced quantum circuit compilation algorithm. 
This algorithm incorporates a decay effect and employs a bidirectional heuristic 
search based on swaps. It achieves a good trade-off between the depth of the 
quantum circuit and the number of gates. QPanda3 has further optimized the 
implementation of Sabre and extensively applies it in the compilation process. 
Sabre plays a crucial role in optimization, layout, and routing, and through 
efficient algorithm design, it can handle complex quantum circuits, improving 
the accuracy and efficiency of compilation. The Sabre algorithm is closely 
integrated with the characteristics of quantum hardware, enabling the compiled 
circuit to better adapt to the hardware's execution environment, thereby ensuring 
efficient execution of the algorithm.

\subsubsection{Independent Optimization Steps}
This step involves isolating certain optimization strategies as a standalone 
procedure, intended for pre-layout and post-routing optimization. The optimization 
strategies concerned encompass merging adjacent quantum gates, eliminating 
redundant operations, simplifying complex gate sequences, and so forth. These 
measures aim to reduce the depth and complexity of the circuit, thereby enhancing 
execution efficiency. The following two simplistic examples illustrate the 
optimization effects. It is noteworthy that the examples utilize gates such as RZ 
and CZ, which are not decomposed or transformed during the preprocessing stage.

\textbf{\\(1) Example 1 - Merging Adjacent Gates}

This circuit presents a scenario that is amenable to optimization, where 
consecutive RZ gates can be consolidated. This scenario can be further generalized 
to the optimization of various types of consecutive single-gate operations. In 
this instance, when QPanda3 compiles the circuit at optimization level 0, no 
optimization strategies are employed. Consequently, the result retains two 
consecutive RZ gate operations, consistent with the quantum logic circuit as 
depicted in the figure. However, when QPanda3 compiles the circuit at optimization 
levels 1 and 2, it merges these two consecutive RZ gates into a single RZ gate. 
The optimized circuit necessitates only the time required to execute one RZ gate, 
whereas the pre-optimized circuit consumes approximately twice the duration.

\begin{figure}[htbp]
    \centering
    \begin{subfigure}[h]{0.45\textwidth}
        \centering
        \includegraphics[width=0.7\linewidth]{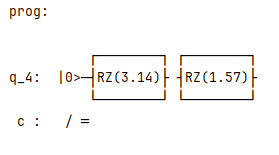}
        \caption{}
    \end{subfigure}
    \hfill
    \begin{subfigure}[h]{0.45\textwidth}
        \centering
        \includegraphics[width=0.7\linewidth]{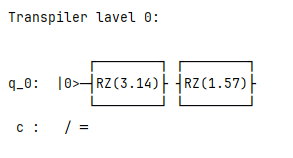}
        \caption{}
    \end{subfigure}

    \begin{subfigure}[h]{0.45\textwidth}
        \centering
        \includegraphics[width=0.7\linewidth]{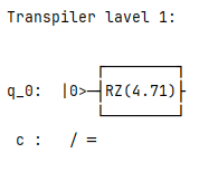}
        \caption{}
    \end{subfigure}
    \hfill
    \begin{subfigure}[h]{0.45\textwidth}
        \centering
        \includegraphics[width=0.7\linewidth]{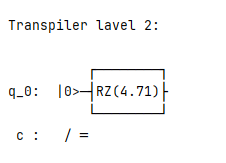}
        \caption{}
    \end{subfigure}
    \caption{Merging Adjacent Gates}
    \label{fig:fig8}
\end{figure}

\textbf{\\(2) Example 2 - Eliminating Redundant Gates}

This circuit showcases a scenario that can be optimized, where consecutive SWAP 
gates exist and can be eliminated. This scenario can be extended to the 
optimization of various types of consecutive two-qubit gate operations. When 
QPanda3 performs quantum circuit compilation at optimization level 2, adjacent 
SWAP gates are removed. The elimination of redundant gates effectively enhances 
the utilization of quantum bits.

\begin{figure}[htbp]
    \centering
    \begin{subfigure}[h]{0.45\textwidth}
        \centering
        \includegraphics[width=0.7\linewidth]{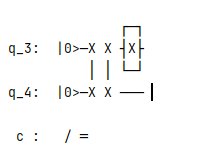}
        \caption{}
    \end{subfigure}
    \hfill
    \begin{subfigure}[h]{0.45\textwidth}
        \centering
        \includegraphics[width=0.7\linewidth]{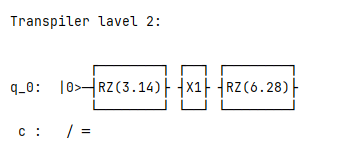}
        \caption{}
    \end{subfigure}
    \caption{Eliminating Redundant Gates}
    \label{fig:fig9}
\end{figure}

\subsubsection{Pre-placement Optimization and Post-routing Optimization}

Prior to the placement stage, QPanda3 conducts a series of preprocessing 
optimizations on the quantum circuit. These optimization operations aim to 
alleviate the burden on subsequent placement and routing processes, thereby 
improving compilation efficiency. For instance, by merging adjacent quantum 
gates, the number of gates that need to be considered during the placement stage 
can be reduced. Additionally, by eliminating ineffective operations, the 
occupation of quantum hardware resources can be minimized. After the routing 
stage is completed, QPanda3 further optimizes the circuit to enhance its 
execution efficiency even more.

\section{Device-Based Quantum Program Analysis}

Whether a quantum program can be executed efficiently on a specific computing 
device and yield desirable results determines its suitability for 
performance-sensitive tasks and its advantage in terms of computational resource 
costs. Efficient utilization of device resources in quantum programs relies on 
device-oriented fine-tuning. On one hand, skillful program design based on 
programming experience can be employed; on the other hand, purposeful adjustments 
can be made according to the runtime performance of the quantum program on the 
computing device. This section focuses on the latter, introducing the tools 
provided by QPanda3 for device-based quantum program analysis.

\subsection{Runtime State Information and Quantum Program Analysis}

In classical computing, runtime state information of various hardware components 
can be directly obtained through dedicated counters, registers, and debug-level 
interfaces during program execution. Due to the unique nature of quantum 
mechanics, however, observing a quantum processor without interrupting its 
operation poses challenges. Typically, one cannot read information from physical 
qubits as one would from bits in a classical register without causing the program 
to halt. Nevertheless, in the NISQ (Noisy Intermediate-Scale Quantum) era, a 
necessary condition for quantum computing is the ability to drive the directed 
evolution of qubits externally. Information such as the type and duration of 
externally initiated drive operations can clearly be acquired and fed back to 
the software layer. Additionally, statistical information on certain 
runtime-inaccessible parameters can be obtained through multiple trials. This 
quantum device information can be published as product specification parameters 
on the one hand, and re-collected as needed on the other. In the context of 
quantum program analysis, the former is suitable for analyzing standard product 
specifications, while the latter is applicable to analyzing the current state of 
the device.

Software is an essential component for program analysis. Statistical information 
about the hardware can be used as input data for direct analysis by the software. 
Alternatively, runtime-related information from the hardware can be collected 
within the software itself to analyze the program's performance on the current 
hardware during the present time period.

QPanda3 is a modern software tool designed to serve quantum computing. It provides 
device-oriented quantum program analysis tools. The open interfaces allow users to 
perform device-specific quantum program analysis based on the specification 
parameters and runtime statistical information of quantum computing devices. 
Integrated within the quantum cloud, QPanda3 operates in high synergy with 
quantum computers to meet the demands of quantum program analysis for various 
objectives. Currently, QPanda3 offers quantum circuit analysis tools tailored to 
device performance and quantum program performance analysis tools based on 
workflow and device information.

\subsection{Quantum Circuit Analysis for Device Performance}

\subsubsection{Overview}

QPanda3 extends the quantum computing performance benchmarks to the quantum 
circuit analysis context. Specifically, QPanda3 incorporates key benchmarks 
proposed by SupermarQ\cite{tomesh2022supermarq}, including Program Communication, 
Critical-Depth, Entanglement-Ratio, Parallelism, and Liveness. These benchmarks 
serve to measure the quantum computing performance of quantum circuits on specific 
devices, providing guidance for users to evaluate and improve quantum programs.

When utilizing the open interfaces provided by QPanda3, users are required to 
supply the compiled circuit as the object of analysis. The compiled circuit can 
be obtained through the circuit compilation tools introduced in Section 5.

QPanda3 packages and returns the quantum program analysis results based on the 
aforementioned benchmarks in the form of Python lists. This approach facilitates 
users in collecting corresponding result data when analyzing a large number of 
circuits. Additionally, QPanda3 provides corresponding visualization tools.

Here, an example is provided to illustrate how to use QPanda3 to obtain the 
performance metrics of a quantum program, including Program Communication, 
Critical-Depth, Entanglement-Ratio, Parallelism, and Liveness. QPanda3 offers 
quantum program analysis functionality through the 
\lstinline{draw_circuit_features} interface in the pyqpanda3.profiling package. 
The following code constructs a simple quantum circuit, which is managed using a 
QCircuit object. Finally, the \lstinline{draw_circuit_features} function is 
utilized to perform the analysis.


\begin{lstlisting}[language=Python, caption=An example Python program that evaluates the properties of quantum states., label=alg_char]
from pyqpanda3.profiling import draw_circuit_features
from pyqpanda3.core import QCircuit, H, CNOT, X
circuit = QCircuit(4)
circuit << H(0) << CNOT(0, 1) << CNOT(1, 2) << X(3)
draw_circuit_features(circuit, True)
\end{lstlisting}

\begin{figure}[h]
    \centering
    \includegraphics[width=0.5\textwidth]{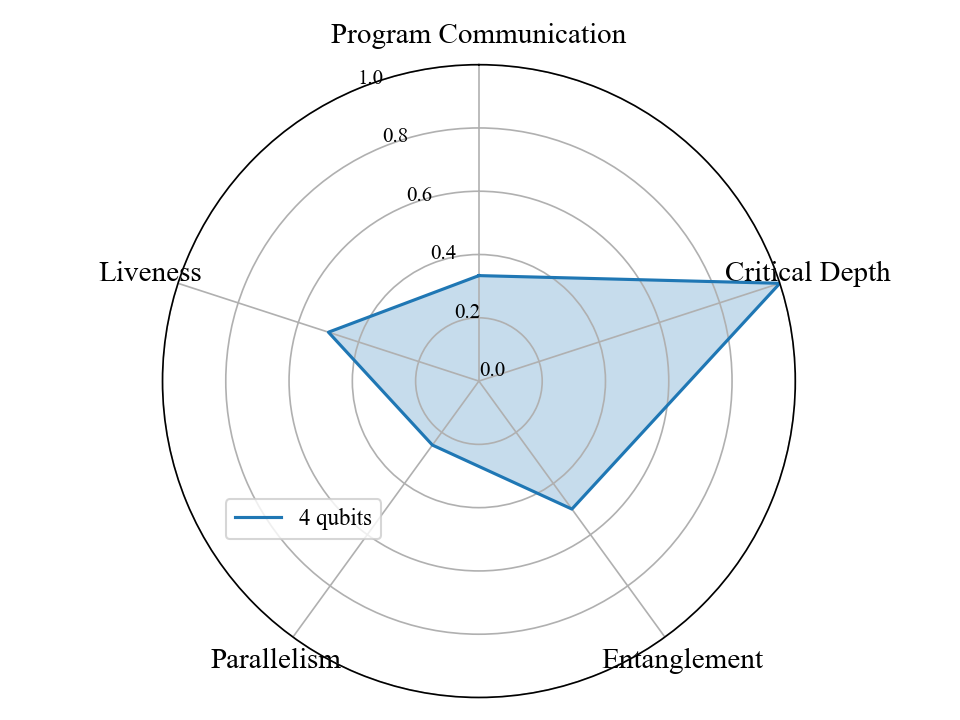}
    \caption{Performance Metrics of A Quantum Program}
    \label{fig:performance_analysis}
\end{figure}

The corresponding visualization results are shown in Figure \ref{fig:performance_analysis}.

\subsection{Performance Analysis of Quantum Programs Based on Flow and Device Information}

Inspired by classical program analysis methods, QPanda3 provides users with a tool 
for analyzing quantum programs based on flow and device information. This tool 
assists users in analyzing the utilization rates and execution time percentages 
of various quantum logic gates within their respective processes in quantum 
programs, providing a basis for users to evaluate and improve their quantum 
programs. This tool is implemented based on gprof. Next, we will first 
demonstrate the usage of relevant interfaces through a code example. Then, 
centering around this example, we will introduce the differences between this 
feature of QPanda3 and classical program analysis, as well as the other software 
components of this functionality.

\subsubsection{An Example}

QPanda3 provides the functionality for performance analysis of quantum programs 
based on flow and device information through the \lstinline{draw_circuit_profile} 
interface in the pyqpanda3.profiling package. In the following code, two quantum 
circuits, cir1 and cir2, are first constructed, and then the 
\lstinline{draw_circuit_profile} function is used to analyze the circuit cir2. 
QPanda3 abstracts circuits consisting solely of quantum gates using the QCircuit 
class. Adding one QCircuit object to another represents the addition of a new 
quantum sub-circuit. The device information used includes the quantum gates 
supported by the quantum processor employed and the average execution time of 
these gates on the quantum processor. This information is passed as the second 
parameter to the \lstinline{draw_circuit_profile} function in the form of a 
Python dictionary.

\begin{lstlisting}[language=Python, caption=An example Python program that evaluates the properties of quantum states., label=alg_char]
from pyqpanda3.profiling import draw_circuit_profile
from pyqpanda3.core import QCircuit, H, CNOT, X
cir1 = QCircuit(4)
cir1 << H(0) << CNOT(0, 1) << CNOT(1, 2) << X(3) << X(3)
cir2 = QCircuit(4)
cir2 << X(1) << cir1
draw_circuit_profile(cir2, {'CNOT': 200, 'H': 60, 'X': 30}, True)
\end{lstlisting}

\begin{figure}[htbp]
    \centering
    \includegraphics[width=0.4\textwidth]{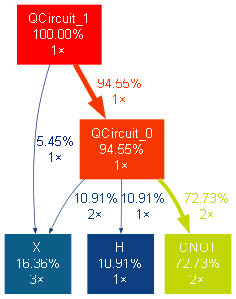}
    \caption{Profile of Circuit}
    \label{fig:circuit_profile}
\end{figure}

In this code, the third parameter of the \lstinline{draw_circuit_profile} function 
controls whether visualization is performed. The corresponding visualization 
result is shown in Figure \ref{fig:circuit_profile}.

\subsubsection{Comparison with Classical Program Analysis}

Quantum sub-circuits bear similarities to called functions (subroutines) in 
classical programs. Both represent the totality of program operations within a 
certain period of time. The former represents all quantum logic gate operations 
within a period, while the latter represents all classical program operations 
within a period. Both can also contain other sequences of operations. A quantum 
sub-circuit can contain other sub-circuits, and a called function (subroutine) 
can call other functions (subroutines). Therefore, both can be effectively 
described using directed graphs. Function (subroutine) calls can be accurately 
described using edges in a directed graph. Similarly, the inclusion relationships 
between sub-circuits can also be accurately described using edges in a directed 
graph. From the example above, it can be seen that QPanda3 accurately describes 
the relationships between sub-circuits and between circuits and quantum gates 
using a directed graph. In Figure 11, QCircuit\_0 corresponds to the first 
instantiated QCircuit object, cir1, and QCircuit\_1 corresponds to the second 
instantiated QCircuit object, cir2. The arrow from QCircuit\_1 to QCircuit\_0 
indicates that the quantum circuit corresponding to QCircuit\_1 (cir2) has the 
quantum circuit corresponding to QCircuit\_0 (cir1) as a direct sub-circuit. The 
arrow from QCircuit\_0 to H indicates that the quantum circuit corresponding to 
QCircuit\_0 (cir1) directly contains the H gate. Obviously, the H gate in the 
diagram is also a sub-circuit of QCircuit\_1 (cir2). For clarity, the term "direct" 
is used here to specifically refer to relationships like those between QCircuit\_0 
and QCircuit\_1 (or between QCircuit\_0 and H).

During execution, there is a fundamental distinction between quantum sub-circuits 
and called functions (subroutines) in classical programs. In a classical 
subroutine, each instruction reutilizes the same hardware computing resources, 
such as the program instruction pointer register, stack register, and program 
status word register. The program status word register is employed to facilitate 
subroutine calls in classical programs. However, to prevent the collapse of the 
quantum system, quantum computing processors that handle entanglement will map 
all operations in a quantum circuit onto the quantum processor in a single step. 
That is, sub-circuits like cir1 are fully expanded when cir2 is executed, rather 
than being invoked and executed through instruction jumps and instruction stack 
maintenance as in classical programs.

Flow-based quantum program analysis holds significant importance. In the actual 
execution of quantum circuits, sub-circuits are unfolded, yet their utilization 
brings great convenience during the design phase of quantum logic circuits. On 
the one hand, some sub-circuits are adaptations of classical subroutines. On the 
other hand, some sub-circuits are sequences of quantum gates designed for 
specific functions. Through flow-based and device-informed quantum program 
performance analysis, users can ascertain the resource usage of various quantum 
sub-circuits and quantum gates during execution, thereby adjusting the 
implementation schemes of corresponding sub-circuits accordingly.

\subsubsection{Output Analysis Results}

The analysis results provided by \lstinline{draw_circuit_profile} offer the 
frequency and time consumption percentage of each sub-circuit and quantum gate 
within the quantum circuit. The percentages in Figure 11 represent the 
corresponding time consumption percentages, and the "Nx" notation indicates the 
frequency value N. When multiple arrows converge on a single node, the respective 
usage frequencies and time consumption percentages are summarized using a 
summation approach.

\begin{figure}[htbp]
    \centering
    \includegraphics[width=0.5\textwidth]{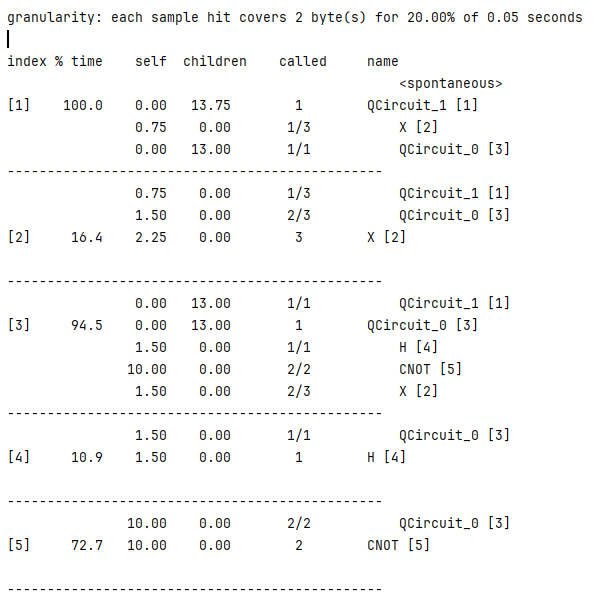}
    \caption{Gprof-Style Report}
    \label{fig:gprof_report}
\end{figure}

QPanda3 also supports outputting results in the form of gprof-style reports. 
These outputs provide more specific and detailed analysis data. Figure 
\ref{fig:gprof_report} presents partial results from the aforementioned example 
code.

\section{Variational Quantum Circuits}

The mathematical formulation of variational quantum circuits in QPanda3 can be described 
as:  $ U \left( \theta \right),\quad \theta \in \mathbb{R}^{d_1 \times \cdots \times d_n} , d_i \in \{1, \dots, D\} $.  
Since certain combinations of basic gates can exhibit complex effects, QPanda3 restricts 
$\theta$ to real-valued space. For $U \left( \theta \right),\quad \theta \in \mathbb{R}^{d} , d \in \{1, \dots, D\} $, 
when the formal parameter \cite{graham1994concrete}\cite{wilf2005generatingfunctionology} 
$\theta$ is assigned specific values 
${\mathbf{v} := (\theta_0, \theta_1, \dots, \theta_{n-1})\in \mathbb{R}^{n}} $, it yields 
a unitary matrix $U \left( \mathbf{v} \right)$ with fixed numerical elements. This unitary 
matrix corresponds to a conventional quantum circuit. When the variational quantum circuit 
consists of multiple single-parameter quantum gates with distinct formal parameters, the 
trial state of VQS described in 
\cite{yuan2019theory} can be obtained as $U(\theta) |0\rangle^{\otimes{m}}$, where:  
$U(\theta) = U_{n-1}(\theta_{n-1}) \dots U_{0}(\theta_{0}), \theta = (\theta_{n-1}, \dots ,\theta_{0})$.  
If the variational quantum circuit contains parameterized gates with repeated formal 
parameters \cite{jones2020efficient}, then:  
$U(\theta) = U_{N}(\theta_{N}) \dots U_{j}(\theta_{k})\dots U_{i}(\theta_{k}) \dots U_{0}(\theta_{0})$  
suffices, where $N \geq j \geq i \geq 0, \theta = (\theta_{N}, \dots, \theta_{0}), N \geq k \geq 0$.  
When the parameters of the quantum gates are expressions of formal parameters, the following 
formulation ($U_F$) remains applicable:  
$U(\theta) = U_{N}(F_{N}(\theta)) \dots U_{1}(F_{1}(\theta))$,  
where $F_i(\theta)$ denotes the expression associated with index $i$, generated from the formal 
parameter $\theta$ through mathematical operations. QPanda3 employs the VQCircuit class to 
provide an abstraction for variational quantum circuits. In Python code, the VQCircuit class 
can be imported as follows:

\begin{lstlisting}[language=Python, caption=An example Python program that update and modify quantum states., label=alg_mod]
from pyqpanda3.vqcircuit import VQCircuit
\end{lstlisting}

\subsection{Constructing Variational Quantum Circuits}

\subsubsection{Setting Formal Parameters $\theta$}

Our recommendation is that formal parameters (or placeholders) with the same name or 
symbol (e.g., $\theta$) should refer to the same parameter object (or variable) within 
the same context, while in different contexts they should refer to distinct parameter 
objects (or variables). It is possible for two different variational quantum circuits to 
both use $\theta$ as the name for their formal parameters, which may introduce certain 
issues. On one hand, naming conflicts commonly occur across various domains. During 
programming, such conflicts may lead to logical confusion in the code and consequently 
cause computational errors. On the other hand, larger-scale software programs may involve 
deeply nested parameter passing, making maintenance challenging due to difficulties in 
determining parameter meanings. These issues are particularly pronounced in weakly typed 
programming languages like Python. QPanda3 binds each variational quantum circuit to a 
formal parameter (corresponding to $\theta$ in $U(\theta)$), and this formal parameter 
is only valid for that specific variational quantum circuit. Through this enforced 
constraint, QPanda3 alleviates the aforementioned problems to some extent. Additionally, 
since $U(\theta)$ is a unary function of $\theta$, the process of explicitly naming the 
formal parameter (which effectively corresponds to $\theta$) can be omitted during 
programming. This not only avoids ambiguity but also simplifies the programming workflow.

For $U \left( \theta \right),\quad \theta \in \mathbb{R}^{d_1 \times \cdots \times d_n} , d_i \in \{1, \dots, D\}$,
in QPanda3 it is necessary to specify the value of each $d_{i}$ to determine the concrete 
form of $\theta$. Here each $d_i$ value also corresponds to the dimensional information of 
the numerical array assigned to $\theta$. For scalar, vector, matrix and multi-dimensional 
array forms of $\theta$, basic code examples are provided below.

(1) Scalar $\theta \in \mathbb{R}$
\begin{lstlisting}[language=Python, caption=An example Python program that update and modify quantum states., label=alg_mod]
from pyqpanda3.vqcircuit import VQCircuit
vqc = VQCircuit()
vqc.set_Param([1])
\end{lstlisting}

(2) Vector
$\theta \in \mathbb{R}^{d},d=3$
\begin{lstlisting}[language=Python, caption=An example Python program that update and modify quantum states., label=alg_mod]
from pyqpanda3.vqcircuit import VQCircuit
vqc = VQCircuit()
vqc.set_Param([3])
\end{lstlisting}

(3) Matrix $\theta \in \mathbb{R}^{d_1 \times d_2}, d_1=3, d_2=2$
\begin{lstlisting}[language=Python, caption=An example Python program that update and modify quantum states., label=alg_mod]
from pyqpanda3.vqcircuit import VQCircuit
vqc = VQCircuit()
vqc.set_Param([3,2])
\end{lstlisting}

(4) Multidimensional array $\theta \in \mathbb{R}^{d_1 \times d_2 \times d_3}, d_1=3, d_2=2, d_3=4$
\begin{lstlisting}[language=Python, caption=An example Python program that update and modify quantum states., label=alg_mod]
from pyqpanda3.vqcircuit import VQCircuit
vqc = VQCircuit()
vqc.set_Param([3,2,4])
\end{lstlisting}

\subsubsection{Adding Quantum Logic Gates}

In QPanda3, general quantum programs are abstracted using the QProg class, while 
quantum circuits without measurements are abstracted using QCircuit. Quantum programs 
or measurement-free quantum circuits are constructed by adding quantum logic gates to 
QProg or QCircuit objects. For QProg, QCircuit, and VQCircuit, QPanda3 employs a unified 
approach to adding quantum gates. To facilitate differentiation, we refer to gate parameters 
in constant value form as fixed parameters, and gate parameters whose values are determined 
by $\theta$ as variable parameters. Quantum gates without variable parameters are called 
non-parameterized quantum gates, while those containing variable parameters are called 
parameterized quantum gates.

(1) Adding a single non-parametric quantum gate or quantum gate containing only fixed parameters
\begin{lstlisting}[language=Python, caption=An example Python program that update and modify quantum states., label=alg_mod]
from pyqpanda3.vqcircuit import VQCircuit
from pyqpanda3.core import X,RX
vqc = VQCircuit()
vqc.set_Param([3])
vqc << X(0)
vqc << RX(0,3.14)
\end{lstlisting}

(2) Adding a measurement-free subcircuit
\begin{lstlisting}[language=Python, caption=An example Python program that update and modify quantum states., label=alg_mod]
from pyqpanda3.vqcircuit import VQCircuit
from pyqpanda3.core import QCircuit,X,RX
vqc = VQCircuit()
vqc.set_Param([3])
cir = QCircuit()
cir << X(0)
cir << RX(0,3.14)
vqc << cir
\end{lstlisting}

(3) Adding parameterized quantum gates

The following code first defines the formal parameters $\theta=(\theta_0,\theta_1,\theta_2)$, 
then adds a parameterized RX gate to the variational quantum circuit vqc. This RX gate 
operates on the qubit with index 0, and its parameter corresponds to $\theta_0$.

\begin{lstlisting}[language=Python, caption=An example Python program that update and modify quantum states., label=alg_mod]
from pyqpanda3.vqcircuit import VQCircuit
from pyqpanda3.core import X,RX,RY
vqc = VQCircuit()
vqc.set_Param([3])
vqc << RX(0,vqc.Param([0]))
\end{lstlisting}

QPanda3 allows assigning names to specific elements of the formal parameter 
$\theta=(\theta_0,\theta_1,\theta_2)$. If an element of $\theta$ is assigned a name, 
that name can subsequently be used to reference the element directly. Note that each 
element can only be named once. The following code first defines the formal parameters 
$\theta=(\theta_0,\theta_1,\theta_2)$, then adds a parameterized RX gate to the 
variational quantum circuit vqc. This RX gate operates on the qubit with index 0, and 
its parameter corresponds to $\theta_0$. During the addition of the RX gate, $\theta_0$ 
is assigned the name "gamma". Finally, a parameterized RY gate is added to the variational 
quantum circuit vqc. This RY gate operates on the qubit with index 1, and its parameter 
corresponds to the element of $\theta$ named 'gamma', i.e., $\theta_0$.

\begin{lstlisting}[language=Python, caption=An example Python program that update and modify quantum states., label=alg_mod]
from pyqpanda3.vqcircuit import VQCircuit
from pyqpanda3.core import X,RX,RY
vqc = VQCircuit()
vqc.set_Param([3])
vqc << RX(0,vqc.Param([0],'gama'))
vqc << RY(1,vqc.Param('gama'))
\end{lstlisting}

\subsubsection{Reusing Variational Quantum Circuit Structures}

QPanda3 supports constructing variational quantum circuits 
\( U(\theta) = U_{n-1} \dotsb U_{0} \) using variational quantum circuits \( U^{(i)}(\theta^{(i)}) \) 
and non-parameterized quantum circuits \( U^{(j)} \), where \( U_{k} = U^{(k)}(\theta_{k}) \) 
or \( U_{k} = U^{(k)} \), \( k \in \{0, n-1\} \). A simple example is constructing 
\( U(\theta) = U^{(0)}(\theta_{0}) \) from \( U^{(0)}(\theta^{(0)}) \). Note that in this 
symbolic construction process, the formal parameter \( \theta^{(0)} \) must be replaced 
with the formal parameter \( \theta_{0} \), where \( \theta_{0} \subseteq \theta \) 
or \( \theta_{0} \in \theta \). In fact, the only difference between 
\( U^{(0)}(\theta^{(0)}) \) and \( U(\theta) \) lies in their formal parameters. This is 
also reflected in their circuit structures-the types of quantum gates and their 
interconnection patterns are identical, except that the variable parameters of the gates 
are determined by different formal parameters. QPanda3 uses the member method \lstinline{append} 
of \lstinline{VQCircuit} to achieve such structural reuse and formal parameter substitution. The 
input parameter \lstinline{sub_vqc} corresponds to the variational quantum circuit 
(e.g., $ U^{(0)}(\theta^{(0)}) $) whose structure is being reused. The input 
parameter \lstinline{placeholder_map} is used to specify the substitution relationship between 
formal parameters. \lstinline{placeholder_map} is a list of Python tuples, where the first element 
of each tuple corresponds to an element of $ \theta^{(k)} $, and the second element 
corresponds to an element of $ \theta $.

A concrete example is shown in the code. Since this corresponds to a single-qubit system, 
the distinction and explanation of qubit indices can be omitted here. For clarity, we 
denote the variational quantum circuit obtained by 
\lstinline{get_vqc1} as $U^{{1}}(\theta^{(1)})=RX(\theta_{1}^{(1)}) \cdot X,\quad \theta^{(1)} = (\theta^{(1)}_{0},\theta^{(1)}_{1})$, 
and the variational quantum circuit obtained by 
\lstinline{get_vqc2} as $U^{{2}}(\theta^{(2)})= RZ(\theta_{2}^{{(2)}}) \cdot U^{{1}}([\theta_{0}^{{2}},\theta_{1}^{{2}}]),\quad \theta^{(2)} = (\theta^{(2)}_{0},\theta^{(2)}_{1},\theta^{(2)}_{2})$. Thus, $U^{{2}}(\theta^{(2)})=RZ(\theta_{2}^{{(2)}}) \cdot RX(\theta_{1}^{(2)}) \cdot X$.

Although this code below is relatively simple, it demonstrates how reusing 
variational quantum circuit structures can better achieve modularity in program code, 
improve code reusability, and facilitate the construction and maintenance of larger-scale 
program systems.

\begin{lstlisting}[language=Python, caption=An example Python program that update and modify quantum states., label=alg_mod]
from pyqpanda3.vqcircuit import VQCircuit
from pyqpanda3.core import QCircuit,X,RX,RZ
 
def get_vqc1():
  vqc = VQCircuit()
  vqc.set_Param([2])
  vqc << X(0)
  vqc << RX(0,vqc.Param([1]))
  return vqc
def get_vqc2():
  vqc2 = VQCircuit()
  vqc2.set_Param([3])
  vqc1= get_vqc1()
  vqc2.append(
    vqc1
    ,[
      (vqc1.Param([0]),vqc2.Param([0]))
      ,(vqc1.Param([1]),vqc2.Param([1]))
    ]
    )
  vqc2 << RZ(2,vqc2.Param([2]))
  return vqc2

vqc = get_vqc2()
\end{lstlisting}

\subsubsection{Expressions as Parameters for Parameterized Quantum Gates}

QPanda3 supports the construction of variational quantum circuits in the form of formula 
\lstinline{U_F}. Here, the expression refers to one derived from the formal parameter 
$\theta$ through mathematical operations. Although this formula does not explicitly mention 
subcircuits without variable parameters, it can clearly be easily extended to variational 
quantum circuits containing subcircuits without variable parameters. It should be noted that 
this case must also satisfy the requirement of deep binding between formal parameters and 
the variational quantum circuit. Therefore, an expression should not be composed of formal 
parameters (or their elements) from multiple variational quantum circuits. Otherwise, when 
such an expression serves as a gate parameter in a variational quantum circuit, it would 
violate this constraint.

Currently, for expressions serving as parameters of parameterized quantum gates, QPanda3 
supports addition, multiplication, and scalar multiplication operations, providing functionality 
for expression evaluation and expression printing. The following code demonstrates the basic 
usage. For the formal parameter $\theta=(p_0,p_1)$ of the variational quantum circuit in the 
code, the expression is $e=3.14p_0 p_1+p_1+4$.

\begin{lstlisting}[language=Python, caption=An example Python program that update and modify quantum states., label=alg_mod]
from pyqpanda3.vqcircuit import VQCircuit
from pyqpanda3.core import RX
vqc = VQCircuit()
vqc.set_Param([2])
p0 = vqc.Param([0],'p0')
p1 = vqc.Param([1],'p1')
e = 3.14*p0*p1+p1+4
print('e:\n',e)
vqc << RX(0,e)
\end{lstlisting}

The corresponding output is:

\begin{lstlisting}[language=Python, caption=An example Python program that update and modify quantum states., label=alg_mod]
e:
 ((3.14*p0*p1+p1)+4)
\end{lstlisting}

\subsubsection{Displaying the Structure of Variational Quantum Circuits}

Currently, QPanda3 provides functionality to display the structure of variational 
quantum circuits in a format similar to OriginIR \cite{QPanda3-Tutorial-OriginIR}, 
which is implemented through the member method \lstinline{display_ansatz} of 
VQCircuit. The output of this method, as observed from the code results, excludes 
the qubit and classical bit declarations required by OriginIR but otherwise largely 
aligns with OriginIR's format except for variable quantum gate parameters. For 
parameter positioning, variable parameters follow the same convention as fixed 
parameters. In terms of output format, fixed parameters are displayed as numerical 
values, while variable parameters are represented by identifier strings. These identifier 
strings for variable parameters use assigned names if available for specific elements 
of the formal parameter; otherwise, they adopt a fixed format "Param(list of indices of 
the element in $\theta$)" to represent the element.

\begin{lstlisting}[language=Python, caption=An example Python program that update and modify quantum states., label=alg_mod]
from pyqpanda3.vqcircuit import VQCircuit
from pyqpanda3.core import RX,RZ
vqc = VQCircuit()
vqc.set_Param([2,2])
p0 = vqc.Param([0,0],'p0')
p1 = vqc.Param([0,1],'p1')
e = (3.14*p0*p1+p1+4)*vqc.Param([1,1])
vqc << RX(0,e)
vqc << RZ(0,vqc.Param([1,0]))
print('vqc:\n')
vqc.display_ansatz()
\end{lstlisting}

The corresponding output is:

\begin{lstlisting}[language=Python, caption=An example Python program that update and modify quantum states., label=alg_mod]
vqc:

RX q[0],(((3.14*p0*p1+p1)+4)*Param([1,1]))
RZ q[0],(Param([1,0]))
\end{lstlisting}

\subsection{Obtaining Quantum States, Expectations, and Gradients}

\subsubsection{Acquiring Parameterized Quantum States for Given Parameters}

The construction of parameterized quantum states (PQS) signifies a profound coupling 
between classical optimization and quantum dynamics. The quantum state preparation 
achieved through the unitary transformation sequence 
$|\psi(\boldsymbol{\theta})\rangle = \prod_k U_k(\theta_k)|0\rangle^{\otimes n}$ 
establishes a continuously tunable quantum state manifold in Hilbert space. This 
construction enables precise characterization of multi-electron wavefunctions in quantum 
chemistry, particularly for resolving the entangled electronic structures in strongly 
correlated systems, surpassing the computational limitations of traditional configuration 
interaction methods \cite{higgott2019variational}. In combinatorial optimization, PQS 
transforms discrete optimization problems into continuous energy landscapes within quantum 
state space, offering novel paradigms for solving NP-hard problems such as Max-Cut 
\cite{goemans1995improved}\cite{farhi2014quantum} and graph coloring 
\cite{guerreschi2017practical} \cite{farhi2014quantum}. Theoretical studies reveal that the 
entanglement hierarchy and expressibility of quantum circuits jointly constitute sufficient 
conditions for quantum advantage, where the coverage density of quantum states in the target 
Hilbert space directly influences algorithmic convergence rates \cite{sim2019expressibility}. 
Recent advances demonstrate that synergistic design of neural quantum states and variational 
quantum circuits can further enhance the representational capacity of quantum states for 
complex systems \cite{gong2024quantum}.

In QPanda3, a concrete quantum circuit represented by QCircuit can be obtained by assigning 
actual values to the formal parameter $\theta$, and the parameterized quantum state with 
given parameter values can be acquired by evolving the resulting quantum circuit. An example 
usage is provided in the code (Appendix \ref{sec:cal_parametrized_qstate}).

\subsubsection{Computing Hamiltonian Expectation Value}

The observable properties of quantum systems are classically extracted through Hamiltonian 
expectation values 
$\langle H\rangle_{\boldsymbol{\theta}} = \langle\psi(\boldsymbol{\theta})|H|\psi(\boldsymbol{\theta})\rangle$, 
an operation that establishes the fundamental connection between quantum information and 
macroscopic physical quantities. In condensed matter physics, this expectation value directly 
corresponds to core physical properties such as order parameters and correlation functions, 
serving as a diagnostic tool for cutting-edge research on high-temperature superconductivity 
and quantum spin liquids. In quantum chemistry, leveraging reduced density matrix theory, the 
expectation value calculation of molecular Hamiltonians 
$H = \sum_{pq}h_{pq}a_p^\dagger a_q + \frac{1}{2}\sum_{pqrs}g_{pqrs}a_p^\dagger a_q^\dagger a_s a_r$ 
is transformed into predictions of electronic structure properties, enabling sub-chemical 
accuracy simulations of reaction pathway energy barriers \cite{cao2019quantum}. In financial 
derivatives pricing, by constructing a quantum version of the Black-Scholes equation, expectation 
value calculations provide an exponentially accelerated solution for high-dimensional path 
integrals \cite{egger2020credit}. The theoretical significance of this operation is further 
highlighted in characterizing the quantum-classical boundary, where its computational complexity 
serves as a critical metric for delineating classes of quantum-solvable problems 
\cite{preskill2018quantum}.

QPanda3 provides the functionality to construct Hamiltonian operators using linear combinations 
of Pauli operations.

\begin{lstlisting}[language=Python, caption=An example Python program that update and modify quantum states., label=alg_mod]
from pyqpanda3.hamiltonian import Hamiltonian # Import the corresponding package name
# Represent a linear combination of Pauli operations as a list of tuples
# The first tuple indicates that the coefficient of this term is 12.36525580995888 + 14.85172018664403j, 
# and the operation applies Pauli Y on the qubit with index 0 and Pauli X on the qubit with index 1
paulis = [
        ('YX', [0, 1], (12.36525580995888 + 14.85172018664403j)),  # (Pauli basis, qubit indices on which the Pauli basis acts, coefficient)
        ('YY', [0, 1], (12.920260765526914 + 26.29613065236354j))  # The qubit indices used should be consistent with those used in the quantum circuit
    ]
ham = Hamiltonian(paulis) # ham is the constructed Hamiltonian operator

\end{lstlisting}

In QPanda3, two methods are available for computing Hamiltonian expectations, corresponding 
to \lstinline{get_hamiltonian_expectation} and \lstinline{get_hamiltonian_expectation2} in 
the code (Appendix \ref{sec:cal_ham}).

\subsubsection{Computing Gradient Values}

In the field of quantum computing, the concept of gradient extends from the classical rate 
of function change to the sensitivity of quantum states to parameterized operations
\cite{mitarai2018quantum}, with its general form defined as the partial derivative of the 
measurement expectation value $\left \langle H_{\theta } \right \rangle $ with respect to 
circuit parameter 
$\theta$, $\nabla \langle \psi ( \theta ) | H  | \psi ( \theta ) \rangle$\cite{schuld2019evaluating}. 
This definition incorporates the unitary constraints of quantum evolution and the 
characteristics of non-commuting operators, introducing statistical uncertainty from 
quantum measurements compared to classical gradients\cite{cerezo2021cost}, serving as 
the optimization engine for variational quantum algorithms. The significance of variational 
quantum gradient computation manifests in multiple dimensions: it directly determines 
algorithmic feasibility, for example, reducing convergence iterations by 60\% when 
overcoming barren plateaus\cite{cerezo2021barren}; in quantum machine learning, it supports 
weight updates for tasks like image classification, improving accuracy by 
23\%\cite{havlivcek2019supervised}; simultaneously, it advances the practicality of NISQ 
devices, expanding solvable problem scales by 10 times\cite{preskill2018quantum}. Quantum 
gradient computation methods have evolved into four mainstream techniques: finite difference 
estimates gradients through minor parameter perturbations\cite{sweke2020stochastic}, simple 
but noise-amplifying and requiring $O(k)$ circuit evaluations; parameter shift rules leverage 
quantum gate properties for analytical solutions\cite{mitarai2018quantum}\cite{schuld2019evaluating}
\cite{banchi2021measuring}, more precise but requiring $O(2k)$ executions; automatic 
differentiation propagates gradients backward through computational graphs in hybrid 
frameworks\cite{bergholm2018pennylane}, $O(1)$ efficient but constrained by quantum no-cloning
\cite{wootters1982single}; adjoint methods\cite{jones2020efficient} enable efficient gradient 
computation through quantum state operations.

The adjoint method leverages the time-reversal symmetry in quantum mechanics to compute 
gradients through the synergy of forward evolution and adjoint evolution: first execute 
the parameterized circuit $U(\theta)$ to generate the final state, then apply the Hamiltonian 
$H$ and reversely execute $U^{\dagger}(\theta)$ to revert to the initial state. The core 
advantage of this method lies in its revolutionary improvement in resource efficiency-regardless 
of the parameter scale, only a constant 2 circuit executions (1 forward + 1 adjoint) are 
required to obtain the full gradient.

The set of rotation gates $G=\{RX,RY,RZ,CRX,CRY,CRZ\}$ constitutes the core parameterized 
gates for variational quantum circuits. In variational quantum circuits within QPanda3, these 
rotation gates can all function as parameterized quantum gates.

Currently, QPanda3 combines classical backpropagation and the adjoint method to support 
gradient computation for variational quantum circuits where expressions serve as variable 
parameters. A complete application example is shown in the code (Appendix \ref{sec:cal_gradient}).

\subsection{Application Examples}

A QAOA application example is provided in Appendix \ref{sec:qaoa_example_max_cut}.

\section{Obtaining Parameterized Quantum States for Given Parameter Values}
\label{sec:cal_parametrized_qstate}

\begin{lstlisting}[language=Python, caption=An example Python program that update and modify quantum states., label=alg_mod]
from pyqpanda3.vqcircuit import VQCircuit
from pyqpanda3.core import QCircuit,X,RX,RY,Y,CPUQVM,QProg
def get_vqc():
    # Prepare the variational quantum circuit U(theta)
    vqc = VQCircuit()
    vqc.set_Param([2])  # 1> Agree that the parameter theta is a one-dimensional vector with a length of 2, 2> The parameter theta is bound to the variational quantum circuit U(theta) and cannot be used for other variational quantum circuits
    vqc << RX(0, vqc.Param([0]))  # Add a parameterized quantum gate RX, whose parameter corresponds to the first element of the vector theta
    vqc << RY(1, vqc.Param([1]))
    return vqc
def get_param_val():
    # Prepare the actual numerical values thetaval for the parameter theta
    param_val = [5.14, 6.14]  # The number of elements should be consistent with the number of elements in the parameter theta (vector)
    return param_val
 
def get_qstate(vqc, param_val):
    # Obtain |psi(thetaval))
    cir = vqc(param_val).at([0])  # Obtain U(thetaval)
    qvm = CPUQVM()
    qvm.run(QProg(cir), 1)  # Perform the evolution
    stv = qvm.result().get_state_vector()
    return stv
 
vqc = get_vqc()  # Prepare the variational quantum circuit U(theta)
param_val = get_param_val()  # Prepare the actual numerical values thetaval for the parameter theta
stv = get_qstate(vqc, param_val)  # Obtain |psi(thetaval))
print('|psi(thetaval)):',stv)
\end{lstlisting}

\section{Example Code for Computing Hamiltonian Expectation}
\label{sec:cal_ham}

\begin{lstlisting}[language=Python, caption=An example Python program that update and modify quantum states., label=alg_mod]
from pyqpanda3.vqcircuit import VQCircuit,DiffMethod
from pyqpanda3.core import QCircuit,X,RX,RY,Y,CPUQVM,QProg
from pyqpanda3.hamiltonian import Hamiltonian
def get_vqc():
    # Prepare the variational quantum circuit U(theta)
    vqc = VQCircuit()
    vqc.set_Param([2])  # 1> Agree that the parameter theta is a one-dimensional vector with a length of 2. 2> The parameter theta is bound to the variational quantum circuit U(theta) and cannot be used for other variational quantum circuits.
    vqc << RX(0, vqc.Param([0]))  # Add a parameterized quantum gate RX, whose parameter corresponds to the first element of the vector theta
    vqc << RY(1, vqc.Param([1]))
    return vqc
 
def get_param_val():
    # Prepare the actual numerical values thetaval for the parameter theta
    param_val = [2.14, 3.14]  # The number of elements should be consistent with the number of elements in the parameter theta (vector)
    return param_val
 
def get_hamiltonian():
    # Prepare the Hamiltonian operator H
    paulis = [
        ('YY', [0, 1], (12.36525580995888 + 14.85172018664403j)),  # (Pauli basis, qubit indices on which the Pauli basis acts, coefficient)
        ('YX', [0, 1], (12.920260765526914 + 26.29613065236354j))  # The qubit indices used should be consistent with those used in the quantum circuit
    ]
 
    return Hamiltonian(paulis)
  
def get_hamiltonian_expectation(vqc, param_val, ham):
    # Obtain the Hamiltonian expectation value (Method 1)
    res = vqc(param_val).expval_hamiltonian(
        ham
        , [0]
        , used_threads=4
        , backend='CPU'
        )
    return res
 
def get_hamiltonian_expectation2(vqc: VQCircuit, param_val, ham):
    # Obtain the Hamiltonian expectation value (Method 2)
    # The interface get_gradients_and_expectation can obtain both the gradient values and the expectation values simultaneously. Here, only the expectation value is returned.
    res = vqc.get_gradients_and_expectation(
        params=param_val
        , observable=ham
        , diff_method=DiffMethod.ADJOINT_DIFF
        ).expectation_val()
    return res
 
vqc = get_vqc()  # Prepare the variational quantum circuit U(theta)
param_val = get_param_val()  # Prepare the actual numerical values thetaval for the parameter theta
ham = get_hamiltonian()  # It is recommended not to use 'H' as the variable name to avoid naming conflicts with the H gate.
ham_expectation = get_hamiltonian_expectation(
    vqc
    , param_val
    , ham
    )  # Obtain the Hamiltonian expectation value (Method 1)
ham_expectation2 = get_hamiltonian_expectation2(
    vqc
    , param_val
    , ham
    )  # Obtain the Hamiltonian expectation value (Method 2)
 
print('ham_expectation:',ham_expectation)
print('ham_expectation2:',ham_expectation2)
\end{lstlisting}

\section{Example Code for Computing Gradients of Variational Quantum Circuits}
\label{sec:cal_gradient}

\begin{lstlisting}[language=Python, caption=An example Python program that update and modify quantum states., label=alg_mod]

from pyqpanda3.vqcircuit import VQCircuit,DiffMethod
from pyqpanda3.core import QCircuit,X,RX,RY,Y,CPUQVM,QProg
from pyqpanda3.hamiltonian import Hamiltonian
 
def get_vqc():
    # Prepare the variational quantum circuit U(theta)
    vqc = VQCircuit()
    vqc.set_Param([2])  # 1> Agree that the parameter theta is a one-dimensional vector with a length of 2. 2> The parameter theta is bound to the variational quantum circuit U(theta) and cannot be used for other variational quantum circuits.
    vqc << RX(0, vqc.Param([0]))  # Add a parameterized quantum gate RX, whose parameter corresponds to the first element of the vector theta
    vqc << RY(1, vqc.Param([1]))
    return vqc
 
def get_param_val():
    # Prepare the actual numerical values thetaval for the parameter theta
    param_val = [2.14, 3.14]  # The number of elements should be consistent with the number of elements in the parameter theta (vector)
    return param_val
 
def get_hamiltonian():
    # Prepare the Hamiltonian operator H
    paulis = [
        ('YY', [0, 1], (12.36525580995888 + 14.85172018664403j)),  # (Pauli basis, qubit indices on which the Pauli basis acts, coefficient)
        ('YX', [0, 1], (12.920260765526914 + 26.29613065236354j))  # The qubit indices used should be consistent with those used in the quantum circuit
    ]
 
    return Hamiltonian(paulis)
 
def get_gradient(vqc, param_val, ham):
    # Obtain the gradient values (Method 1)
    return vqc.get_gradients(
        params=param_val
        , observable=ham
        , diff_method=DiffMethod.ADJOINT_DIFF
        )
 
def get_gradient2(vqc: VQCircuit, param_val, ham):
    # Obtain the gradient values (Method 2)
    # The interface get_gradients_and_expectation can obtain both the gradient values and the expectation values simultaneously.
    return vqc.get_gradients_and_expectation(
        params=param_val
        , observable=ham
        , diff_method=DiffMethod.ADJOINT_DIFF
        )
 
vqc = get_vqc()  # Prepare the variational quantum circuit U(theta)
param_val = get_param_val()  # Prepare the actual numerical values thetaval for the parameter theta
ham = get_hamiltonian()  # It is recommended not to use 'H' as the variable name to avoid naming conflicts with the H gate.
res1 = get_gradient(vqc, param_val, ham)  # Obtain the gradient values (Method 1)
res2 = get_gradient2(vqc, param_val, ham)  # Obtain the gradient values (Method 2)
 
print('res1:',res1)
print('res1 gradient:',res1.gradients())
print('res2:',res2)
print('res2 gradient:',res2.gradients())
\end{lstlisting}

\section{A QAOA Application Example (Max Cut Problem)}
\label{sec:qaoa_example_max_cut}

\subsection{Problem Definition} 
    Given an undirected graph $G=(V,E)$, where: 
    \begin{itemize} 
        \item $V$ is the vertex set ($|V|=n$) 
        \item $E$ is the edge set, with each edge having weight $w_{ij}$ 
        \item Objective: Partition vertices into two mutually exclusive subsets $S$ and $V\setminus S$ to maximize the cut value: 
            $C(S) = \sum_{\substack{(i,j)\in E \\ i\in S, j\notin S}} w_{ij}$ 
        \item Quantum encoding of this problem 
        \begin{itemize} 
            \item Assign qubit to each vertex $i$, with state $|0\rangle$ or $|1\rangle$ indicating subset membership 
            \item Problem Hamiltonian: $H_C = \frac{1}{2}\sum_{(i,j)\in E} w_{ij}(1 - Z_i Z_j)$ 
            \item Minimizing $\langle H_C \rangle$ is equivalent to maximizing cut value 
        \end{itemize} 
    \end{itemize}

\subsection{A Specific Problem Instance (Max Cut Problem for a Square Graph)} 
    \begin{itemize} 
        \item Vertex set V: \{0, 1, 2, 3\} 
        \item Edge set E: see Table \ref{tab:max_cut_E} 
        \begin{table}[H] 
            \centering 
            \caption{Max Cut Edge Set E} 
            \label{tab:max_cut_E} 
            \begin{tabular}{l c r} 
                \hline 
                \textbf{Edge} & \textbf{Weight} & \textbf{Note} \\ 
                \hline 
                (0,1) & w=1.0 & {} \\ 
                (1,2) & w=1.0 & {} \\ 
                (2,3) & w=1.0 & {} \\ 
                (3,0) & w=1.0 & {} \\ 
                (0,2) & w=0.5 & Diagonal \\ 
                \hline 
            \end{tabular} 
        \end{table} 
        \item Theoretical optimal solution: vertex partition \{0,2\} and \{1,3\} 
    \end{itemize}

\subsection{Solution Using VQCircuit} 
\subsubsection{Step List} 
    \begin{enumerate} 
        \item Construct the problem Hamiltonian $H_C$ 
        \item Design the QAOA parameterized quantum circuit 
        \item Optimize parameters using a classical optimizer 
        \item Measure the optimal quantum state to obtain the solution 
    \end{enumerate} 
    
\subsubsection{Example Code Implementation} 
    \begin{lstlisting}[language=Python, caption=An example Python program that evaluates the properties of quantum states., label=alg_char]

from pyqpanda3.hamiltonian import Hamiltonian
from pyqpanda3.vqcircuit import VQCircuit, DiffMethod
from pyqpanda3.core import H, RZ, RX, CNOT, QProg, CPUQVM, measure
import numpy as np
import matplotlib.pyplot as plt
import networkx as nx

# =====================
# 1 Problem definition and Hamiltonian construction
# =====================

# Create graph structure (vertices 0-3)
edges = [
    (0, 1, 1.0),  # (vertex i, vertex j, weight)
    (1, 2, 1.0),
    (2, 3, 1.0),
    (3, 0, 1.0),
    (0, 2, 0.5)  # diagonal
]

# Correctly construct Max Cut Hamiltonian H_C = 1/2 SUM(w_ij (I - Z_i Z_j)) 
ham_terms = []
for i, j, w in edges:
    term = "ZZ"
    # Coefficient should be positive since we need to maximize the cut value
    ham_terms.append((term, [i, j], 0.5 * w))

maxcut_hamiltonian = Hamiltonian(ham_terms)
print("Problem Hamiltonian:", maxcut_hamiltonian)


# =====================
# 2. Construct QAOA quantum circuit
# =====================

def build_qaoa_circuit(num_qubits, p=1):
    """
    Build p-layer QAOA circuit
    :param num_qubits: Number of qubits (vertices)
    :param p: QAOA layers
    :return: VQCircuit object
    """
    vqc = VQCircuit()
    # Set parameters: p layers, 2 parameters per layer (gamma, beta)
    vqc.set_Param([p, 2])

    # Initial Hadamard gates
    for q in range(num_qubits):
        vqc << H(q)

    # QAOA layers
    for layer in range(p):
        # Problem unitary operator U_C(gamma) = e^{-i*gamma*H_C}
        for (i, j, w) in edges:
            gamma = vqc.Param([layer, 0])  # gamma parameter
            vqc << CNOT(i, j) << RZ(j, 2 * gamma * w) << CNOT(i, j)

        # Mixing unitary operator U_B(beta) = e^{-i*beta*SUM(X_i)}
        beta = vqc.Param([layer, 1], f'beta_{layer}')  # beta parameter
        for q in range(num_qubits):
            vqc << RX(q, 2 * beta)  # RX(2*beta) implements e^{-i*beta*X}

    return vqc


# Build 2-layer QAOA circuit
num_vertices = 4
qaoa_layers = 2
vqc = build_qaoa_circuit(
    num_vertices
    , p=qaoa_layers
    )
print("\nQAOA circuit structure:")
vqc.display_ansatz()


# =====================
# 3. Parameter optimization
# =====================

def optimize_parameters(vqc, hamiltonian):
    """
    Optimize QAOA parameters
    :return: Optimized parameter values
    """
    # Random initial parameters
    init_params = np.random.uniform(
            0
            , np.pi
            , size=qaoa_layers * 2
        )

    # Use BFGS optimizer
    from scipy.optimize import minimize

    def objective(params):
        """Objective function: Minimize Hamiltonian expectation"""
        params_2d = params.reshape((qaoa_layers, 2))
        result = vqc.get_gradients_and_expectation(
            params_2d,
            hamiltonian,
            diff_method=DiffMethod.ADJOINT_DIFF
        )
        return result.expectation_val(), np.array(result.gradients()).flatten()

    # Optimization process
    res = minimize(
        objective,
        init_params,
        jac=True,  # Use gradient information
        method='L-BFGS-B',
        bounds=[(0, 2 * np.pi)] * len(init_params),
        options={'maxiter': 50}
    )

    if res.success:
        print(f"\nOptimization successful! Expectation value: {res.fun:.4f}")
        return res.x.reshape((qaoa_layers, 2))
    else:
        raise RuntimeError("Parameter optimization failed: " + res.message)


# Execute parameter optimization
optimal_params = optimize_parameters(vqc, maxcut_hamiltonian)
print("Optimal parameter values:")
print(optimal_params)


# =====================
# 4. Result measurement and analysis
# =====================

def measure_solution(vqc, params):
    """Measure optimal quantum state to obtain solution"""
    # Get optimized quantum circuit
    prog = QProg()
    prog << vqc(params).at([0])

    # Quantum virtual machine execution
    qvm = CPUQVM()

    # Add measurement gates
    for i in range(num_vertices):
        prog << measure(i, i)

    # Sample 1000 times
    qvm.run(prog, shots=5000)
    results = qvm.result().get_prob_dict()

    # Parse results
    solutions = []
    for bitstr, prob in results.items():
        bitstr_full = bitstr.zfill(num_vertices)
        # Reverse bit string to match vertex order
        bitstr_full = bitstr_full[::-1]

        cut_value = 0
        # Calculate cut value
        for i, j, w in edges:
            if bitstr_full[i] != bitstr_full[j]:
                cut_value += w
        solutions.append((bitstr_full, cut_value, prob))

    # Sort by cut value
    solutions.sort(key=lambda x: x[1], reverse=True)
    return solutions


# Get and print results
solutions = measure_solution(
    vqc
    , optimal_params
    )
print("\nTop 5 solutions:")
print("Bitstring | Cut Value | Probability")
print("-----------------------------------")
for bitstr, cut_val, prob in solutions[:5]:
    print(f"{bitstr}   | {cut_val:.1f}       | {prob:.4f}")

# =====================
# 5. Visualization of results
# =====================

# Extract cut value distribution
tmp_dict = {}
for _,cut_val,prob in solutions:
    if cut_val in tmp_dict.keys():
        tmp_dict[cut_val]+=prob
    else:
        tmp_dict[cut_val]=prob
cut_values = []
probabilities = []

for val,prob in tmp_dict.items():
    cut_values.append(val)
    probabilities.append(prob)

plt.figure(figsize=(10, 6))
plt.bar(cut_values, probabilities, width=0.1)
plt.xlabel('Cut Value')
plt.ylabel('Probability')
plt.title('Max Cut Solution Distribution')
plt.xticks(sorted(set(cut_values)))
plt.grid(alpha=0.2)
plt.savefig('cut_distribution.png')  # Save image
plt.show()

# Visualize optimal solution
optimal_solution = solutions[0][0]
# Assign colors using reversed bit string
node_colors = ['red' if bit == '0' else 'blue' for bit in optimal_solution]

# Draw graph structure
G = nx.Graph()
G.add_edges_from([(i, j, {'weight': w}) for i, j, w in edges])

pos = nx.spring_layout(G)
nx.draw(
    G
    , pos
    , node_color=node_colors
    , with_labels=True
    ,node_size=800
    , font_size=16
    , edge_color='gray'
    )
edge_labels = {(i, j): f"w={w}" for i, j, w in edges}
nx.draw_networkx_edge_labels(
    G
    , pos
    , edge_labels=edge_labels
    )
plt.title(f'Optimal Cut: {solutions[0][1]}')
plt.savefig('optimal_cut.png')  # Save image
plt.show()

\end{lstlisting}

\subsection{Explanation of Running Results} 

\subsubsection{One Run Result with 700 shots}

\begin{lstlisting}[language=Python, caption=An example Python program that update and modify quantum states., label=alg_mod]
Problem Hamiltonian: { qbit_total = 4, pauli_with_coef_s = { 'Z0 Z1 ':0.5 + 0j, 'Z1 Z2 ':0.5 + 0j, 'Z2 Z3 ':0.5 + 0j, 'Z0 Z3 ':0.5 + 0j, 'Z0 Z2 ':0.25 + 0j, } }

QAOA circuit structure:


H q[0]
H q[1]
H q[2]
H q[3]
CNOT q[0],q[1]
RZ q[1],(1*2*Param([0,0]))
CNOT q[0],q[1]
CNOT q[1],q[2]
RZ q[2],(1*2*Param([0,0]))
CNOT q[1],q[2]
CNOT q[2],q[3]
RZ q[3],(1*2*Param([0,0]))
CNOT q[2],q[3]
CNOT q[3],q[0]
RZ q[0],(1*2*Param([0,0]))
CNOT q[3],q[0]
CNOT q[0],q[2]
RZ q[2],(0.5*2*Param([0,0]))
CNOT q[0],q[2]
RX q[0],(2*beta_0)
RX q[1],(2*beta_0)
RX q[2],(2*beta_0)
RX q[3],(2*beta_0)
CNOT q[0],q[1]
RZ q[1],(1*2*Param([1,0]))
CNOT q[0],q[1]
CNOT q[1],q[2]
RZ q[2],(1*2*Param([1,0]))
CNOT q[1],q[2]
CNOT q[2],q[3]
RZ q[3],(1*2*Param([1,0]))
CNOT q[2],q[3]
CNOT q[3],q[0]
RZ q[0],(1*2*Param([1,0]))
CNOT q[3],q[0]
CNOT q[0],q[2]
RZ q[2],(0.5*2*Param([1,0]))
CNOT q[0],q[2]
RX q[0],(2*beta_1)
RX q[1],(2*beta_1)
RX q[2],(2*beta_1)
RX q[3],(2*beta_1)

Optimization successful! Expectation value: -0.3581
Optimal parameter values:
[[3.68782485 2.28451761]
 [0.68693742 1.32432699]]

Top 5 solutions:
Bitstring&Cut Value&Probability
-----------------------------------
1010  &4.0      &0.1400
0101  &4.0      &0.1443
1000  &2.5      &0.0986
1100  &2.5      &0.0100
0010  &2.5      &0.0814
    
\end{lstlisting}

\begin{figure*}[htbp]
    \centering
    \includegraphics[width=0.8\textwidth]{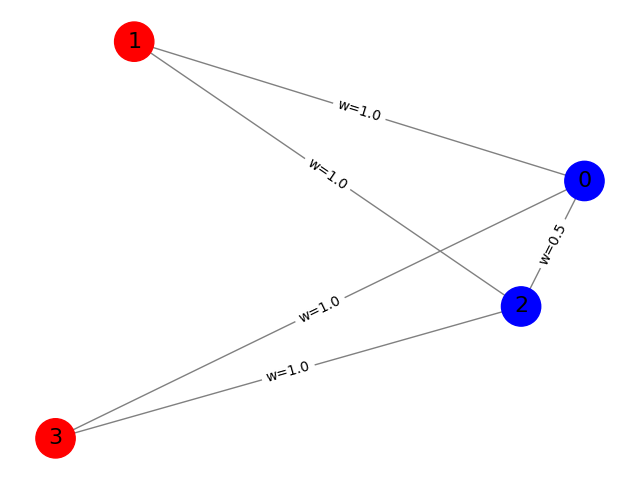} 
    \caption{Optimal Cut Result for 700 Shots}
    \label{fig:optimal_cut_700}    
\end{figure*}

\begin{figure*}[htbp]
    \centering
    \includegraphics[width=0.8\textwidth]{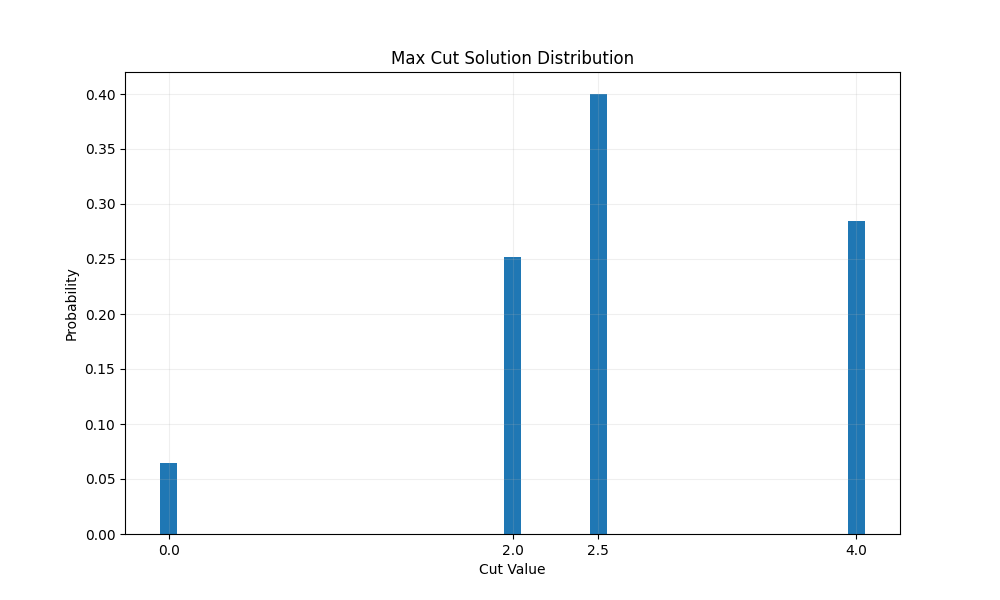}
    \caption{Optimal Cut Distribution for 700 Shots} 
    \label{fig:cut_distribution_700}   
\end{figure*}

\subsubsection{One Run Result with 5000 shots}

\begin{lstlisting}[language=Python, caption=An example Python program that update and modify quantum states., label=alg_mod]
Problem Hamiltonian: { qbit_total = 4, pauli_with_coef_s = { 'Z0 Z1 ':0.5 + 0j, 'Z1 Z2 ':0.5 + 0j, 'Z2 Z3 ':0.5 + 0j, 'Z0 Z3 ':0.5 + 0j, 'Z0 Z2 ':0.25 + 0j, } }

QAOA circuit structure:
H q[0]
H q[1]
H q[2]
H q[3]
CNOT q[0],q[1]
RZ q[1],(1*2*Param([0,0]))
CNOT q[0],q[1]
CNOT q[1],q[2]
RZ q[2],(1*2*Param([0,0]))
CNOT q[1],q[2]
CNOT q[2],q[3]
RZ q[3],(1*2*Param([0,0]))
CNOT q[2],q[3]
CNOT q[3],q[0]
RZ q[0],(1*2*Param([0,0]))
CNOT q[3],q[0]
CNOT q[0],q[2]
RZ q[2],(0.5*2*Param([0,0]))
CNOT q[0],q[2]
RX q[0],(2*beta_0)
RX q[1],(2*beta_0)
RX q[2],(2*beta_0)
RX q[3],(2*beta_0)
CNOT q[0],q[1]
RZ q[1],(1*2*Param([1,0]))
CNOT q[0],q[1]
CNOT q[1],q[2]
RZ q[2],(1*2*Param([1,0]))
CNOT q[1],q[2]
CNOT q[2],q[3]
RZ q[3],(1*2*Param([1,0]))
CNOT q[2],q[3]
CNOT q[3],q[0]
RZ q[0],(1*2*Param([1,0]))
CNOT q[3],q[0]
CNOT q[0],q[2]
RZ q[2],(0.5*2*Param([1,0]))
CNOT q[0],q[2]
RX q[0],(2*beta_1)
RX q[1],(2*beta_1)
RX q[2],(2*beta_1)
RX q[3],(2*beta_1)

Optimization successful! Expectation value: -0.3581
Optimal parameter values:
[[1.20130278 0.61696788]
 [1.97099634 2.90861432]]

Top 5 solutions:
Bitstring&Cut Value&Probability
-----------------------------------
1010  &4.0      &0.4126
0101  &4.0      &0.4328
1000  &2.5      &0.0050
1100  &2.5      &0.0312
0010  &2.5      &0.0054
\end{lstlisting}

\begin{figure*}[htbp]
    \centering
    \includegraphics[width=0.8\textwidth]{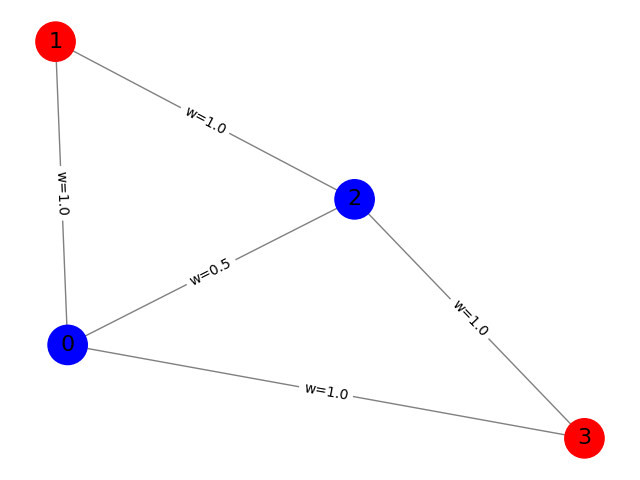}
    \caption{Optimal Cut Result for 5000 Shots}
    \label{fig:optimal_cut_5000} 
\end{figure*}

\begin{figure*}[htbp]
    \centering
    \includegraphics[width=0.8\textwidth]{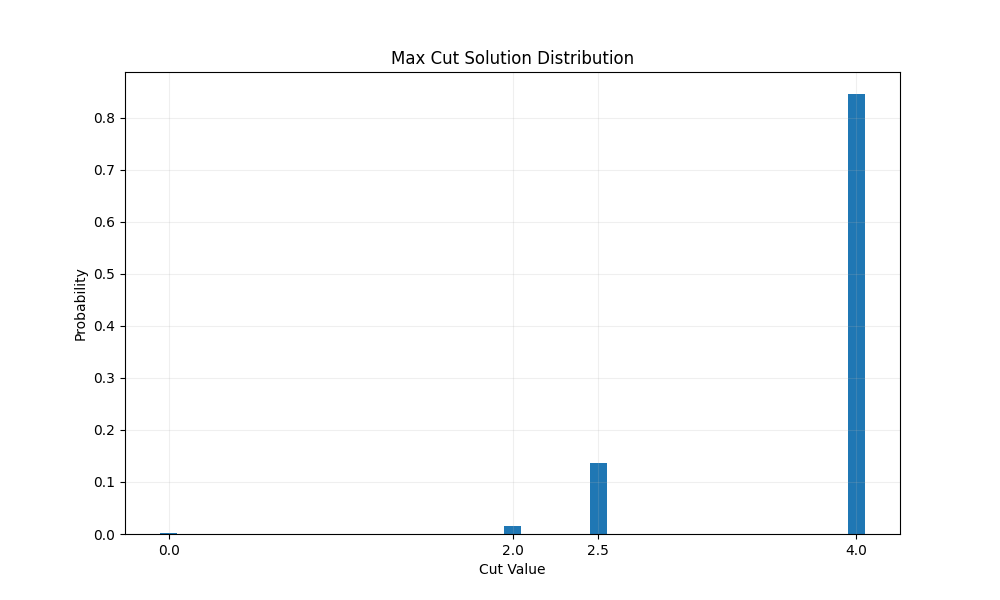}
    \caption{Optimal Cut Distribution for 5000 Shots}
    \label{fig:cut_distribution_5000}
\end{figure*}

The theoretical formula $H_C = \frac{1}{2}\sum_{(i,j)\in E} w_{ij}(I - Z_i Z_j)$ is exactly 
equivalent to the code implementation. The theoretical formula can be decomposed into a 
constant term $\frac{1}{2}\sum w_{ij}I$ and a ZZ operator term $-\frac{1}{2}\sum w_{ij}Z_i Z_j$, 
where the constant term is reasonably omitted in quantum optimization because it only shifts 
the energy value globally without changing the relative relationships of the eigenstates. 
In the code, the ZZ term $\frac{1}{2}w_{ij}Z_i Z_j$ is implemented with a positive coefficient 
$0.5 * w$, which is equivalent to the negative term $-\frac{1}{2}w_{ij}Z_i Z_j$ in the 
theoretical formula, because minimizing the expectation value of $\frac{1}{2}w_{ij}Z_i Z_j$ 
is equivalent to maximizing the theoretical objective.

\begin{table}[h]  
    \centering          
    \caption{Correspondence between Vertex States and Energy Contributions} 
    \label{tab:max_cut_res_analysis_1}   
    \begin{tabular}{p{2cm} p{0.3cm} p{2cm} p{2.7cm}}  
        \hline
        \textbf{Vertex Relationship}&$Z_i Z_j$&\textbf{Theoretical Cut Contribution}&\textbf{Code Hamiltonian Contribution}\\
        \hline
        Same Side&$+1$&0&$+\frac{1}{2}w_{ij}$\\
        Opposite Sides&$-1$&$w_{ij}$&$-\frac{1}{2}w_{ij}$\\
        \hline 
    \end{tabular}
\end{table}

Physically, essentially, when vertices are on opposite sides (cut contribution $w_{ij}$), $Z_i Z_j = -1$ 
causes the code term $\frac{1}{2}w_{ij}Z_i Z_j$ to become $-\frac{1}{2}w_{ij}$ (negative value); 
when on the same side (no contribution), it yields a positive value $\frac{1}{2}w_{ij}$. 
Therefore, minimizing the expectation value of the code Hamiltonian naturally drives 
the system toward the state with maximum cut contribution, which is completely consistent 
with the theoretical objective. The key code implementation is as follows:

\begin{lstlisting}[language=Python, caption=An example Python program that update and modify quantum states., label=alg_mod]
for i, j, w in edges:
    ham_terms.append(("ZZ", [i, j], 0.5 * w))  # Positive coefficient achieves physical equivalence to the negative term
\end{lstlisting}

Using the QAOA algorithm, we successfully solved the maximum cut problem for a 4-vertex weighted 
graph featuring a ring structure (vertices 0-1-2-3-0) with an additional diagonal edge (0-2). 
The ring edges carried a weight of 1.0, while the diagonal edge had a weight of 0.5. Two independent 
experiments demonstrated that the algorithm consistently identified the theoretical maximum cut 
value of 4.0, corresponding to two optimal solutions: `1010` and `0101`. These solutions represent 
the optimal partitioning scheme where vertices are divided into groups {0,2} and {1,3}. Increasing 
the number of quantum measurements (shots) significantly enhanced result precision, raising the 
probability of obtaining the optimal solution from 29.7\% to 82.58\%, thereby verifying the 
probabilistic nature of quantum algorithms.

\begin{table}[h]  
    \centering          
    \caption{QAOA Experimental Results Comparison} 
    \label{tab:max_cut_res_analysis_2}   
    \begin{tabular}{p{2cm} p{1cm} p{1cm} p{3cm}}  
        \hline
        \textbf{Evaluation Metric} & \textbf{Shots=700} & \textbf{Shots=5000} & \textbf{Change Analysis} \\ 
        \hline 
        Expectation Value     & -0.3581           & -1.5075           & More negative values indicate better optimization      \\
        Optimal Parameters  & [[3.69 2.28],[0.69 1.32]] & [[1.20 0.62],[1.97 2.91]] & Multiple equivalent optima in parameter space    \\
        Total Optimal Solution Probability & 28.4\%             & 84.54\%            & Increased sampling enhances accuracy        \\
        Suboptimal Solution Probability & 40.0\%               & 15\%                & Probability distribution concentrates on optima        \\
        Optimal Cut Value & 4.0               & 4.0               & Consistently achieves theoretical optimum              \\
        \hline 
    \end{tabular}
\end{table}

As shown in Table \ref{tab:max_cut_res_analysis_2}, different parameter sets consistently converge 
to the same optimal solution, demonstrating the algorithm's robustness. Increasing shots from 700 
to 5000 elevates the probability of obtaining the optimal solution from 28.4\% to 84.54\% (row 3 
of Table \ref{tab:max_cut_res_analysis_2}), consistent with the law of large numbers. The diversity 
in parameter space (row 2 of Table \ref{tab:max_cut_res_analysis_2}) reveals multiple equivalent 
optimal parameter combinations in QAOA, all leading to identical optimal solutions. These results 
fully validate QAOA's effectiveness for combinatorial optimization problems, achieving the theoretical 
optimum at p=2 layers with over 80\% probability and demonstrating quantum computing's potential 
for optimization challenges.

\end{appendices}

\end{document}